\documentclass{aa}
\usepackage{longtable}
\usepackage{multirow}
\usepackage{booktabs,fixltx2e}
\usepackage[flushleft]{threeparttable}
\usepackage{graphicx,natbib,txfonts,color}

\begin{document}

\title{Detailed homogeneous abundance studies of 14 Galactic \textit{s}-process enriched post-AGB stars: In search of lead (Pb).
\thanks{Based on observations collected with the Very Large Telescope
  at the ESO Paranal Observatory (Chili) of programme numbers 066.D-0171, 073.D-0241 and 094.D-0067.}
\thanks{Based on observations made with the Mercator Telescope, operated on the 
island of La Palma by the Flemish Community, at the Spanish Observatorio del Roque de los Muchachos 
of the Instituto de Astrofísica de Canarias.} }

\author{K. De Smedt\inst{1}
\and H. Van Winckel\inst{1}
\and D. Kamath\inst{1}
\and L. Siess\inst{2}
\and S. Goriely\inst{2}
\and A. I. Karakas\inst{3}
\and R. Manick\inst{1}
}

\offprints{H. Van Winckel, Hans.VanWinckel@ster.kuleuven.be}

\institute{ Instituut voor Sterrenkunde, KU Leuven, Celestijnenlaan 200D,
B-3001 Leuven, Belgium
\and Institut d'Astronomie et d'Astrophysique, Universit\'{e} Libre de Bruxelles, ULB, CP 226, 1050 Brussels, Belgium
\and Research School of Astronomy and Astrophysics, Australian National University, Canberra, ACT 2611, Australia
}

\date{Received  / Accepted}

\authorrunning{K. De Smedt et al.}
\titlerunning{Pb in Galactic \textit{s}-process enriched post-AGB stars}

\abstract
{This paper is part of a larger project in which we systematically study the chemical abundances 
 of Galactic and extragalactic post-asymptotic giant branch (post-AGB)
 stars.}
{Lead (Pb) is the final product of the \textit{s}-process nucleosynthesis and
 is predicted to have large overabundances with respect to other
 \textit{s}-process elements in AGB stars of low metallicities. However, Pb
 abundance studies of \textit{s}-process enriched post-AGB stars in the
 Magellanic Clouds show a discrepancy between observed and predicted
 Pb abundances. The determined upper limits based on spectral
 studies are much lower than what is predicted.  }
{We used high-resolution UVES and HERMES spectra for detailed spectral
 abundance studies of our sample of 14 Galactic post-AGB stars.  None of
 the sample stars display clear Pb lines, and we only deduced upper
 limits of the Pb abundance by using spectrum synthesis in the
 spectral ranges of the strongest Pb lines.}
{ We do not find any clear evidence of Pb overabundances in our
 sample. The derived upper limits are strongly correlated with the effective temperature of
 the stars with increasing upper limits for increasing effective
 temperatures. We confirm the \textit{s}-process 
 enrichment and carbon enhancement of two unstudied 21 $\mu$m sources IRAS 13245-6428 and IRAS 14429-4539.
 The mildly \textit{s}-process enhanced post-AGB star IRAS
 17279-1119 is part of a binary system.
 }
{Stars with T$_{\textrm{eff}}$ >
7500 K do not provide strong constraints on the Pb abundance. We conclude 
that the discrepancy between theory and observation increases towards lower metallicities.
The model predictions 
are consistent with the deduced upper limits on the Pb abundances for all stars with [Fe/H] > -0.7 dex. 
For stars with [Fe/H] < -0.7 dex, however, the model predictions overestimate the Pb abundances with respect to the 
other \textit{s}-process elements.   All
objects, except IRAS 17279-1119, confirm the relation between
neutron exposure [hs/ls] and third dredge-up efficiency [s/Fe], whereas no relation
between metallicity and neutron exposure is detected within the
metallicity range of our total sample ($-$1.4\,<\, [Fe/H]\,<\,$-$0.2). 
The mild enrichment of IRAS 17279-1119 can probably be attributed to a
cut-off of the AGB evolution due to binary interactions.
To our knowledge, IRAS 17279-1119 is the first s-process enhanced Galactic 
post-AGB star known in a binary system and is a possible precursor of the 
extrinsic Ba dwarf stars.  Lead-rich stars have yet to be found in post-AGB stars.  }

\maketitle


\section{Introduction}\label{sect:intro}

The final evolution of low- to intermediate-mass single stars (M
$\lesssim$ 7 M$_{\odot}$) is characterised by a fast transition from
the asymptotic giant branch (AGB) over the post-AGB track towards the
planetary nebula phase. During late stages of AGB evolution,
convective thermal pulses (TPs) occur in the intershell, which are possibly
followed by third dredge-ups (TDUs) that transport newly synthesised
material from the stellar interior to the surface. The main dredged-up 
element is  $^{12}$C as primary product of the triple alpha reaction. 
AGB stars are thought to be very important
contributors to the total carbon and nitrogen enrichment of galaxies
\citep[e.g.][]{romano10,kobayashi11}.

Apart from $^{12}$C, the TDU also brings elements created by neutron
synthesis to the stellar surface. The slow-neutron capture process 
(or \textit{s}-process) is at the origin of approximately half of all 
cosmic abundances past the iron peak. It is generally acknowledged
that the $^{13}$C($\alpha$,n)$^{16}$O reaction is the main neutron source
in low-mass AGB stars (1-3 M$_{\odot}$) 
\citep[e.g.][]{straniero95,gallino98,mowlavi98,abia02,karakas14,neyskens15}. It
is also widely accepted that a $^{13}$C-pocket is produced by 
the transport of protons from the convective envelope into the He-rich intershell.
The neutrons created in the $^{13}$C-pocket by the alpha capture
reaction can then be used for the creation 
of heavy elements by the slow neutron capture process (\textit{s}-process).
Although observations confirm that heavy elements 
can indeed be created by AGB stars, the physical  
mechanisms behind the AGB internal nucleosynthesis and associated
dredge-up processes  are poorly understood.

Along the valley of stability, the isotopes with the lowest cross-section for neutron capture are located around the nuclei with the magic number N=50 (the light \textit{s}-elements or ls) and N=80 (the heavy \textit{s}-elements or hs).  
The end product of the \textit{s}-process nucleosynthesis chain is the doubly
magic $^{208}$Pb isotope. The ratio [hs/ls] is an intrinsic indicator of the total neutron irradiation. The \textit{s}-process chain is limited to the ls elements for moderate irradiations, while the hs elements are being produced for increasing neutron irradiation.
When the neutron irradiation increases even further, the elements beyond the hs peak are produced up to the final product Pb. As the neutron production is thought to be largely independent of the initial metallicity, there are more neutrons available per iron seed and, hence at low metallicity  Pb is predicted to have large
overabundances with respect to other \textit{s}-elements \citep[see e.g.][and references therein]{gallino98, goriely00, lugaro12}.

These Pb abundance predictions are confirmed in some low metallicity extrinsically enriched metal-poor objects, where
strong Pb enhancements are indeed observed \citep[e.g.][]{vaneck01, vaneck03, behara10}. 
These objects are polluted with matter 
from an evolved binary companion when it passed the TP-AGB phase but that is now 
a dim white dwarf. However, not all metal-deficient objects with
\textit{s}-process enrichment show this strong Pb overabundance
\citep[e.g.][]{aoki01, vaneck03, bisterzo12}. A very wide range of neutron irradiations are needed at a given metallicity to explain the abundance spread of extrinsically enhanced low metallicity objects \citep{bisterzo10}, the physical orgin of which is not clear. A complicating factor in modelling the extrinsically enhanced objects is that the original enriched AGB star is now a cool white dwarf that is not detected.

Here we focus on intrinsically enriched objects. 
The AGB photospheres themselves are dominated by molecular transitions that hinder the study 
of individual elements \citep[e.g.][]{abia08}. This problem does not occur in 
post-AGB stars, for which the photospheres are hotter and are dominated by atomic 
transitions. The spectra of post-AGB stars allow for extensive chemical studies of individual elements.
Post-AGB atmospheres display the outcome of chemical enrichment from internal 
nucleosynthesis and
dredge-up processes during the entire stellar evolution.  
This makes post-AGB stars ideal probes to study AGB nucleosynthesis \citep{vanwinckel03}.

In our recent studies of post-AGB stars in the Magellanic Clouds \citep{vanaarle11,kamath14,kamath15}, we
focused on the intrinsic \textit{s}-process enriched post-AGB star
J004441.04-732136.4 (J004441) in the Small Magellanic Cloud 
\citep[SMC;][]{ desmedt12,desmedt14}. In \cite{desmedt14}, we compared the
observed abundance results with fine-tuned, theoretical
state-of-the-art AGB models. Although this star is metal-poor and
strongly \textit{s}-process enriched, we found a strong discrepancy
between the observed and predicted Pb overabundance. The best-fitting
model overestimates the Pb overabundance by more than 2 dex. Moreover,
the same Pb discrepancy was detected in three other metal-deficient
post-AGB stars in the LMC \citep{desmedt14}. The extreme overabundances in combination with a negative [Pb/hs] value
may point to a neutron irradiation that is different from the currently accepted AGB s-process scenarios where the neutron irradiation occurs in the radiative layers of the intershell in between thermal pulses. One possible alternative is proton 
ingestion directly into the convective thermal pulse, but preliminary model calculations \citep{lugaro15} are as yet inconclusive and not able to model the full abundance ratios of J004441.

This low Pb content was
also detected by \citet{reyniers07a} for the LMC star MACHO
47.2496.8. For all of these objects with intrinsic enrichment, the
upper limits of the Pb abundances are equal or smaller than the
overabundances of the other \textit{s}-elements.  In our most recent
study \citep{desmedt15}, we found a similar discrepancy for two newly
identified \textit{s}-process enriched post-AGB stars in the LMC. We
concluded that the low Pb abundance seems to be a common feature in
$s$-process rich, post-AGB stars in the Magellanic Clouds.
Furthermore, we find that all the objects studied until now have low
initial masses and low metallicities. Therefore, the low observed Pb
abundances strongly contradict theoretical nucleosynthetic predictions 
of Pb.



In this contribution, we extend our Pb studies and focus on Galactic \textit{s}-process rich objects. We present  a homogeneous study of the neutron irradiation as traced by the s-process element distribution of 14 post-AGB stars. We especially focus  on the Pb abundances. In combination with the studied Magellanic Cloud objects, we aim at a systematic study of the distribution of \textit{s}-elements detected in post-AGB stars, which cover a wide range of metallicities. We provide new constraints for the models of AGB nucleosynthesis and associated processes.


The selection of our Galactic sample stars and the observations are
described in Sect. \ref{sect:obs}.  The spectral analyses are
discussed in Sect. \ref{sect:analysis} followed by the abundance results 
of all elements lighter than Pb in Sect. \ref{sect:abun}. The abundance results 
of Pb are presented in Sect. \ref{subsect:pb_abun}. The neutron irradiation of our
sample of stars is discussed in Sect.  \ref{sect:discussion}.
In Sect. \ref{sect:iras17279}, we specifically focus on the evolutionary status 
of one of the sample stars, IRAS 17279-1119. We end this
paper with the conclusions in Sect. \ref{sect:conclusion}.

\section{Sample selection and observations}\label{sect:obs}

\begin{table*}[tb!]
\caption{\label{table:obs} Overview of the sample: name(s), observational logs, references of previous studies, and radial velocities.}
\begin{threeparttable}
\begin{tabular}{llcccccc} \hline\hline
IRAS       & Other       &    Date      &  UT   &                  Exp. time$^{a}$                   & Telescope+       &  References  & r$_{\textrm{v}}$ \\
           & name       &              & start &                     (s)                      & Spectrograph     &              &     (km/s)       \\
\hline
05113+1347$^{d}$ &             &  2014-10-07  & 07:55 &                  Blue: 1 $\times$ 2676      & VLT + UVES       &      1       &      6 $\pm$ 2  \\
           &             &  2014-11-23  & 05:03 &                  Blue: 1 $\times$ 2676      & VLT + UVES       &              &      5 $\pm$ 1  \\
           &             &              &       &                  Red: 2 $\times$ 1200       & VLT + UVES       &              &      5 $\pm$ 1  \\           
\hline
05341+0852$^{d}$ &             &  2014-10-06  & 07:52 &                  All: 1 $\times$ 1826       & VLT + UVES       &      2       &     28 $\pm$ 1  \\ 
           &             &  2014-10-06  & 08:28 &                  All: 1 $\times$ 1826       & VLT + UVES       &              &     28 $\pm$ 1  \\
\hline
06530-0213 &             &  2014-12-24  & 04:03 &                  All: 1 $\times$ 2427       & VLT + UVES       &      3       &     52 $\pm$ 1  \\
           &             &  2014-12-24  & 04:50 &                  All: 1 $\times$ 2427       & VLT + UVES       &              &     52 $\pm$ 1  \\
\hline
07134+1005$^{d}$ & HD 56126    &  2014-09-05  & 05:50 &                       1 $\times$  920                & Mercator + HERMES &     2       &     92  $\pm$ 1 \\    
           &             &  2014-09-08  & 05:39 &                       1 $\times$ 1000                & Mercator + HERMES &             &     94  $\pm$ 1 \\
\hline
07430+1115 &             &  2014-12-24  & 05:33 &                  Blue: 1 $\times$ 2176      & VLT + UVES       &      4       &     41  $\pm$ 2 \\ 
           &             &              &       &                  Red: 2 $\times$ 1000       & VLT + UVES       &              &     41  $\pm$ 1 \\
\hline    
08143-4406$^{b}$ &             &  2001-01-16  & 06:41 &            Blue437: 2 $\times$ 1800      & VLT + UVES       &      3       &     52  $\pm$ 2 \\
           &             &  2001-01-16  & 07:16 &                  Red860:  3 $\times$ 500       & VLT + UVES       &              &     52  $\pm$ 1 \\
           &             &  2001-02-01  & 04:21 &                  Red:  1 $\times$ 1800      & VLT + UVES       &              &     52  $\pm$ 1 \\
\hline
08281-4850 &             &  2015-01-25  & 04:18 &                  All:  1 $\times$ 2400      & VLT + UVES       &        5     &    117  $\pm$ 1 \\
           &             &  2015-01-25  & 03:47 &                  All:  1 $\times$ 2600      & VLT + UVES       &              &    116  $\pm$ 1 \\
           &             &  2015-01-31  & 02:39 &                  All:  1 $\times$ 2116      & VLT + UVES       &              &    117  $\pm$ 1 \\
\hline
13245-5036$^{d}$ &             &  2015-03-06  & 04:53 &                  All:  1 $\times$ 346      & VLT + UVES       &              &     54  $\pm$ 1 \\
\hline
14325-6428$^{c}$ &             &  2004-05-13  & 07:13 &                  Blue437: 1 $\times$ 1800  & VLT + UVES       &      5       &   -83  $\pm$ 1 \\
           &             &              &       &                  Red860:  1 $\times$ 1800  & VLT + UVES       &              &    -83  $\pm$ 1 \\
           &             &  2004-05-13  & 06:39 &                  Red:  1 $\times$ 1800  & VLT + UVES       &              &    -82  $\pm$ 1 \\
\hline
14429-4539$^{b,d}$ &       &  2001-02-03  & 08:24 &                  Red: 1 $\times$ 1800      & VLT + UVES       &             &      0  $\pm$ 2 \\
                 &       &              & 08:55 &                  Red: 1 $\times$ 1200      & VLT + UVES       &             &      0  $\pm$ 2 \\
                 &       &  2001-02-06  & 07:53 &                  Red: 1 $\times$ 1800      & VLT + UVES       &             &      0  $\pm$ 2 \\
                 &       &  2001-02-12  & 07:54 &                  Blue437: 3 $\times$ 2000      & VLT + UVES       &             &     4  $\pm$ 2 \\
                 &       &              &       &                  Red860: 3 $\times$ 2000      & VLT + UVES       &             &      4  $\pm$ 2 \\
\hline
17279-1119       &  HD158616  &  2014-09-18 & 00:00 &              All: 1 $\times$ 65      & VLT + UVES       &      6      &     62  $\pm$ 1 \\
\hline
19500-1709$^{d}$     &  HD187885  &  2009-07-16 & 00:20 &                      2  $\times$  2000              & Mercator + HERMES &      2      &     15  $\pm$ 1 \\
\hline
22223+4327$^{d}$     &  V448 Lac  & 2009-07-31  & 02:50 &                   4  $\times$  1800    & Mercator + HERMES &      2,6     &    -41  $\pm$ 1 \\
\hline
22272+5435$^{d}$     &  HD235858  & 2014-08-29 & 01:55 &                    3  $\times$  720    & Mercator + HERMES &      1     &    -38  $\pm$ 1 \\
\hline
\end{tabular}
    \begin{tablenotes}
      \small
      \item $^1$\citet{reddy02}, $^2$\citet{vanwinckel00}, $^3$\citet{reyniers04}, $^4$\citet{reddy99}, $^5$\citet{reyniers07c}, $^6$\citet{rao11}
      \item $^{a}$ Exposure times of UVES spectra are split up into three categories: exposure times for the Blue arm, Red arm, or both arms; the latter is indicated with 'All'.
      Terms 'Blue' and 'Red' refer to observations with the Blue390 and Red580 setting. For older observations, the Blue437 and Red860 settings have also been used.
      \item $^{b}$ For more information about these observations, see \citet{reyniers_thesis}.
      \item $^{c}$ For more information about the observations of IRAS 14325-6428, see \citet{reyniers07c}.
      \item $^{d}$ Identified 21 $\mu$m source.       
    \end{tablenotes}
\end{threeparttable}
\end{table*}

Our sample consists of 12 Galactic post-AGB stars, which are known to
be \textit{s}-process rich, and two new post-AGB candidates, IRAS 13245-5036 and IRAS 14429-4539,
for which no abundance studies have been reported yet.
The majority of the already studied Galactic post-AGB stars are 21 $\mu$m objects, named for the strong
solid-state feature at 21 $\mu$m \citep{kwok89}. To date, all 21
$\mu$m objects are acknowledged to be post-AGB stars with carbon and \textit{s}-process
enhancements in their photospheres \citep[e.g.][]{hrivnak09}. Since
both IRAS 13245-5036 and IRAS 14429-4539 are post-AGB candidates with a clear 
21 $\mu$m feature \citep{cerrigone11}, we added these stars to our sample.

We use high-resolution spectra of two different spectrographs. The
first spectrograph is the Ultraviolet and Visual Echelle Spectrograph
(UVES; \citet{dekker00}), the echelle spectrograph mounted on the 8m
UT2 Kueyen Telescope of the VLT array at the Paranal Observatory of
ESO in Chili. The second spectrograph is the High Efficiency and
Resolution Mercator Echelle Spectrograph (HERMES; \citep{raskin11}),
the spectrograph mounted on the 1.2m Mercator telescope at the Roque de los
Muchachos Observatory on La Palma.

For all objects, we first checked whether optical high-resolution
spectra with sufficient signal-to-noise (S/N) around 4058 \AA{}, the
spectral region of the strongest identified Pb I line, were already
available for UVES in the ESO archive and for HERMES in the archive of
the Institute of Astronomy (KULeuven). If not, we requested and performed
observations for those objects for which a S/N of 25 around 4058 \AA{} 
would be reached within an hour. All objects, observational details and previous abundance studies are
listed in Table \ref{table:obs}. 
The last column shows the radial velocity of the observed spectra.

\subsection{UVES spectra}

The UVES spectra obtained in the period 2014-2015 (see Table \ref{table:obs}) were observed 
using the same setting.  We selected the dichroic beam splitter resulting in 
a wavelength coverage for the blue UVES arm 
from approximately 3280 to 4530 \AA{}, and for the lower and upper part of
the mosaic CCD chip from approximately 4780 to 5770 \AA{} and
from 5800 to 6810 \AA,{} respectively. A slit width of 1 arcsecond 
was used to gain an optimal compromise between spectral resolution 
and slit-loss minimisation. 

We used the reduced UVES spectra from
\citet{reyniers04} for our analysis for IRAS\,08143-4406. These spectra were observed with 
different settings, resulting in two sets of spectra representing 
different wavelength coverages. There is a time gap of 16 days between 
the observations of both sets of data. This gap was 
large enough to observe significant changes in the photosphere of IRAS 
08143-4406, resulting in two different sets of atmospheric parameters 
in the analysis by \citet{reyniers04}. The UVES spectra
of post-AGB candidate IRAS 14429-4539 were obtained using the same settings as 
for IRAS 08143-4406. For details about the observations of IRAS 08143-4406 and
IRAS14429-4539, we refer to \citet{reyniers_thesis}. For  
IRAS 14325-6428, we use the observed spectra described in \citet{reyniers07c}
for the spectral analysis.

All UVES spectra in Table \ref{table:obs}, except for IRAS 08143-4406
and IRAS 14429-4539, were reduced using the UVES pipeline in the Reflex
environment of ESO
\footnote{https://www.eso.org/sci/software/reflex/}.  The reduction
scheme includes extraction of frames, determinations of wavelength
calibration and applying this scale to flat-field divided data. Cosmic
clipping was also included. We chose the standard reduction
parameters of the UVES pipeline as these gave the best signal-to-noise
(S/N) ratio of the final spectrum. The spectra of IRAS 08143-4406 and IRAS 
14429-4539 were already reduced with older versions of the UVES pipeline. 

\subsection{HERMES spectra}
All HERMES spectra were obtained in high-resolution mode ($\lambda$/$\Delta \lambda$ = 85.000). 
These spectra cover a wavelength range from about 3770 \AA{} up to about 9000 \AA{}, although 
some gaps in the spectra are present at large wavelengths, as the
reddest orders are too long for the CCD.
The specific reduction pipeline of HERMES \citep{raskin11} includes similar steps as the UVES pipeline
and is used for the full reduction of the spectra. 

\subsection{Normalisation and merging}

When necessary, the individual reduced spectra of each object are first
corrected for radial velocity differences (see Table 1). The radial velocities are determined for UVES
spectra using the positions
of individual spectral lines. The radial velocities are determined for HERMES
spectra using a
cross-correlation routine specific to the HERMES pipeline \citep{raskin11}.

Thereafter, weighted mean spectra are calculated for the individual
wavelength ranges of UVES and for the total HERMES spectra. For the
normalisation of the spectra, we subdivided the
weighted mean spectra into fixed wavelength ranges of
120 \AA{} where the first and last 10 \AA{} overlap with the previous
and subsequent spectral parts. Specific care was taken to conserve the
Balmer profiles during the normalisation procedure. Each subspectrum 
is normalised individually by fitting a fifth order polynomial through interactively
defined continuum points. After the individual normalisations, all
normalised subspectra are then merged into one large spectrum, which is
used for the spectral abundance studies. 

\section{Spectral analyses}\label{sect:analysis}

\begin{figure}[t!]
\resizebox{\hsize}{!}{\includegraphics{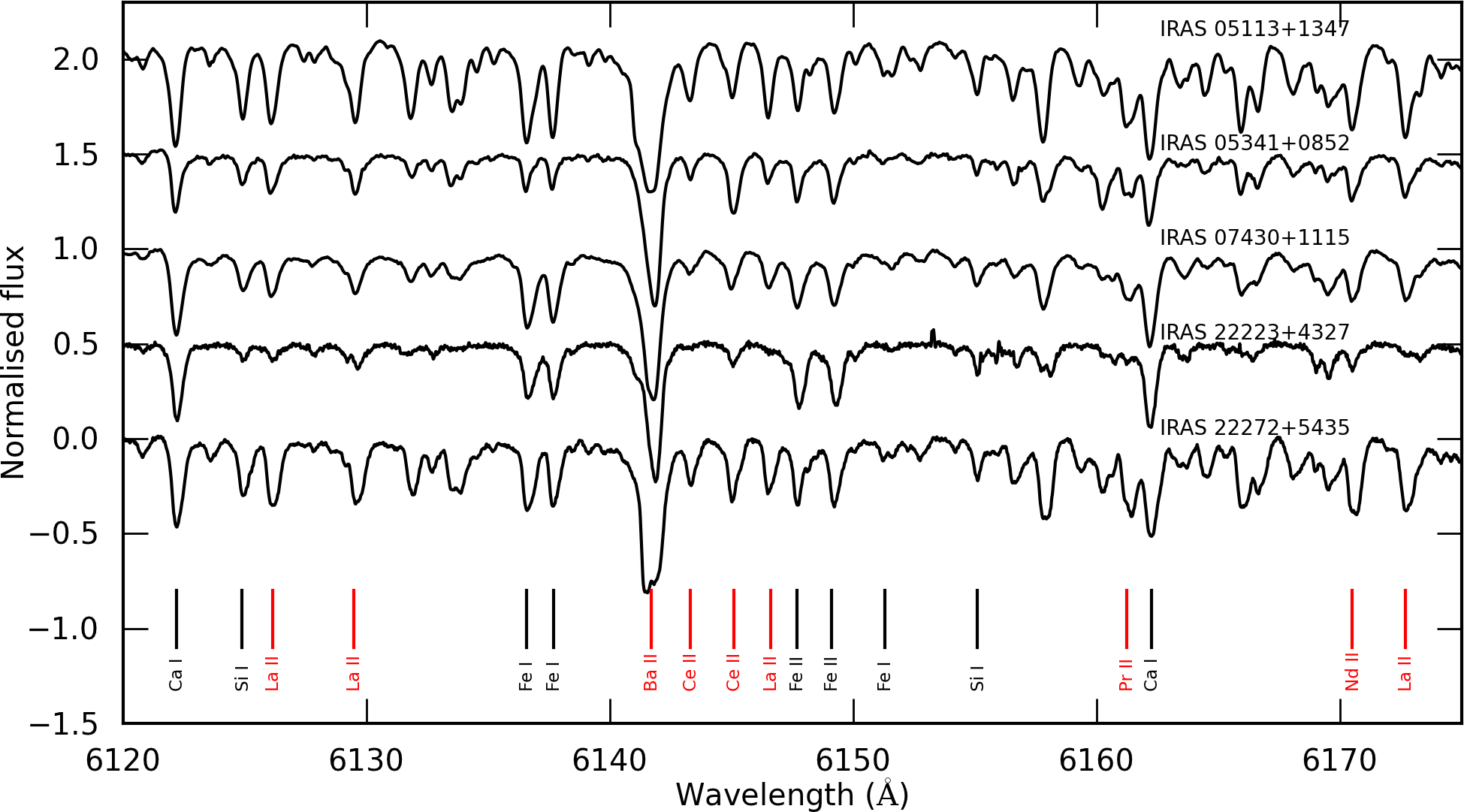}}
\caption{Comparison of the normalised spectra of all sample stars with T$_{\textrm{eff}}$ < 7000 K. 
The spectra have been shifted in flux for clarity and all spectra are shifted to zero velocity. 
Red and black vertical lines indicate positions of \textit{s} nuclei and non \textit{s} nuclei, respectively. 
For more information, see text.}\label{fig:comp1}
\end{figure}

\begin{figure}[t!]
\resizebox{\hsize}{!}{\includegraphics{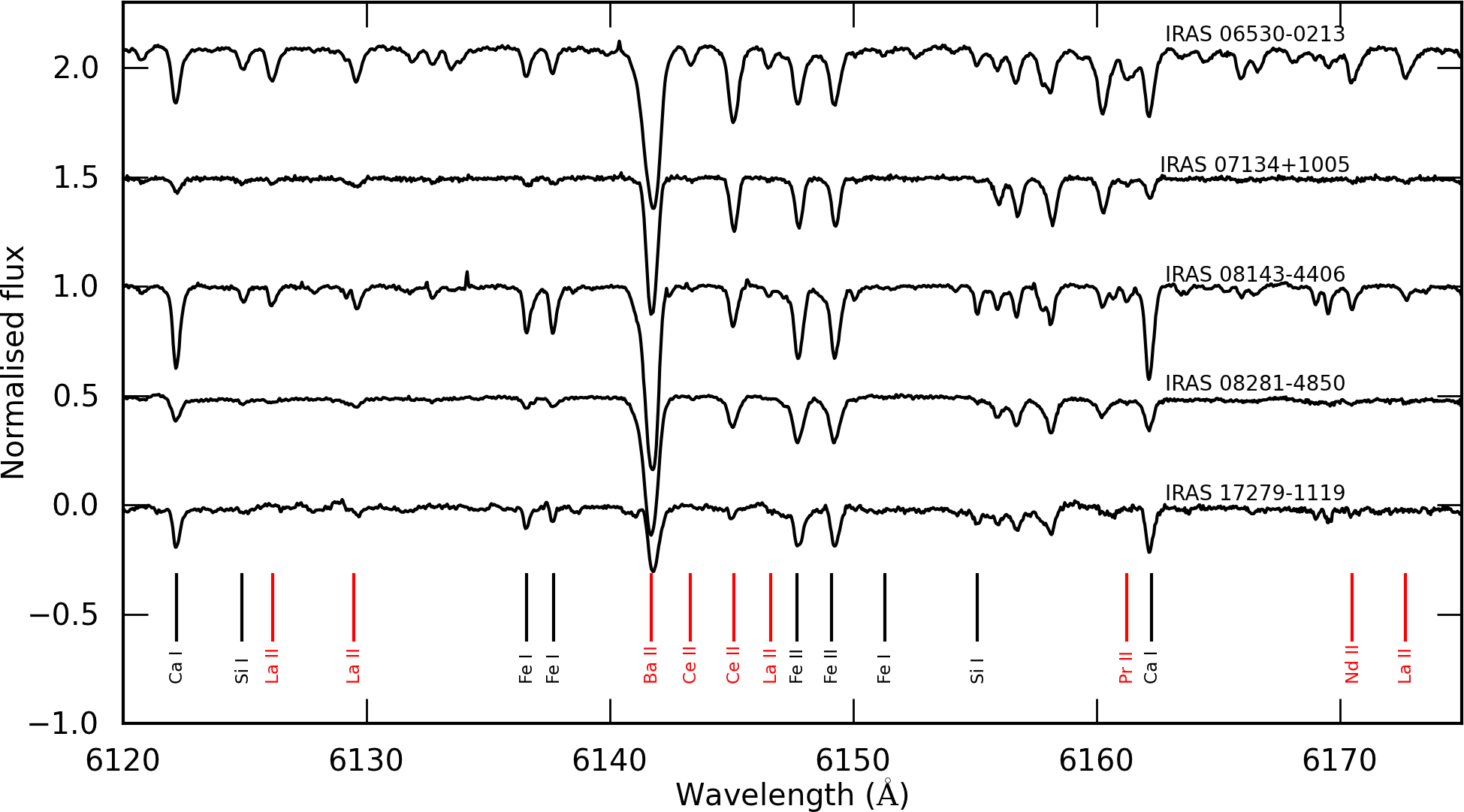}}
\caption{Similar to Fig. \ref{fig:comp1}, except for the sample stars with 7000 K $\leqslant$ T$_{\textrm{eff}}$ < 8000 K.}\label{fig:comp2}
\end{figure}

\begin{figure}[t!]
\resizebox{\hsize}{!}{\includegraphics{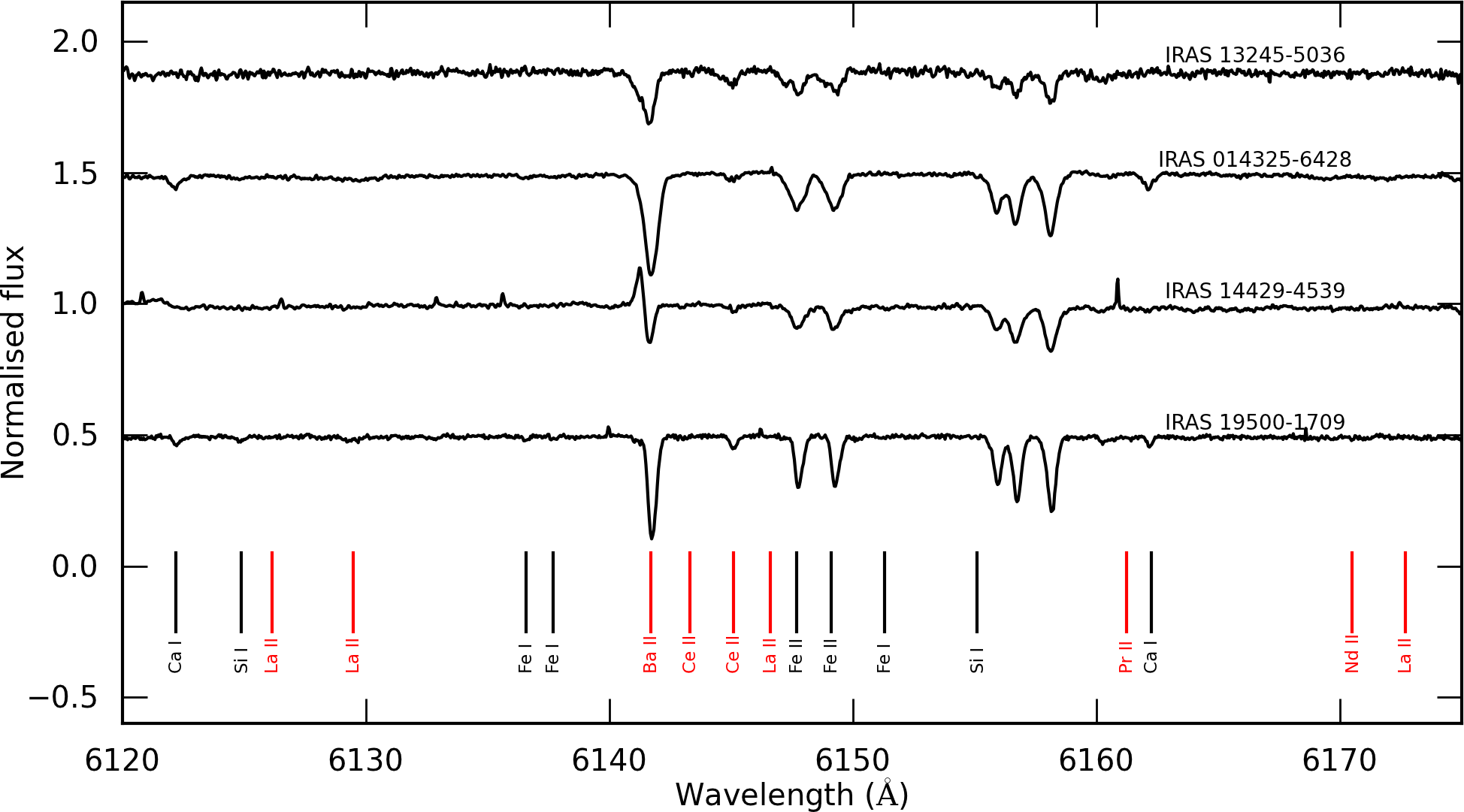}}
\caption{Similar to Figs. \ref{fig:comp1} and \ref{fig:comp2}, except for the sample stars with T$_{\textrm{eff}}$ $\geqslant$ 8000 K.}\label{fig:comp3}
\end{figure}

Figs. \ref{fig:comp1}, \ref{fig:comp2}, and \ref{fig:comp3} presents
the spectral region around the Ba II line at 6141.813 \AA{} together with
the identification of several spectral lines. Each figure corresponds to a
different range in effective temperature T$_{\textrm{eff}}$ (see
Sect. \ref{subsect:atmos}). Fig. \ref{fig:comp1} shows our sample
stars with T$_{\textrm{eff}}$ < 7000 K, Fig. \ref{fig:comp2} shows
sample stars with 7000 K $\leqslant$ T$_{\textrm{eff}}$ < 8000 K, and
Fig. \ref{fig:comp3} shows the sample stars with T$_{\textrm{eff}}$
$\geqslant$ 8000 K. The figures illustrate how the 
number of strong spectral lines decreases with 
increasing effective temperature for all elements.

The overall S/N of the red spectra of UVES (see Table \ref{table:obs}) 
and the optical spectral
regions of HERMES is very high and ranges from 80 to 200. 
We perform a comparative systematic analysis of all objects and hence decide not to rely on the previous
results in the literature, but perform an independent full spectral
analysis using the same method for all objects.  Depending on the temperature of the star, the S/N of
spectra below 3700 \AA\, are too poor for both spectrographs,
so these regions are not used for our 
analysis. For the UVES data, we use the full red spectral
region sampled in our setting as well as the spectral region from
3900 \AA{} redwards. For hotter stars, we could extend this region towards bluer wavelengths. 

\begin{table*}[tb!]
\caption{\label{table:atmos} Model atmospheres of the sample of stars. The errors for [Fe/H] include line-to-line scatter and model uncertainty (see Sect. \ref{subsect:abun}).
N$_{\textrm{FeI}}$ and N$_{\textrm{FeII}}$ show the number of lines used
for Fe I and Fe II, respectively.}
\begin{threeparttable}
\begin{tabular}{lccccccc} \hline\hline
Object          &  T$_{\textrm{eff}}$ (K)  &  $\log g$ (dex) &  $\xi_t$ (km/s) &  $\textrm{[FeI/H]}$  &  $\textrm{[FeII/H]}$ &  N$_{\textrm{FeI}}$  & N$_{\textrm{FeII}}$ \\
                &    $\pm$ 125    &  $\pm$ 0.25  &    $\pm$ 0.25    &                               &                      &                      &                    \\
\hline        
IRAS 05113+1347 &      5500       &      0.50    &       5.00       &       -0.49 $\pm$ 0.15        &    -0.54  $\pm$ 0.17 &          21          &        11           \\
IRAS 05341+0852 &      6750       &      1.25    &       3.50       &       -0.70 $\pm$ 0.15        &    -0.54  $\pm$ 0.11 &          47          &        20           \\
IRAS 06530-0213 &      7375       &      1.25    &       4.00       &       -0.38 $\pm$ 0.20        &    -0.32  $\pm$ 0.11 &          24          &        54           \\
IRAS 07134+1005 &      7250       &      0.50    &       3.25       &       -0.96 $\pm$ 0.31        &    -0.91  $\pm$ 0.20 &          42          &        32           \\
IRAS 07430+1115 &      6000       &      1.00    &       3.25       &       -0.31 $\pm$ 0.15        &    -0.35  $\pm$ 0.15 &          43          &        17           \\
IRAS 08143-4406$^{a}$ &      7000       &      1.50    &       5.50       &       -0.45 $\pm$ 0.16        &    -0.43  $\pm$ 0.11 &          17          &        8      \\
IRAS 08143-4406$^{b}$ &      7250       &      1.50    &       5.00       &       -0.37 $\pm$ 0.17        &    -0.36  $\pm$ 0.11 &          70          &        23     \\
IRAS 08281-4850 &      7875       &      1.25    &       5.50       &       -0.29 $\pm$ 0.31        &    -0.26  $\pm$ 0.11 &          44          &        23           \\
IRAS 13245-5036 &      9500       &      2.75    &       4.50       &       -0.35 $\pm$ 0.20        &    -0.30  $\pm$ 0.10 &          21          &        41           \\
IRAS 14325-6428 &      8000       &      1.00    &       5.75       &       -0.55 $\pm$ 0.33        &    -0.56  $\pm$ 0.10 &          26          &        58           \\
IRAS 14429-4539 &      9375       &      2.50    &       4.75       &       -0.26 $\pm$ 0.21        &    -0.18  $\pm$ 0.11 &          26          &        52           \\
IRAS 17279-1119 &      7250       &      1.25    &       3.00       &       -0.51 $\pm$ 0.19        &    -0.64  $\pm$ 0.12 &          44          &        30           \\
IRAS 19500-1709 &      8000       &      1.00    &       6.00       &       -0.58 $\pm$ 0.42        &    -0.59  $\pm$ 0.10 &          33          &        32           \\
IRAS 22223+4327 &      6500       &      1.00    &       4.75       &       -0.32 $\pm$ 0.14        &    -0.30  $\pm$ 0.11 &          80          &        25           \\
IRAS 22272+5435 &      5750       &      0.50    &       4.25       &       -0.77 $\pm$ 0.14        &    -0.77  $\pm$ 0.12 &          35          &        24           \\
\hline
\end{tabular}
    \begin{tablenotes}
      \small
      \item $^{a}$ Results of the Blue437 and Red860 observations of IRAS 08143-4406 in Table \ref{table:obs}.
      \item $^{b}$ Results of the Red580 observations of IRAS 08143-4406 in Table \ref{table:obs}.    
    \end{tablenotes}
\end{threeparttable}
\end{table*}

\begin{figure}[t!]
\includegraphics[width=8.0cm,height=6.0cm]{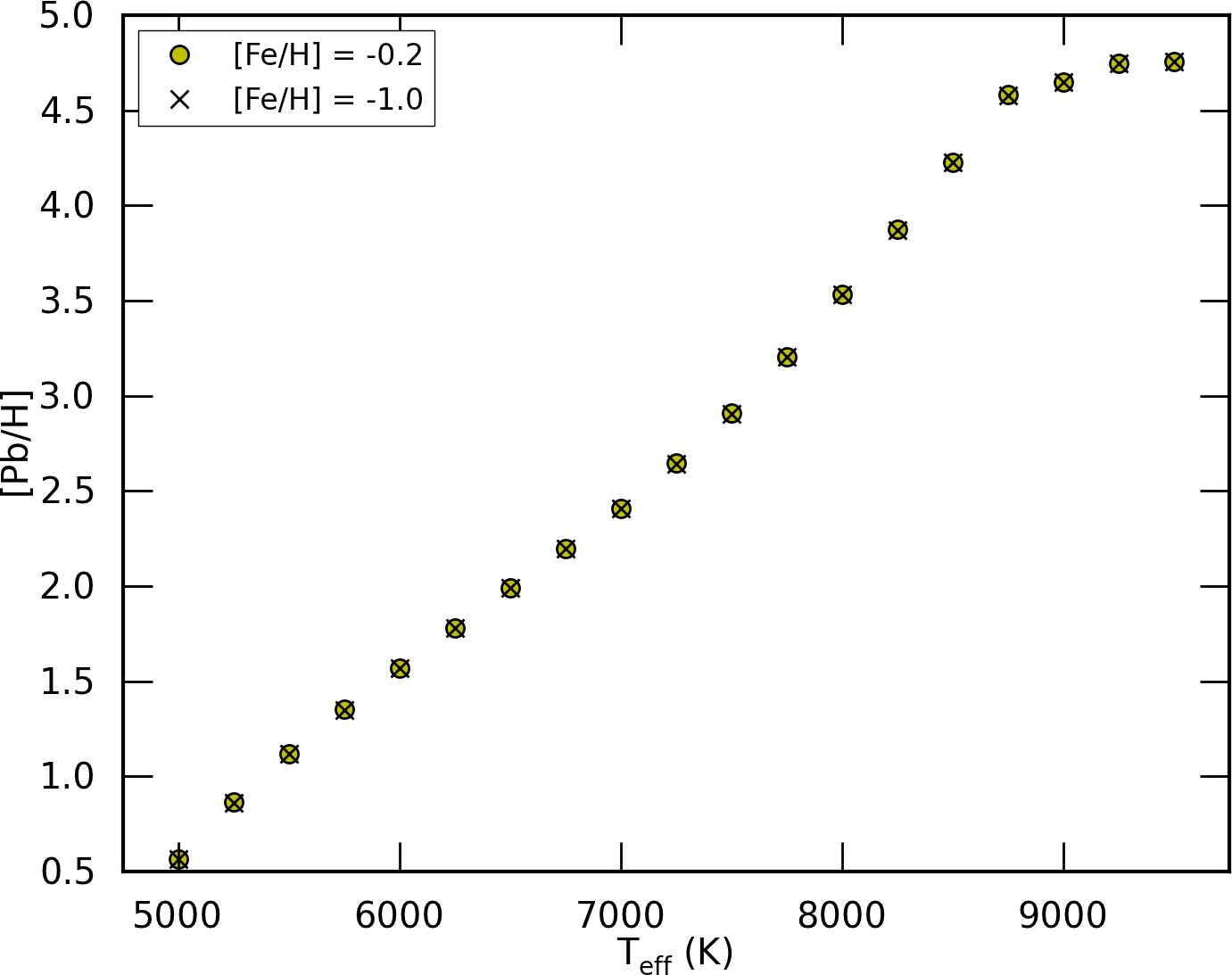}
\caption{[Pb/H] abundance needed to obtain a Pb I or Pb II spectral line of 5 m\AA{} as a function of effective temperature T$_{\textrm{eff}}$ 
and metallicity [Fe/H]. For more information, see text.}\label{fig:teff_Pb}
\end{figure}

In this contribution we focus on Pb but unfortunately, the optical
spectrum of Pb is poor in spectral lines for the typical atmospheric
parameters of our sample stars.
The strongest Pb I lies in the blue part of the spectrum ($\lambda$4057.807 \AA{}), a spectral
region with low S/N and a large number of blends. 
For some hotter stars, the strongest line of Pb II at 5608.853 \AA{}
is used to constrain the Pb abundance.
To investigate the possible strength of Pb, we calculated which input [Pb/H] abundance
is needed to create a spectral line of 5 m\AA{}. This value is the lower
limit for distinguishing spectral lines from noise in the
spectra. Fig. \ref{fig:teff_Pb} shows these theoretical Pb abundances
for different effective temperatures and for [Fe/H] = -1.0 dex and
[Fe/H] = -0.2 dex. These abundances are calculated for a $\log g$ = 1.5 dex for all
stars with T$_{\textrm{eff}}$ < 9250 K and for $\log \ g$ = 2.0 dex for 9250 and 9500
K.  Fig. \ref{fig:teff_Pb} shows that the needed Pb abundance
for a spectral line detection increases significantly with rising
temperature.  The figure also shows that the effect of
metallicity is negligible in the temperature and metallicity range of
the programme stars.

In order to trace Pb we have to synthesise the spectral regions around
the Pb lines and, hence, we need a complete set of abundances. For
these, we used the spectral regions with good S/N. 
We use the same methods for the atmospheric parameter determinations and abundance studies for
all programme stars.  Then, we used these abundances to make detailed
spectral synthesis spectra in the regions of the strongest Pb
lines. In this way, we obtained the best constraints on the Pb
abundances.

We have written a Python
wrapper (PyMOOG) around the local thermal equilibrium (LTE) abundance
calculation routine MOOG \citep[version June 2014][]{sneden73}, which
is combined with LTE Kurucz-Castelli atmosphere models
\citep{castelli04}. Spectral line identification is based on 
linelists from the Vienna atomic line database (VALD) \citep{vald} in combination with a
list of spectral lines gathered at the Institute of Astronomy 
(KULeuven) for the chemical analysis of A, F, and G stars 
\citep{vanwinckel00}. The implemented linelists cover a wavelength
range from 3000 up to 11000 \AA{}, covering the full wavelength
coverage of the UVES and HERMES spectra. We included linelists
for about 160 ions, ranging from helium (He, Z=2) up to uranium (U, Z=92).  
Neutral and firstly ionised lines are available for most elements. For some
\textit{s}-elements, the second ionisation is also included and is used
for the spectral abundance studies of the hottest stars in the sample.

The equivalent width (EW) of spectral lines are interactively 
measured via direct integration in PyMOOG. The abundances are
computed via an iterative process in which the theoretical EWs of
single lines are computed for given abundances and matched to the
observed EWs. Blended lines are avoided as much as possible, although
for some species, Pb in particular, we  sometimes could only use blended lines in our strongly \textit{s}-process
enriched stars. 

\subsection{Atmospheric parameters}\label{subsect:atmos}

The atmospheric parameters are determined using Fe I and Fe II lines
in the standard spectroscopic method: the effective temperature
T$_{\textrm{eff}}$ is determined by imposing the iron abundance to be
independent of lower excitation potential; surface gravity $\log g$ by imposing
ionisation equilibrium between the individual Fe I and Fe II
abundances; microturbulent velocity by imposing 
the iron abundance to be independent of reduced equivalent width, which we 
define as EW/$\lambda$.

To decrease the abundance uncertainty imposed by the step size of the atmospheric parameters, 
we use linear interpolation to calculate
atmospheric models that lie within the parameter steps of the
Kurucz-Castelli models. We choose temperature steps of 125 K, 
surface gravity steps of 0.25 dex, and microturbulent velocity steps of 0.25 km/s.

The individual atmospheric parameter results for the sample of stars
are listed in Table \ref{table:atmos}. The shown uncertainties for [FeI/H]
and [FeII/H] include both line-to-line scatter and atmospheric parameter uncertainties
as described below in Sect. \ref{subsect:abun}. Similar to
\cite{reyniers04}, we have determined the atmospheric parameters
separately for the different spectra sets of IRAS 08143-4406 (see Sect. \ref{sect:obs}). 
Within the time gap between the observations, the spectral lines of 
IRAS 08143-4406 had changed significantly, resulting
in different sets of atmospheric parameters. These changing spectral line shapes are
expected as post-AGB stars cross the Population II Cepheid
instability strip during their evolution, and,  hence, pulsations may strongly
change the atmospheric parameters. For the other stars, the
spectral lines did not show any visible changes within the time of the
observations.

Typically, the uncertainties for Fe I lines are larger than those of
Fe II because of the larger sensitivity of Fe I lines to temperature. 
For the two new 21 $\mu$m post-AGB
stars IRAS 13245-5036 and IRAS 14429-4539, we find higher
temperatures and higher surface gravities with respect to the other sample
stars, which points to a more evolved phase of the post-AGB
evolutionary stage. 

\begin{table*}[tb!]
\caption{\label{table:abun_cold} Abundance results of IRAS 05113+1347, IRAS 05341+0852, IRAS 07430+1115, and IRAS 22272+5435.}
\begin{threeparttable}
\setlength{\tabcolsep}{0.10cm}
\begin{tabular}{| lc | ccccc | ccccc | ccccc | ccccc| } \hline
  &  &  \multicolumn{5}{c|}{IRAS05113}  &  \multicolumn{5}{c|}{IRAS05341}  
  &  \multicolumn{5}{c|}{IRAS07430} &  \multicolumn{5}{c|}{IRAS22272} \\
  &  &  \multicolumn{5}{c|}{[Fe/H] = -0.49}  &  \multicolumn{5}{c|}{[Fe/H] = -0.54}  
  &  \multicolumn{5}{c|}{[Fe/H] = -0.31} &  \multicolumn{5}{c|}{[Fe/H] = -0.77}  \\
 &  &  \multicolumn{5}{c|}{T$_{\textrm{eff}}$= 5500 K}  &  \multicolumn{5}{c|}{ T$_{\textrm{eff}}$ = 6750 K}  
  &  \multicolumn{5}{c|}{ T$_{\textrm{eff}}$ = 6000 K} &  \multicolumn{5}{c|}{T$_{\textrm{eff}}$ = 5750 K}  \\
  &  &  \multicolumn{5}{c|}{ $\log g$ = 0.5 dex}  &  \multicolumn{5}{c|}{ $\log g$ =  1.25 dex}  
  &  \multicolumn{5}{c|}{ $\log g$ = 1.00 dex} &  \multicolumn{5}{c|}{ $\log g$ = 0.5 dex}  \\ 
\hline
Ion & log $\epsilon_{\odot}$ 
& N  &  log $\epsilon$  &  $\sigma_{\textrm{l2l}}$  &  [X/Fe] &  $\sigma_{tot}$  
& N  &  log $\epsilon$  &  $\sigma_{\textrm{l2l}}$  &  [X/Fe] &  $\sigma_{tot}$ 
& N  &  log $\epsilon$  &  $\sigma_{\textrm{l2l}}$  &  [X/Fe] &  $\sigma_{tot}$
& N  &  log $\epsilon$  &  $\sigma_{\textrm{l2l}}$  &  [X/Fe] &  $\sigma_{tot}$ \\
\hline
C I & 8.43
 & 6 & 8.87 & 0.11 & 0.65 & 0.16
 & 16 & 8.73 & 0.10 & 1.03 & 0.10
 & 6 & 8.87 & 0.06 & 0.79 & 0.13
 & 14 & 8.71 & 0.07 & 1.05 & 0.12
 \\
O I & 8.69
 & 1 & 8.49 & 0.20 & 0.01 & 0.27
 & 2 & 8.71 & 0.07 & 0.75 & 0.11
 & 2 & 8.64 & 0.20 & 0.30 & 0.22
 & 2 & 8.55 & 0.02 & 0.63 & 0.09
 \\
 \hline
Na I & 6.24
 & 2 & 5.88 & 0.01 & 0.17 & 0.12
 & 1 & 6.00 & 0.20 & 0.46 & 0.23
 & 2 & 6.20 & 0.09 & 0.27 & 0.12
 & 2 & 5.90 & 0.00 & 0.42 & 0.09
 \\
Mg I & 7.60
 & 1 & 7.20 & 0.20 & 0.13 & 0.24
 & 1 & 6.88 & 0.20 & -0.02 & 0.24
 & 2 & 7.26 & 0.01 & -0.03 & 0.11
 & 3 & 6.75 & 0.20 & -0.08 & 0.17
 \\
Al I & 6.45
 &   &   &   &   &  
 &   &   &   &   &  
 & 2 & 6.37 & 0.13 & 0.24 & 0.15
 & 1 & 6.07 & 0.20 & 0.39 & 0.23
 \\
Si I & 7.51
 & 4 & 7.23 & 0.06 & -0.07 & 0.20
 & 5 & 7.04 & 0.10 & 0.26 & 0.18
 & 5 & 7.52 & 0.13 & 0.36 & 0.19
 & 4 & 7.06 & 0.05 & 0.32 & 0.16
 \\
S I & 7.12
 & 1 & 6.97 & 0.20 & 0.06 & 0.25
 & 3 & 6.65 & 0.09 & 0.25 & 0.15
 & 2 & 6.69 & 0.02 & -0.08 & 0.13
 & 2 & 6.64 & 0.02 & 0.28 & 0.11
 \\
Ca I & 6.34
 & 3 & 5.66 & 0.03 & -0.14 & 0.10
 & 7 & 5.83 & 0.05 & 0.20 & 0.08
 & 7 & 6.03 & 0.09 & -0.0 & 0.08
 & 8 & 5.81 & 0.10 & 0.24 & 0.08
 \\
Ca II & 6.34
 &   &   &   &   &  
 & 3 & 5.97 & 0.09 & 0.36 & 0.12
 &   &   &   &   &  
 & 1 & 5.79 & 0.20 & 0.22 & 0.24
 \\
Sc II & 3.15
 & 6 & 2.87 & 0.09 & -0.08 & 0.13
 & 6 & 2.72 & 0.09 & 0.30 & 0.09
 & 4 & 3.00 & 0.06 & 0.20 & 0.10
 & 3 & 2.75 & 0.20 & 0.37 & 0.18
 \\
Ti I & 4.95
 & 4 & 4.47 & 0.09 & 0.05 & 0.13
 &   &   &   &   &  
 & 5 & 4.90 & 0.11 & 0.27 & 0.11
 & 2 & 4.50 & 0.05 & 0.32 & 0.10
 \\
Ti II & 4.95
 & 3 & 4.41 & 0.05 & -0.34 & 0.12
 & 2 & 4.50 & 0.20 & 0.28 & 0.19
 & 9 & 4.86 & 0.09 & 0.26 & 0.09
 & 2 & 4.47 & 0.03 & 0.29 & 0.11
 \\
V I & 3.93
 & 3 & 3.71 & 0.11 & 0.32 & 0.15
 &   &   &   &   &  
 & 1 & 4.00 & 0.20 & 0.38 & 0.24
 &   &   &   &   &  
 \\
V II & 3.93
 &   &   &   &   &  
 &   &   &   &   &  
 & 2 & 3.99 & 0.03 & 0.41 & 0.13
 & 2 & 3.21 & 0.03 & 0.05 & 0.11
 \\
Cr I & 5.64
 & 4 & 5.02 & 0.03 & -0.09 & 0.10
 & 2 & 4.99 & 0.06 & 0.05 & 0.12
 & 2 & 5.20 & 0.02 & -0.13 & 0.09
 & 4 & 4.91 & 0.06 & 0.03 & 0.08
 \\
Cr II & 5.64
 &   &   &   &   &  
 & 10 & 5.03 & 0.06 & 0.12 & 0.07
 & 6 & 5.16 & 0.07 & -0.13 & 0.08
 & 9 & 4.94 & 0.06 & 0.06 & 0.07
 \\
Mn I & 5.43
 & 1 & 4.60 & 0.20 & -0.3 & 0.24
 &   &   &   &   &  
 & 2 & 4.96 & 0.06 & -0.15 & 0.11
 & 1 & 4.58 & 0.20 & -0.08 & 0.22
 \\
Mn II & 5.43
 &   &   &   &   &  
 & 1 & 5.40 & 0.20 & 0.70 & 0.22
 &   &   &   &   &  
 & 1 & 4.60 & 0.20 & -0.06 & 0.22
 \\
Fe I & 7.50
 & 21 & 6.96 & 0.11 & 0.00 & 0.08
 & 47 & 6.80 & 0.10 & 0.00 & 0.06
 & 43 & 7.19 & 0.08 & 0.00 & 0.05
 & 35 & 6.73 & 0.07 & 0.00 & 0.05
 \\
Fe II & 7.50
 & 10 & 7.29 & 0.13 & 0.00 & 0.11
 & 20 & 6.77 & 0.07 & 0.00 & 0.05
 & 17 & 7.15 & 0.08 & 0.00 & 0.06
 & 24 & 6.73 & 0.08 & 0.00 & 0.05
 \\
Co I & 4.99
 & 2 & 4.59 & 0.01 & 0.13 & 0.13
 &   &   &   &   &  
 &   &   &   &   &  
 &   &   &   &   &  
 \\
Ni I & 6.22
 & 5 & 5.78 & 0.10 & 0.10 & 0.11
 & 6 & 5.81 & 0.06 & 0.29 & 0.08
 & 11 & 5.89 & 0.12 & -0.01 & 0.09
 & 8 & 5.66 & 0.07 & 0.21 & 0.07
 \\
Cu I & 4.19
 & 1 & 3.94 & 0.20 & 0.28 & 0.24
 & 1 & 4.25 & 0.20 & 0.76 & 0.23
 & 2 & 3.92 & 0.07 & 0.05 & 0.12
 & 1 & 3.80 & 0.20 & 0.37 & 0.22
 \\
Zn I & 4.56
 & 1 & 4.15 & 0.20 & -0.2 & 0.32
 &   &   &   &   &  
 & 1 & 4.46 & 0.20 & 0.25 & 0.28
 & 1 & 4.06 & 0.20 & 0.27 & 0.29
 \\
 \hline
Y II & 2.21
 & 2 & 3.30 & 0.20 & 1.29 & 0.22
 & 4 & 3.47 & 0.10 & 1.98 & 0.12
 & 1 & 3.25 & 0.20 & 1.39 & 0.24
 & 4 & 3.14 & 0.08 & 1.69 & 0.12
 \\
Zr II & 2.58
 & 3 & 3.73 & 0.11 & 1.36 & 0.15
 & 2 & 3.61 & 0.06 & 1.76 & 0.10
 & 2 & 3.45 & 0.12 & 1.22 & 0.15
 & 3 & 3.35 & 0.08 & 1.54 & 0.11
 \\
 \hline
La II & 1.10
 & 6 & 2.69 & 0.16 & 1.80 & 0.16
 & 8 & 2.82 & 0.08 & 2.45 & 0.11
 & 11 & 2.50 & 0.12 & 1.75 & 0.12
 & 6 & 2.65 & 0.20 & 2.32 & 0.15
 \\
Ce II & 1.58
 & 4 & 3.18 & 0.10 & 1.80 & 0.14
 & 5 & 3.12 & 0.06 & 2.27 & 0.10
 & 5 & 2.71 & 0.09 & 1.47 & 0.11
 & 4 & 2.66 & 0.06 & 1.84 & 0.11
 \\
Pr II & 0.72
 & 6 & 2.20 & 0.14 & 1.68 & 0.16
 & 7 & 2.27 & 0.09 & 2.28 & 0.14
 & 11 & 1.94 & 0.13 & 1.57 & 0.14
 & 13 & 1.81 & 0.07 & 1.86 & 0.13
 \\
Nd II & 1.42
 & 8 & 2.88 & 0.05 & 1.66 & 0.13
 & 23 & 3.18 & 0.11 & 2.49 & 0.13
 & 17 & 2.64 & 0.08 & 1.57 & 0.12
 & 7 & 2.61 & 0.05 & 1.96 & 0.13
 \\
Sm II & 0.96
 & 3 & 2.10 & 0.09 & 1.34 & 0.16
 & 2 & 1.99 & 0.12 & 1.76 & 0.16
 & 4 & 2.05 & 0.13 & 1.43 & 0.15
 & 3 & 1.67 & 0.09 & 1.47 & 0.14
 \\
Eu II & 0.52
 & 2 & 1.21 & 0.10 & 0.90 & 0.16
 & 2 & 1.25 & 0.09 & 1.46 & 0.14
 & 2 & 0.98 & 0.07 & 0.81 & 0.13
 & 2 & 0.89 & 0.06 & 1.14 & 0.12
 \\
 \hline
Gd II & 1.07
 & 3 & 2.29 & 0.09 & 1.43 & 0.16
 & 2 & 2.17 & 0.08 & 1.82 & 0.12
 & 3 & 2.00 & 0.07 & 1.28 & 0.13
 & 13 & 1.99 & 0.12 & 1.69 & 0.16
 \\
Tb II & 0.30
 &   &   &   &   &  
 &   &   &   &   &  
 & 1 & 1.70 & 0.20 & 1.75 & 0.27
 &   &   &   &   &  
 \\
Dy II & 1.10
 & 2 & 2.34 & 0.01 & 1.44 & 0.14
 & 2 & 2.25 & 0.20 & 1.88 & 0.20
 & 2 & 2.18 & 0.00 & 1.43 & 0.13
 & 8 & 2.13 & 0.09 & 1.80 & 0.11
 \\
Er II & 0.92
 & 4 & 2.15 & 0.07 & 1.43 & 0.14
 & 2 & 2.23 & 0.05 & 2.04 & 0.12
 &   &   &   &   &  
 & 3 & 1.74 & 0.04 & 1.58 & 0.10
 \\
Yb II & 0.84
 &   &   &   &   &  
 & 1 & 2.50 & 0.20 & 2.39 & 0.24
 &   &   &   &   &  
 & 1 & 2.00 & 0.20 & 1.93 & 0.24
 \\
Lu II & 0.10
 & 2 & 1.34 & 0.20 & 1.45 & 0.23
 & 2 & 1.74 & 0.01 & 2.36 & 0.12
 & 1 & 0.85 & 0.20 & 1.10 & 0.25
 & 2 & 1.25 & 0.20 & 1.92 & 0.21
 \\
Hf II & 0.85
 & 2 & 2.19 & 0.09 & 1.55 & 0.16
 & 1 & 2.15 & 0.20 & 2.02 & 0.23
 &   &   &   &   &  
 & 2 & 1.85 & 0.20 & 1.77 & 0.19
 \\
W II & 0.85
 &   &   &   &   &  
 & 1 & 3.07 & 0.20 & 2.95 & 0.25
 &   &   &   &   &  
 & 1 & 2.30 & 0.20 & 2.21 & 0.25
 \\
\hline
\end{tabular}
\begin{tablenotes}
 \item  \textbf{Notes:} For each ion, the table lists    
the used number of lines (N) for the abundance determination, 
the determined abundance (log $\epsilon$), the uncertainty on this abundance due to line-to-line scatter ($\sigma_{\textrm{l2l}}$), 
the element over iron ratio ([X/Fe]) and the total uncertainty on [X/Fe] ($\sigma_{tot}$) which includes line-to-line scatter 
and atmospheric parameter uncertainty (see text for details). 
 \item The solar abundances (log $\epsilon_{\odot}$) in the second column are retrieved from \citet{asplund09}.
 \item For all elemental abundances determined via only one spectral line or via spectrum synthesis, we adopt a line-to-line scatter of 
 0.20 dex.
\end{tablenotes}
\end{threeparttable}
\end{table*}

\begin{table*}[tb!]
\caption{\label{table:abun_middle1} Abundance results of IRAS 06530-0213, IRAS 22223+4327, 
and both observations of IRAS08143-4406. The list of IRAS08143 BLUE437 contains the results 
for the Blue437 and Red860 spectra, the list of 
IRAS08143 RED580 shows the results for the Red580 spectra (see Sect. \ref{sect:obs}).}
\begin{threeparttable}
\setlength{\tabcolsep}{0.10cm}
\begin{tabular}{| lc | ccccc | ccccc | ccccc | ccccc| } \hline
  &  &  \multicolumn{5}{c|}{IRAS06530}  &  \multicolumn{5}{c|}{IRAS08143 BLUE437}
  &  \multicolumn{5}{c|}{IRAS08143 RED580} &  \multicolumn{5}{c|}{IRAS22223} \\
  &  &  \multicolumn{5}{c|}{[Fe/H] = -0.32}  &  \multicolumn{5}{c|}{[Fe/H] = -0.43}  
  &  \multicolumn{5}{c|}{[Fe/H] = -0.36} &  \multicolumn{5}{c|}{[Fe/H] = -0.30}  \\
 &  &  \multicolumn{5}{c|}{T$_{\textrm{eff}}$= 7375K}  &  \multicolumn{5}{c|}{ T$_{\textrm{eff}}$ = 7000 K}  
  &  \multicolumn{5}{c|}{T$_{\textrm{eff}}$ = 7250 K} &  \multicolumn{5}{c|}{T$_{\textrm{eff}}$ = 6500 K}  \\
  &  &  \multicolumn{5}{c|}{$\log g$ = 1.25 dex}  &  \multicolumn{5}{c|}{ $\log g$ =  1.5 dex}  
  &  \multicolumn{5}{c|}{$\log g$ = 1.50 dex} &  \multicolumn{5}{c|}{$\log g$ = 1.00 dex}  \\   
  
\hline
Ion & log $\epsilon_{\odot}$ 
& N  &  log $\epsilon$  &  $\sigma_{\textrm{l2l}}$  &  [X/Fe] &  $\sigma_{tot}$  
& N  &  log $\epsilon$  &  $\sigma_{\textrm{l2l}}$  &  [X/Fe] &  $\sigma_{tot}$ 
& N  &  log $\epsilon$  &  $\sigma_{\textrm{l2l}}$  &  [X/Fe] &  $\sigma_{tot}$
& N  &  log $\epsilon$  &  $\sigma_{\textrm{l2l}}$  &  [X/Fe] &  $\sigma_{tot}$ \\
\hline
C I & 8.43
 & 10 & 8.94 & 0.07 & 0.83 & 0.13
 & 15 & 8.71 & 0.06 & 0.71 & 0.10
 & 11 & 8.74 & 0.09 & 0.67 & 0.12
 & 19 & 8.72 & 0.06 & 0.59 & 0.07
 \\
N I & 7.83
 &   &   &   &   &  
 & 7 & 7.41 & 0.30 & 0.01 & 0.22
 &   &   &   &   &  
 & 2 & 7.68 & 0.30 & 0.15 & 0.21
 \\
O I & 8.69
 & 6 & 8.72 & 0.08 & 0.35 & 0.11
 &   &   &   &   &  
 & 3 & 8.52 & 0.04 & 0.19 & 0.13
 & 5 & 8.70 & 0.06 & 0.31 & 0.13
 \\
 \hline 
Na I & 6.24
 & 1 & 6.32 & 0.20 & 0.46 & 0.22
 &   &   &   &   &  
 & 2 & 6.35 & 0.08 & 0.48 & 0.11
 & 2 & 6.20 & 0.04 & 0.28 & 0.08
 \\
Mg I & 7.60
 & 1 & 7.22 & 0.20 & -0.0 & 0.23
 & 4 & 7.21 & 0.06 & 0.05 & 0.09
 & 2 & 7.22 & 0.04 & -0.01 & 0.10
 & 1 & 7.30 & 0.20 & 0.02 & 0.22
 \\
Al I & 6.45
 &   &   &   &   &  
 & 2 & 6.40 & 0.01 & 0.40 & 0.09
 & 1 & 6.42 & 0.20 & 0.34 & 0.22
 & 2 & 6.10 & 0.04 & -0.03 & 0.09
 \\
Si I & 7.51
 & 2 & 7.34 & 0.13 & 0.15 & 0.22
 & 4 & 7.54 & 0.06 & 0.46 & 0.17
 & 7 & 7.66 & 0.08 & 0.51 & 0.18
 & 14 & 7.42 & 0.10 & 0.21 & 0.17
 \\
Si II & 7.51
 & 1 & 7.36 & 0.20 & 0.17 & 0.24
 &   &   &   &   &  
 &   &   &   &   &  
 &   &   &   &   &  
 \\
P I & 5.41
 &   &   &   &   &  
 & 1 & 5.80 & 0.20 & 0.82 & 0.26
 &   &   &   &   &  
 &   &   &   &   &  
 \\
S I & 7.12
 & 3 & 7.00 & 0.05 & 0.21 & 0.18
 & 7 & 7.10 & 0.08 & 0.41 & 0.14
 & 3 & 7.10 & 0.01 & 0.34 & 0.15
 & 5 & 6.94 & 0.08 & 0.12 & 0.12
 \\
Ca I & 6.34
 & 12 & 6.18 & 0.09 & 0.22 & 0.09
 & 3 & 6.15 & 0.07 & 0.26 & 0.09
 & 12 & 6.16 & 0.09 & 0.18 & 0.08
 & 8 & 6.01 & 0.08 & -0.01 & 0.07
 \\
Ca II & 6.34
 & 2 & 6.08 & 0.01 & 0.06 & 0.10
 &   &   &   &   &  
 & 4 & 6.11 & 0.07 & 0.13 & 0.10
 & 3 & 6.00 & 0.07 & -0.04 & 0.10
 \\
Sc II & 3.15
 & 5 & 3.15 & 0.03 & 0.33 & 0.09
 &   &   &   &   &  
 & 2 & 3.09 & 0.04 & 0.30 & 0.10
 & 4 & 2.95 & 0.10 & 0.10 & 0.11
 \\
Ti I & 4.95
 &   &   &   &   &  
 & 5 & 5.03 & 0.13 & 0.53 & 0.11
 & 5 & 5.06 & 0.07 & 0.47 & 0.09
 &   &   &   &   &  
 \\
Ti II & 4.95
 & 5 & 4.79 & 0.05 & 0.16 & 0.09
 &   &   &   &   &  
 & 5 & 5.01 & 0.09 & 0.42 & 0.10
 & 5 & 4.54 & 0.12 & -0.11 & 0.11
 \\
V II & 3.93
 &   &   &   &   &  
 & 2 & 3.75 & 0.01 & 0.25 & 0.12
 & 1 & 3.74 & 0.20 & 0.17 & 0.24
 & 4 & 3.64 & 0.17 & 0.01 & 0.15
 \\
Cr I & 5.64
 & 2 & 5.45 & 0.03 & 0.19 & 0.09
 & 4 & 5.33 & 0.07 & 0.13 & 0.09
 & 4 & 5.41 & 0.11 & 0.13 & 0.11
 & 6 & 5.22 & 0.11 & -0.1 & 0.09
 \\
Cr II & 5.64
 & 11 & 5.36 & 0.07 & 0.04 & 0.07
 & 2 & 5.31 & 0.06 & 0.10 & 0.11
 & 13 & 5.31 & 0.06 & 0.03 & 0.07
 & 11 & 5.29 & 0.06 & -0.05 & 0.06
 \\
Mn I & 5.43
 & 2 & 4.99 & 0.20 & -0.07 & 0.19
 & 3 & 5.05 & 0.03 & 0.07 & 0.08
 & 2 & 5.05 & 0.20 & -0.02 & 0.19
 & 2 & 4.98 & 0.03 & -0.13 & 0.08
 \\
Mn II & 5.43
 &   &   &   &   &  
 & 1 & 5.08 & 0.20 & 0.08 & 0.23
 & 1 & 5.05 & 0.20 & -0.01 & 0.23
 &   &   &   &   &  
 \\
Fe I & 7.50
 & 54 & 7.12 & 0.08 & 0.00 & 0.05
 & 17 & 7.05 & 0.08 & 0.00 & 0.06
 & 70 & 7.13 & 0.08 & 0.00 & 0.04
 & 80 & 7.18 & 0.06 & 0.00 & 0.04
 \\
Fe II & 7.50
 & 24 & 7.18 & 0.08 & 0.00 & 0.06
 & 8 & 7.07 & 0.10 & 0.00 & 0.10
 & 23 & 7.14 & 0.09 & 0.00 & 0.06
 & 25 & 7.20 & 0.07 & 0.00 & 0.05
 \\
Co II & 4.99
 &   &   &   &   &  
 &   &   &   &   &  
 &   &   &   &   &  
 & 1 & 4.71 & 0.20 & 0.02 & 0.23
 \\
Ni I & 6.22
 & 5 & 6.07 & 0.05 & 0.23 & 0.08
 & 13 & 6.06 & 0.08 & 0.28 & 0.07
 & 4 & 6.08 & 0.05 & 0.22 & 0.08
 & 13 & 5.87 & 0.07 & -0.03 & 0.06
 \\
Ni II & 6.22
 & 1 & 6.15 & 0.20 & 0.25 & 0.23
 &   &   &   &   &  
 &   &   &   &   &  
 &   &   &   &   &  
 \\
Cu I & 4.19
 & 1 & 4.39 & 0.20 & 0.58 & 0.22
 &   &   &   &   &  
 & 1 & 4.32 & 0.20 & 0.50 & 0.22
 & 1 & 4.09 & 0.20 & 0.22 & 0.22
 \\
Zn I & 4.56
 & 1 & 4.70 & 0.20 & 0.46 & 0.30
 &   &   &   &   &  
 & 1 & 4.60 & 0.20 & 0.40 & 0.29
 & 3 & 4.36 & 0.09 & 0.10 & 0.19
 \\
 \hline
Y II & 2.21
 & 2 & 3.79 & 0.11 & 1.91 & 0.16
 & 3 & 3.63 & 0.04 & 1.85 & 0.11
 & 2 & 3.65 & 0.02 & 1.80 & 0.12
 & 3 & 3.24 & 0.08 & 1.33 & 0.11
 \\
Zr II & 2.58
 & 2 & 3.86 & 0.04 & 1.60 & 0.10
 & 5 & 3.85 & 0.13 & 1.70 & 0.12
 & 4 & 3.85 & 0.10 & 1.63 & 0.11
 & 3 & 3.63 & 0.06 & 1.35 & 0.10
 \\
Mo II & 1.88
 &   &   &   &   &  
 &   &   &   &   &  
 &   &   &   &   &  
 & 1 & 2.89 & 0.20 & 1.31 & 0.23
 \\
 \hline
Ba II & 2.18
 &   &   &   &   &  
 & 1 & 3.45 & 0.20 & 1.70 & 0.27
 &   &   &   &   &  
 & 1 & 2.89 & 0.20 & 1.01 & 0.26
 \\
La II & 1.10
 & 7 & 3.06 & 0.05 & 2.28 & 0.16
 & 18 & 2.47 & 0.05 & 1.80 & 0.11
 & 14 & 2.46 & 0.10 & 1.72 & 0.14
 & 14 & 1.87 & 0.11 & 1.07 & 0.12
 \\
Ce II & 1.58
 & 10 & 3.31 & 0.12 & 2.05 & 0.15
 & 7 & 2.61 & 0.04 & 1.46 & 0.11
 & 5 & 2.56 & 0.10 & 1.35 & 0.14
 & 5 & 2.30 & 0.10 & 1.02 & 0.13
 \\
Pr II & 0.72
 & 5 & 2.75 & 0.07 & 2.35 & 0.19
 & 6 & 2.15 & 0.07 & 1.85 & 0.16
 & 9 & 2.10 & 0.09 & 1.74 & 0.17
 & 5 & 1.44 & 0.07 & 1.02 & 0.15
 \\
Nd II & 1.42
 & 16 & 3.31 & 0.07 & 2.21 & 0.18
 & 12 & 2.70 & 0.06 & 1.71 & 0.14
 & 19 & 2.67 & 0.09 & 1.61 & 0.16
 & 13 & 1.96 & 0.11 & 0.85 & 0.15
 \\
Sm II & 0.96
 & 2 & 2.25 & 0.04 & 1.61 & 0.17
 & 16 & 1.90 & 0.04 & 1.37 & 0.12
 & 6 & 1.91 & 0.08 & 1.31 & 0.15
 & 8 & 1.25 & 0.10 & 0.59 & 0.14
 \\
Eu II & 0.52
 & 2 & 1.56 & 0.01 & 1.36 & 0.17
 &   &   &   &   &  
 & 2 & 1.10 & 0.01 & 0.94 & 0.14
 & 2 & 0.78 & 0.14 & 0.56 & 0.16
 \\
 \hline
Gd II & 1.07
 & 5 & 2.59 & 0.11 & 1.84 & 0.13
 & 8 & 2.00 & 0.06 & 1.36 & 0.09
 & 2 & 2.02 & 0.01 & 1.31 & 0.11
 & 1 & 1.67 & 0.20 & 0.90 & 0.23
 \\
Dy II & 1.10
 &   &   &   &   &  
 &   &   &   &   &  
 &   &   &   &   &  
 & 1 & 1.80 & 0.20 & 1.00 & 0.24
 \\
Er II & 0.92
 & 2 & 2.88 & 0.07 & 2.29 & 0.15
 & 2 & 2.20 & 0.20 & 1.71 & 0.20
 & 1 & 2.20 & 0.20 & 1.64 & 0.24
 & 1 & 1.18 & 0.20 & 0.56 & 0.23
 \\
Tm II & 0.10
 & 1 & 1.45 & 0.20 & 1.67 & 0.24
 &   &   &   &   &  
 &   &   &   &   &  
 &   &   &   &   &  
 \\
Yb II & 0.84
 & 1 & 2.72 & 0.20 & 2.21 & 0.25
 & 1 & 2.15 & 0.20 & 1.74 & 0.25
 & 1 & 2.12 & 0.20 & 1.64 & 0.25
 &   &   &   &   &  
 \\
Lu II & 0.10
 & 3 & 1.85 & 0.20 & 2.07 & 0.19
 &   &   &   &   &  
 & 1 & 1.00 & 0.20 & 1.26 & 0.25
 &   &   &   &   &  
 \\
Hf II & 0.85
 & 1 & 2.43 & 0.20 & 1.91 & 0.22
 &   &   &   &   &  
 &   &   &   &   &  
 &   &   &   &   &  
 \\
W II & 0.85
 & 1 & 2.98 & 0.20 & 2.45 & 0.25
 &   &   &   &   &  
 & 1 & 2.15 & 0.20 & 1.66 & 0.25
 & 1 & 1.45 & 0.20 & 0.90 & 0.25
 \\
\hline
\end{tabular}
\begin{tablenotes}
  \item \textbf{Notes:} Same as for Table \ref{table:abun_cold}.
\end{tablenotes}
\end{threeparttable}
\end{table*}

\begin{table*}[tb!]
\caption{\label{table:abun_middle2} Abundance results of IRAS 07134+1005, IRAS 08281-4850, IRAS 17279-1119, and IRAS 19500-1709.}
\begin{threeparttable}
\setlength{\tabcolsep}{0.10cm}
\begin{tabular}{| lc | ccccc | ccccc | ccccc | ccccc| } \hline
  &  &  \multicolumn{5}{c|}{IRAS07134}  &  \multicolumn{5}{c|}{IRAS08281}  
  &  \multicolumn{5}{c|}{IRAS17279} &  \multicolumn{5}{c|}{IRAS19500} \\
  &  &  \multicolumn{5}{c|}{[Fe/H] = -0.91}  &  \multicolumn{5}{c|}{[Fe/H] = -0.26}  
  &  \multicolumn{5}{c|}{[Fe/H] = -0.64} &  \multicolumn{5}{c|}{[Fe/H] = -0.59}  \\
 &  &  \multicolumn{5}{c|}{T$_{\textrm{eff}}$= 7250 K}  &  \multicolumn{5}{c|}{T$_{\textrm{eff}}$ = 7875 K}  
  &  \multicolumn{5}{c|}{T$_{\textrm{eff}}$ = 7250 K} &  \multicolumn{5}{c|}{ T$_{\textrm{eff}}$ = 8000 K}  \\
  &  &  \multicolumn{5}{c|}{ $\log g$ = 0.5 dex}  &  \multicolumn{5}{c|}{ $\log g$ =  1.25 dex}  
  &  \multicolumn{5}{c|}{ $\log g$ = 1.25 dex} &  \multicolumn{5}{c|}{$\log g$ = 1.00 dex}  \\  
\hline
Ion & log $\epsilon_{\odot}$ 
& N  &  log $\epsilon$  &  $\sigma_{\textrm{l2l}}$  &  [X/Fe] &  $\sigma_{tot}$  
& N  &  log $\epsilon$  &  $\sigma_{\textrm{l2l}}$  &  [X/Fe] &  $\sigma_{tot}$ 
& N  &  log $\epsilon$  &  $\sigma_{\textrm{l2l}}$  &  [X/Fe] &  $\sigma_{tot}$
& N  &  log $\epsilon$  &  $\sigma_{\textrm{l2l}}$  &  [X/Fe] &  $\sigma_{tot}$ \\
\hline
C I & 8.43
 & 47 & 8.68 & 0.08 & 1.16 & 0.22
 & 16 & 8.91 & 0.05 & 0.75 & 0.21
 & 17 & 8.36 & 0.09 & 0.47 & 0.14
 & 15 & 8.82 & 0.06 & 0.99 & 0.33
 \\
N I & 7.83
 & 3 & 7.50 & 0.30 & 0.57 & 0.19
 &   &   &   &   &  
 & 1 & 7.27 & 0.30 & -0.07 & 0.35
 & 3 & 7.65 & 0.30 & 0.41 & 0.17
 \\
O I & 8.69
 & 9 & 8.59 & 0.06 & 0.81 & 0.19
 & 5 & 8.54 & 0.06 & 0.12 & 0.11
 & 2 & 8.39 & 0.05 & 0.24 & 0.16
 & 8 & 8.81 & 0.04 & 0.72 & 0.09
 \\
 \hline
Na I & 6.24
 &   &   &   &   &  
 & 1 & 6.65 & 0.20 & 0.70 & 0.22
 & 2 & 6.25 & 0.05 & 0.52 & 0.11
 &   &   &   &   &  
 \\
Mg I & 7.60
 & 4 & 6.89 & 0.06 & 0.26 & 0.13
 & 3 & 7.38 & 0.09 & 0.07 & 0.10
 & 2 & 7.25 & 0.06 & 0.16 & 0.12
 & 2 & 7.31 & 0.01 & 0.29 & 0.07
 \\
Mg II & 7.60
 &   &   &   &   &  
 &   &   &   &   &  
 &   &   &   &   &  
 & 3 & 7.29 & 0.00 & 0.28 & 0.10
 \\
Al I & 6.45
 &   &   &   &   &  
 &   &   &   &   &  
 &   &   &   &   &  
 & 1 & 5.51 & 0.20 & -0.36 & 0.21
 \\
Si I & 7.51
 &   &   &   &   &  
 & 2 & 7.50 & 0.20 & 0.25 & 0.31
 & 6 & 7.33 & 0.10 & 0.36 & 0.20
 & 1 & 7.03 & 0.20 & 0.11 & 0.49
 \\
Si II & 7.51
 &   &   &   &   &  
 &   &   &   &   &  
 & 1 & 7.33 & 0.20 & 0.36 & 0.26
 & 2 & 7.01 & 0.06 & 0.10 & 0.13
 \\
S I & 7.12
 & 1 & 6.59 & 0.20 & 0.38 & 0.35
 & 3 & 7.24 & 0.03 & 0.38 & 0.24
 & 2 & 6.91 & 0.02 & 0.33 & 0.18
 &   &   &   &   &  
 \\
K I & 5.03
 & 2 & 4.68 & 0.07 & 0.61 & 0.12
 &   &   &   &   &  
 &   &   &   &   &  
 &   &   &   &   &  
 \\
Ca I & 6.34
 & 11 & 5.44 & 0.05 & 0.07 & 0.11
 & 5 & 6.29 & 0.08 & 0.25 & 0.15
 & 7 & 6.00 & 0.07 & 0.17 & 0.09
 & 3 & 6.08 & 0.06 & 0.32 & 0.23
 \\
Ca II & 6.34
 & 3 & 5.47 & 0.03 & 0.04 & 0.18
 & 2 & 6.18 & 0.03 & 0.10 & 0.16
 & 1 & 6.00 & 0.20 & 0.20 & 0.25
 & 3 & 5.92 & 0.08 & 0.18 & 0.26
 \\
Sc II & 3.15
 & 8 & 2.41 & 0.06 & 0.17 & 0.09
 & 5 & 3.19 & 0.04 & 0.30 & 0.13
 & 7 & 2.71 & 0.09 & 0.10 & 0.12
 & 8 & 2.78 & 0.11 & 0.23 & 0.22
 \\
Ti I & 4.95
 &   &   &   &   &  
 &   &   &   &   &  
 & 3 & 4.62 & 0.06 & 0.18 & 0.12
 &   &   &   &   &  
 \\
Ti II & 4.95
 & 24 & 4.18 & 0.09 & 0.14 & 0.08
 & 12 & 4.74 & 0.04 & 0.05 & 0.09
 & 16 & 4.65 & 0.13 & 0.24 & 0.10
 & 33 & 4.54 & 0.09 & 0.19 & 0.14
 \\
V II & 3.93
 & 5 & 3.32 & 0.10 & 0.30 & 0.11
 & 1 & 3.68 & 0.20 & 0.02 & 0.24
 & 5 & 3.92 & 0.10 & 0.53 & 0.13
 & 11 & 3.62 & 0.10 & 0.28 & 0.12
 \\
Cr I & 5.64
 & 3 & 4.70 & 0.14 & 0.03 & 0.16
 & 1 & 5.40 & 0.20 & 0.05 & 0.22
 & 2 & 5.17 & 0.03 & 0.04 & 0.11
 & 3 & 5.12 & 0.10 & 0.06 & 0.11
 \\
Cr II & 5.64
 & 14 & 4.85 & 0.09 & 0.12 & 0.10
 & 12 & 5.33 & 0.05 & -0.05 & 0.06
 & 13 & 5.14 & 0.11 & 0.04 & 0.10
 & 24 & 5.14 & 0.10 & 0.10 & 0.07
 \\
Mn I & 5.43
 & 2 & 4.56 & 0.04 & 0.10 & 0.09
 &   &   &   &   &  
 &   &   &   &   &  
 & 2 & 4.75 & 0.03 & -0.09 & 0.09
 \\
Mn II & 5.43
 & 4 & 4.52 & 0.08 & 0.00 & 0.16
 &   &   &   &   &  
 & 2 & 5.03 & 0.10 & 0.13 & 0.15
 & 5 & 4.91 & 0.04 & 0.07 & 0.06
 \\
Fe I & 7.50
 & 42 & 6.53 & 0.07 & 0.00 & 0.05
 & 44 & 7.21 & 0.07 & 0.00 & 0.04
 & 44 & 6.99 & 0.10 & 0.00 & 0.06
 & 33 & 6.92 & 0.06 & 0.00 & 0.04
 \\
Fe II & 7.50
 & 32 & 6.59 & 0.09 & 0.00 & 0.06
 & 23 & 7.24 & 0.07 & 0.00 & 0.05
 & 30 & 6.96 & 0.13 & 0.00 & 0.08
 & 32 & 6.91 & 0.03 & 0.00 & 0.03
 \\
Co I & 4.99
 &   &   &   &   &  
 &   &   &   &   &  
 & 2 & 4.28 & 0.14 & -0.2 & 0.17
 &   &   &   &   &  
 \\
Co II & 4.99
 &   &   &   &   &  
 &   &   &   &   &  
 &   &   &   &   &  
 & 1 & 4.70 & 0.20 & 0.31 & 0.22
 \\
Ni I & 6.22
 & 5 & 5.34 & 0.08 & 0.09 & 0.09
 &   &   &   &   &  
 & 4 & 6.13 & 0.11 & 0.43 & 0.12
 & 3 & 5.79 & 0.08 & 0.15 & 0.08
 \\
Ni II & 6.22
 & 1 & 5.32 & 0.20 & 0.01 & 0.26
 & 2 & 5.90 & 0.20 & -0.06 & 0.19
 & 4 & 6.07 & 0.07 & 0.39 & 0.11
 & 2 & 5.80 & 0.07 & 0.18 & 0.09
 \\
Zn I & 4.56
 &   &   &   &   &  
 &   &   &   &   &  
 & 1 & 4.25 & 0.20 & 0.23 & 0.31
 &   &   &   &   &  
 \\
 \hline
Y II & 2.21
 & 12 & 2.97 & 0.12 & 1.67 & 0.13
 & 7 & 3.66 & 0.06 & 1.72 & 0.19
 & 10 & 2.38 & 0.12 & 0.71 & 0.12
 & 11 & 3.02 & 0.11 & 1.40 & 0.29
 \\
Zr II & 2.58
 & 15 & 3.28 & 0.09 & 1.61 & 0.09
 & 12 & 3.74 & 0.10 & 1.42 & 0.11
 & 10 & 2.82 & 0.11 & 0.77 & 0.12
 & 22 & 3.32 & 0.10 & 1.34 & 0.19
 \\
Mo II & 1.88
 & 1 & 2.68 & 0.20 & 1.71 & 0.27
 & 1 & 3.35 & 0.20 & 1.73 & 0.23
 &   &   &   &   &  
 & 2 & 2.88 & 0.08 & 1.60 & 0.12
 \\
 \hline
Ba II & 2.18
 & 2 & 3.09 & 0.02 & 1.82 & 0.30
 & 1 & 3.80 & 0.20 & 1.88 & 0.36
 & 1 & 2.80 & 0.20 & 1.16 & 0.31
 & 1 & 3.00 & 0.20 & 1.42 & 0.49
 \\
La II & 1.10
 & 29 & 2.04 & 0.09 & 1.85 & 0.24
 & 17 & 2.70 & 0.11 & 1.87 & 0.28
 & 7 & 1.48 & 0.13 & 0.92 & 0.17
 & 5 & 1.87 & 0.09 & 1.37 & 0.43
 \\
Ce II & 1.58
 & 41 & 2.27 & 0.11 & 1.59 & 0.20
 & 21 & 2.85 & 0.08 & 1.54 & 0.24
 & 11 & 2.02 & 0.10 & 0.98 & 0.15
 & 7 & 2.28 & 0.12 & 1.29 & 0.39
 \\
Pr II & 0.72
 & 5 & 1.43 & 0.08 & 1.62 & 0.31
 & 3 & 2.40 & 0.20 & 1.94 & 0.34
 & 5 & 1.35 & 0.11 & 1.17 & 0.21
 &   &   &   &   &  
 \\
Nd II & 1.42
 & 29 & 2.14 & 0.08 & 1.63 & 0.29
 & 15 & 2.83 & 0.09 & 1.67 & 0.30
 & 8 & 1.64 & 0.07 & 0.75 & 0.19
 & 5 & 2.31 & 0.03 & 1.48 & 0.44
 \\
Sm II & 0.96
 & 13 & 1.49 & 0.07 & 1.44 & 0.26
 &   &   &   &   &  
 & 3 & 1.46 & 0.08 & 1.03 & 0.19
 &   &   &   &   &  
 \\
Eu II & 0.52
 & 2 & 0.20 & 0.20 & 0.59 & 0.32
 & 2 & 1.00 & 0.20 & 0.74 & 0.33
 &   &   &   &   &  
 & 1 & 1.10 & 0.20 & 1.18 & 0.48
 \\
 \hline
Gd II & 1.07
 & 6 & 1.33 & 0.08 & 1.17 & 0.19
 & 2 & 1.98 & 0.02 & 1.18 & 0.18
 &   &   &   &   &  
 & 2 & 1.52 & 0.18 & 1.05 & 0.34
 \\
Dy II & 1.10
 & 2 & 1.14 & 0.20 & 0.95 & 0.28
 & 1 & 1.90 & 0.20 & 1.06 & 0.33
 &   &   &   &   &  
 &   &   &   &   &  
 \\
Er II & 0.92
 & 2 & 1.41 & 0.00 & 1.40 & 0.22
 & 1 & 2.19 & 0.20 & 1.54 & 0.30
 &   &   &   &   &  
 &   &   &   &   &  
 \\
Lu II & 0.10
 & 4 & 0.80 & 0.20 & 1.61 & 0.22
 & 1 & 1.30 & 0.20 & 1.46 & 0.25
 & 1 & 0.50 & 0.20 & 0.94 & 0.27
 & 2 & 0.64 & 0.10 & 1.13 & 0.21
 \\
Hf II & 0.85
 &   &   &   &   &  
 & 2 & 2.35 & 0.20 & 1.76 & 0.20
 & 1 & 1.04 & 0.20 & 0.72 & 0.25
 & 1 & 1.53 & 0.20 & 1.27 & 0.25
 \\
\hline
\end{tabular}
\begin{tablenotes}
  \item \textbf{Notes:} Same as for Table \ref{table:abun_cold}.
\end{tablenotes}
\end{threeparttable}
\end{table*}

\begin{table*}[tb!]
\begin{center}
\caption{\label{table:abun_hot} Abundance results of IRAS 13245-5036, IRAS 14325-6428, and IRAS 14429-4539.}
\begin{threeparttable}
\setlength{\tabcolsep}{0.10cm}
\begin{tabular}{| lc | ccccc | ccccc | ccccc |} \hline
  &  &  \multicolumn{5}{c|}{IRAS13245}  & \multicolumn{5}{c|}{IRAS14325}  & \multicolumn{5}{c|}{IRAS14429}  \\
  &  &  \multicolumn{5}{c|}{[Fe/H] = -0.30}  & \multicolumn{5}{c|}{[Fe/H] = -0.56} & \multicolumn{5}{c|}{[Fe/H] = -0.18}  \\
 &  &  \multicolumn{5}{c|}{T$_{\textrm{eff}}$= 9500 K}  &  \multicolumn{5}{c|}{T$_{\textrm{eff}}$ = 8000 K}  
  &  \multicolumn{5}{c|}{T$_{\textrm{eff}}$ = 9375 K}   \\
  &  &  \multicolumn{5}{c|}{$\log g$ = 2.75 dex}  &  \multicolumn{5}{c|}{ $\log g$ =  1.00 dex}  
  &  \multicolumn{5}{c|}{ $\log g$ = 2.50 dex} \\
\hline
Ion & log $\epsilon_{\odot}$ 
& N  &  log $\epsilon$  &  $\sigma_{\textrm{l2l}}$  &  [X/Fe] &  $\sigma_{tot}$
& N  &  log $\epsilon$  &  $\sigma_{\textrm{l2l}}$  &  [X/Fe] &  $\sigma_{tot}$
& N  &  log $\epsilon$  &  $\sigma_{\textrm{l2l}}$  &  [X/Fe] &  $\sigma_{tot}$  \\
\hline
C I & 8.43
 & 4 & 8.70 & 0.00 & 0.57 & 0.21
 & 24 & 9.06 & 0.10 & 1.18 & 0.23
 & 3 & 8.93 & 0.08 & 0.68 & 0.23
 \\
N I & 7.83
 &   &   &   &   &  
 & 7 & 7.45 & 0.30 & 0.18 & 0.20
 &   &   &   &   &  
 \\
O I & 8.69
 & 5 & 8.65 & 0.12 & 0.26 & 0.13
 & 10 & 8.70 & 0.06 & 0.57 & 0.09
 & 6 & 8.82 & 0.04 & 0.31 & 0.12
 \\
 \hline
Na I & 6.24
 &   &   &   &   &  
 & 1 & 6.50 & 0.20 & 0.81 & 0.22
 &   &   &   &   &  
 \\
Mg I & 7.60
 & 5 & 7.16 & 0.07 & -0.09 & 0.09
 & 4 & 7.51 & 0.05 & 0.47 & 0.08
 & 3 & 7.32 & 0.04 & -0.01 & 0.09
 \\
Mg II & 7.60
 & 2 & 7.20 & 0.07 & -0.11 & 0.11
 & 5 & 7.37 & 0.13 & 0.33 & 0.14
 & 3 & 7.29 & 0.03 & -0.13 & 0.10
 \\
Al I & 6.45
 & 1 & 5.75 & 0.20 & -0.35 & 0.23
 & 1 & 5.70 & 0.20 & -0.2 & 0.21
 & 2 & 5.89 & 0.01 & -0.3 & 0.09
 \\
Al II & 6.45
 &   &   &   &   &  
 &   &   &   &   &  
 & 1 & 5.97 & 0.20 & -0.3 & 0.26
 \\
Si I & 7.51
 & 1 & 7.19 & 0.20 & -0.02 & 0.33
 & 1 & 6.85 & 0.20 & -0.1 & 0.38
 & 1 & 7.10 & 0.20 & -0.23 & 0.33
 \\
Si II & 7.51
 & 3 & 7.19 & 0.07 & -0.02 & 0.12
 & 2 & 6.83 & 0.00 & -0.12 & 0.18
 & 1 & 7.15 & 0.20 & -0.18 & 0.25
 \\
S I & 7.12
 &   &   &   &   &  
 & 1 & 7.20 & 0.20 & 0.64 & 0.32
 &   &   &   &   &  
 \\
Ca I & 6.34
 &   &   &   &   &  
 & 5 & 6.23 & 0.10 & 0.44 & 0.21
 &   &   &   &   &  
 \\
Ca II & 6.34
 &   &   &   &   &  
 & 2 & 6.26 & 0.03 & 0.48 & 0.20
 &   &   &   &   &  
 \\
Sc II & 3.15
 & 6 & 3.29 & 0.11 & 0.44 & 0.18
 & 11 & 2.97 & 0.11 & 0.38 & 0.19
 & 8 & 3.38 & 0.12 & 0.41 & 0.20
 \\
Ti II & 4.95
 & 22 & 4.96 & 0.10 & 0.31 & 0.12
 & 28 & 4.57 & 0.07 & 0.18 & 0.12
 & 42 & 5.03 & 0.10 & 0.26 & 0.12
 \\
V II & 3.93
 & 2 & 3.73 & 0.03 & 0.10 & 0.11
 & 11 & 3.60 & 0.09 & 0.23 & 0.12
 & 4 & 3.99 & 0.09 & 0.24 & 0.13
 \\
Cr I & 5.64
 &   &   &   &   &  
 & 3 & 5.28 & 0.09 & 0.19 & 0.10
 &   &   &   &   &  
 \\
Cr II & 5.64
 & 11 & 5.44 & 0.07 & 0.10 & 0.07
 & 22 & 5.19 & 0.07 & 0.11 & 0.05
 & 28 & 5.51 & 0.09 & 0.05 & 0.07
 \\
Mn I & 5.43
 &   &   &   &   &  
 & 1 & 4.94 & 0.20 & 0.06 & 0.23
 &   &   &   &   &  
 \\
Mn II & 5.43
 &   &   &   &   &  
 &   &   &   &   &  
 & 2 & 5.30 & 0.04 & 0.05 & 0.10
 \\
Fe I & 7.50
 & 21 & 7.15 & 0.10 & 0.00 & 0.07
 & 26 & 6.95 & 0.06 & 0.00 & 0.04
 & 26 & 7.24 & 0.09 & 0.00 & 0.06
 \\
Fe II & 7.50
 & 41 & 7.20 & 0.06 & 0.00 & 0.04
 & 58 & 6.94 & 0.06 & 0.00 & 0.04
 & 52 & 7.32 & 0.09 & 0.00 & 0.05
 \\
Ni I & 6.22
 &   &   &   &   &  
 & 3 & 5.78 & 0.06 & 0.12 & 0.08
 &   &   &   &   &  
 \\
Ni II & 6.22
 & 4 & 5.93 & 0.07 & 0.01 & 0.09
 & 4 & 5.74 & 0.10 & 0.08 & 0.10
 & 4 & 5.97 & 0.10 & -0.07 & 0.11
 \\
 \hline
Y II & 2.21
 & 11 & 3.31 & 0.04 & 1.40 & 0.23
 & 10 & 2.99 & 0.12 & 1.34 & 0.26
 & 10 & 3.15 & 0.10 & 1.13 & 0.25
 \\
Zr II & 2.58
 & 17 & 4.00 & 0.08 & 1.72 & 0.15
 & 17 & 3.18 & 0.12 & 1.16 & 0.16
 & 13 & 3.86 & 0.10 & 1.46 & 0.17
 \\
 \hline
Ba II & 2.18
 & 1 & 4.00 & 0.20 & 2.12 & 0.34
 & 1 & 3.10 & 0.20 & 1.48 & 0.42
 &   &   &   &   &  
 \\
La II & 1.10
 & 3 & 3.02 & 0.10 & 2.22 & 0.28
 & 7 & 1.85 & 0.06 & 1.31 & 0.35
 & 4 & 2.61 & 0.11 & 1.69 & 0.29
 \\
Ce II & 1.58
 & 4 & 3.40 & 0.20 & 2.12 & 0.28
 & 3 & 2.29 & 0.20 & 1.27 & 0.35
 &   &   &   &   &  
 \\
Pr III & 0.72
 & 2 & 2.83 & 0.08 & 2.41 & 0.11
 &   &   &   &   &  
 &   &   &   &   &  
 \\
Nd II & 1.42
 &   &   &   &   &  
 & 2 & 2.25 & 0.20 & 1.39 & 0.40
 &   &   &   &   &  
 \\
Nd III & 1.42
 & 12 & 3.30 & 0.07 & 2.18 & 0.07
 &   &   &   &   &  
 & 15 & 2.94 & 0.09 & 1.70 & 0.08
 \\
Sm II & 0.96
 &   &   &   &   &  
 & 1 & 1.75 & 0.20 & 1.35 & 0.41
 &   &   &   &   &  
 \\
Eu II & 0.52
 &   &   &   &   &  
 & 1 & 0.75 & 0.20 & 0.79 & 0.41
 &   &   &   &   &  
 \\
 \hline
Gd II & 1.07
 &   &   &   &   &  
 & 1 & 1.76 & 0.20 & 1.25 & 0.34
 &   &   &   &   &  
 \\
Dy II & 1.10
 &   &   &   &   &  
 & 1 & 1.48 & 0.20 & 0.94 & 0.39
 &   &   &   &   &  
 \\
Dy III & 1.10
 & 1 & 2.79 & 0.20 & 1.99 & 0.23
 &   &   &   &   &  
 &   &   &   &   &  
 \\
Lu II & 0.10
 &   &   &   &   &  
 & 1 & 0.85 & 0.20 & 1.31 & 0.28
 &   &   &   &   &  
 \\
Hf II & 0.85
 &   &   &   &   &  
 & 1 & 1.53 & 0.20 & 1.24 & 0.25
 & 1 & 2.54 & 0.20 & 1.87 & 0.25
 \\
\hline
\end{tabular}
\begin{tablenotes}
  \item \textbf{Notes:} Same as for Table \ref{table:abun_cold}.
\end{tablenotes}
\end{threeparttable}
\end{center}
\end{table*}

\subsection{Abundance determination}\label{subsect:abun}

The abundance analysis of the individual programme star is performed
with our preferred atmospheric parameter sets listed in Table
\ref{table:atmos}. We only use spectral lines with EWs smaller than
150 m\AA{}, as stronger lines are saturated. We also avoid spectral 
lines with EWs smaller than 5 m\AA{} as these may be 
confused with noise in the spectra.

Ideally, we would only use isolated lines but the strong enrichments and/or 
high metallicities of the programme stars make this difficult. 
Therefore, we use spectrum synthesis to check if suspicious spectral lines 
are part of identified blends. For some elements, we could only determine the 
abundance with spectral synthesis.

The number of spectral lines for the carbon abundance determination 
varies strongly with temperature, while for oxygen only a limited number 
of spectral lines is available in the temperature range of the programme stars. 
For almost all stars, the O abundances are determined from the 
high excitation oxygen multiplet at 6156 \AA{} and/or
the forbidden oxygen lines at 6300 and 6363 \AA{}. These  lines
are not affected by possible non-LTE effects \citep[see e.g.][]{kiselman02}. For
some stars with HERMES spectra, we could determine the nitrogen
abundances for spectral lines in the wavelength range between 7000 and 9000 \AA{}. 
These lines are known to be non-LTE sensitive, however, and
therefore we apply the temperature-sensitive, non-LTE corrections of 
\cite{lyubimkov11} (see Sect. \ref{sect:abun}).

The number of lines available for elements ranging from sodium (Na,
Z=11) up to sulphur (S, Z=16) is limited. Therefore, these abundances
cannot be determined in all stars. For the iron-peak
elements, ranging from calcium (Ca, Z=20) up to zinc (Zn, Z=30), we
find a wide range of elemental abundances for the majority of stars. 

In all programme stars, we find spectral lines of \textit{s}-process 
elements of the strontium peak (Sr, Z=38) and barium peak (Ba, Z=56),
respectively known as the light \textit{s}-process (ls) elements and heavy 
\textit{s}-process (hs) elements. Unfortunately, we could not determine the abundance
of Sr in any object as the available Sr lines are always strongly saturated.
For the other ls elements yttrium (Y, Z=39) and zirconium (Zr, Z=40),
we find useful spectral lines in each object to determine their abundance.
For some stars, we also detect one or two lines for
molybdenum (Mo, Z=42). 

For the hs elements, we can always determine the abundances of  
lanthanum (La, Z=57) and neodymium
(Nd, Z=60), and for most of the stars, strong
cerium (Ce, Z=58) lines are present. In a few objects, we are able 
to determine the abundance of Ba with a weak Ba II line. In the other 
stars, the available Ba lines are highly saturated.
We can also determine the abundances of a range of elements 
beyond the Ba-peak such as europium (Eu, Z=63), gadolinium (Gd,
Z=64), dysprosium (Dy, Z=66), erbium (Er, Z=68), ytterbium (Yb, Z=70),
lutetium (Lu, Z=71), hafnium (Hf, Z=72), and tungsten (W, Z=74). Most
of the abundances of these elements are determined with spectral
synthesis. We present here a very homogeneous determination of these very heavy elements in all objects with the same atomic data and methodology. Despite the fact that the abundances are often determined with only one or a few lines, we consider the abundances reliable. The high abundances are a clear indication that several elements beyond the Ba-peak are also efficiently synthesised in these objects.

\begin{figure}[t!]
\resizebox{\hsize}{!}{\includegraphics{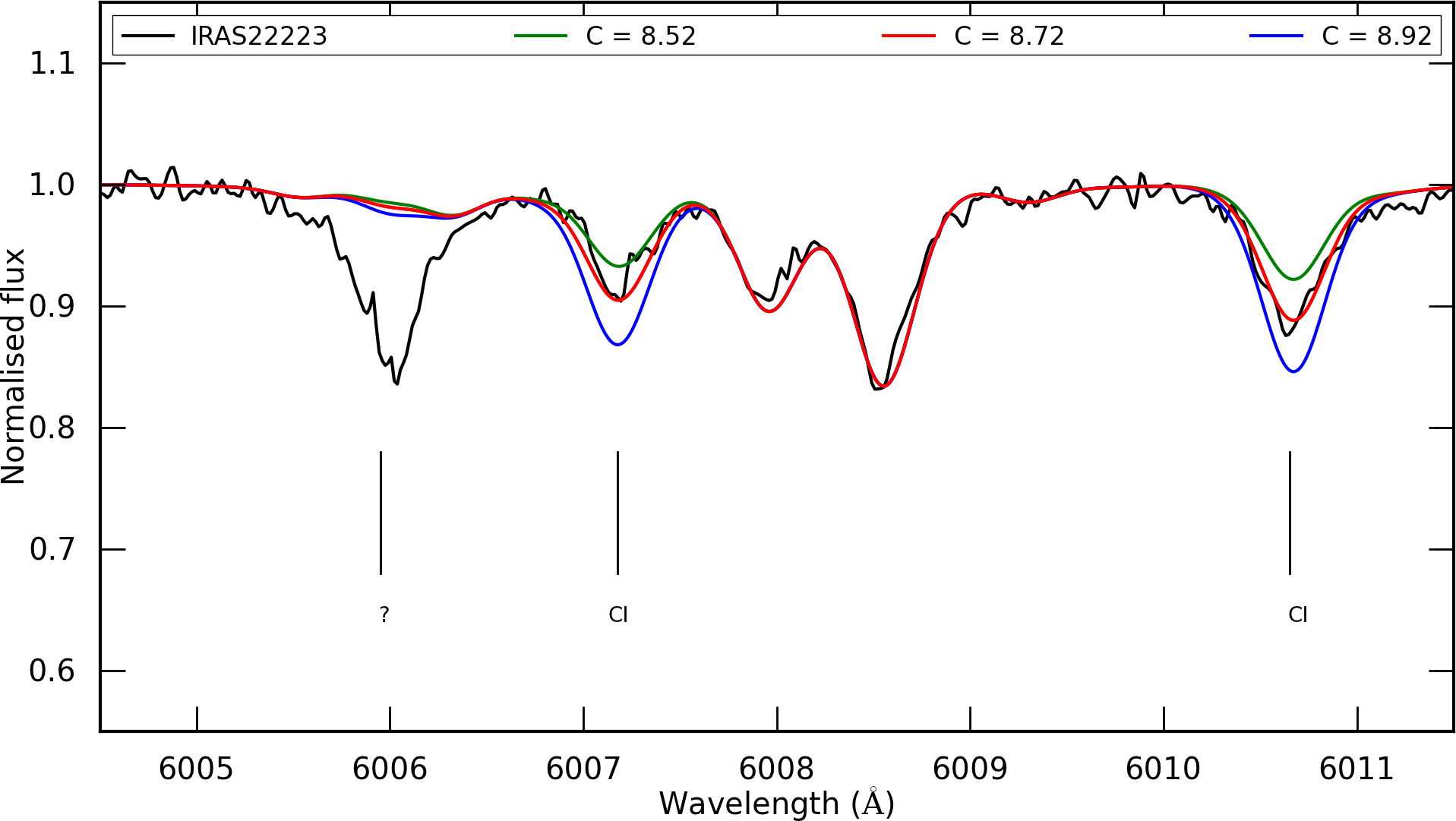}}
\caption{Spectrum synthesis of the C I lines at 6007.176 and 6010.675 \AA{} for IRAS 22223+4327. The black spectrum is the observed HERMES spectrum; the coloured spectra represent synthetic spectra with different C abundances. The red line represents the determined C abundance; 
 the green and blue spectra represent synthetic spectra with a C abundance of -0.2 dex and +0.2 dex, respectively.}\label{fig:iras22223_synth}
\end{figure}

\begin{figure}[t!]
\resizebox{\hsize}{!}{\includegraphics{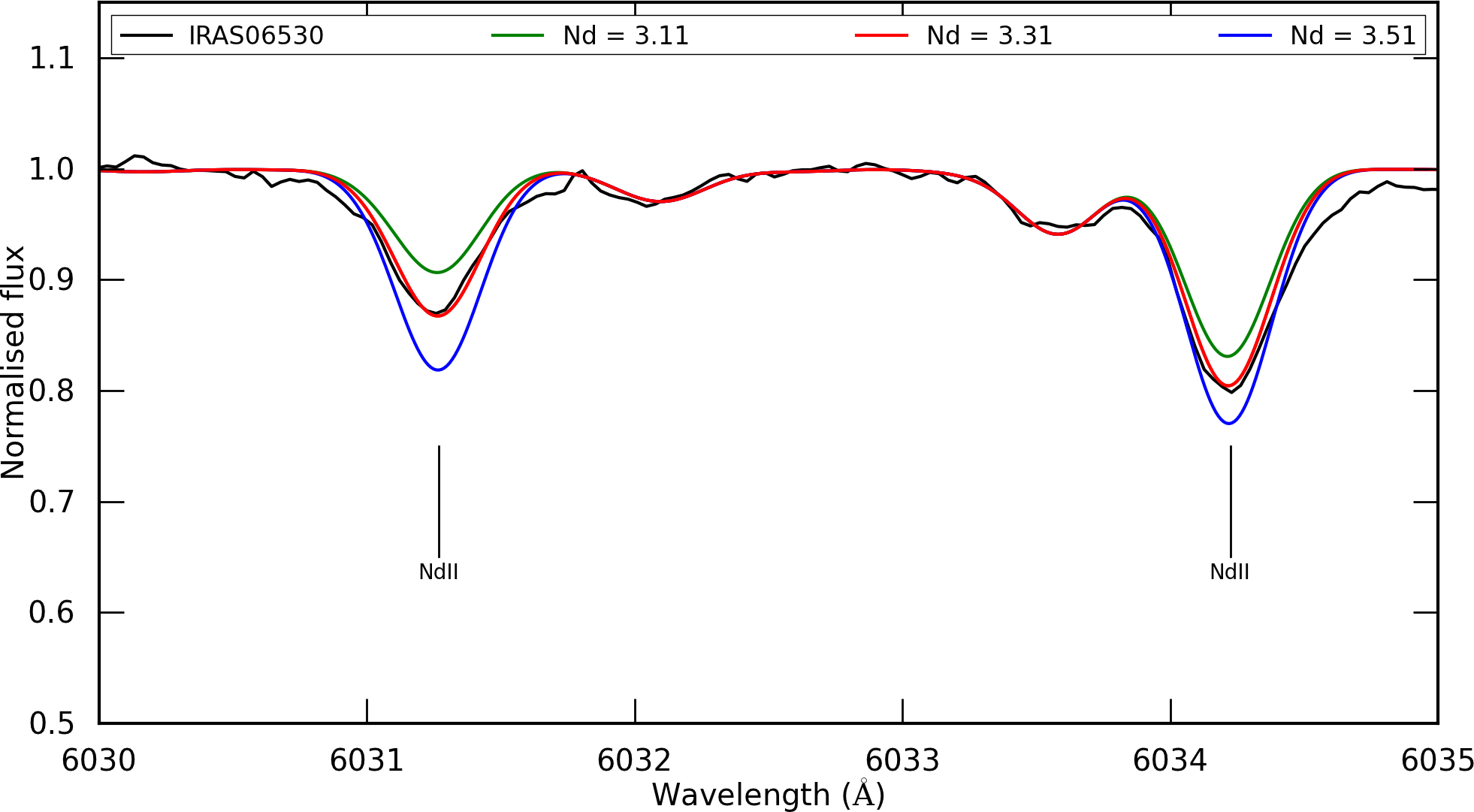}}
\caption{Spectrum synthesis of the Nd II lines at 6031.270 and 6034.228 \AA{} for IRAS 06530-0213. The black spectrum is the observed UVES spectrum;
the coloured spectra represent synthetic spectra with different Nd abundances. The red line represents the determined Nd abundance; 
the green and blue spectra represent synthetic spectra with a Nd abundance of -0.2 dex and +0.2 dex, respectively.}\label{fig:iras06530_synth}
\end{figure}

\begin{figure}[t!]
\resizebox{\hsize}{!}{\includegraphics{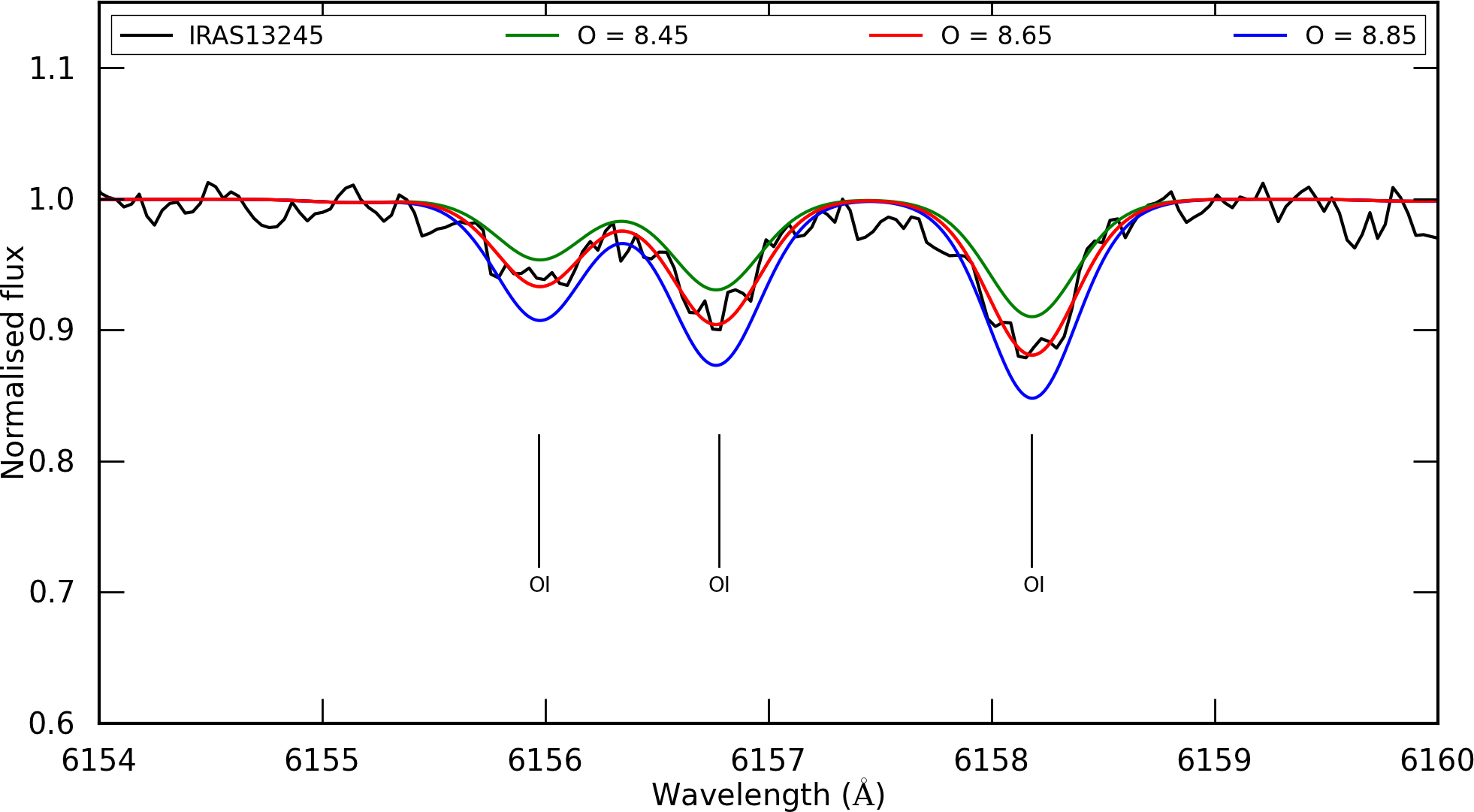}}
\caption{Spectrum synthesis of the O I triplet at 6156 \AA{} for IRAS 13245-5036. The black spectrum is the observed UVES spectrum,
the coloured spectra represent synthetic spectra with different O abundances. The red line represents the determined O abundance;
the green and blue spectra represent synthetic spectra with an O abundance of -0.2 dex and +0.2 dex, respectively.}\label{fig:iras13245_synth}
\end{figure}

\section{Abundance results to Z < 82 for all sample stars }\label{sect:abun}

\begin{figure*}[t!]
\centering
\includegraphics[width=18.5cm]{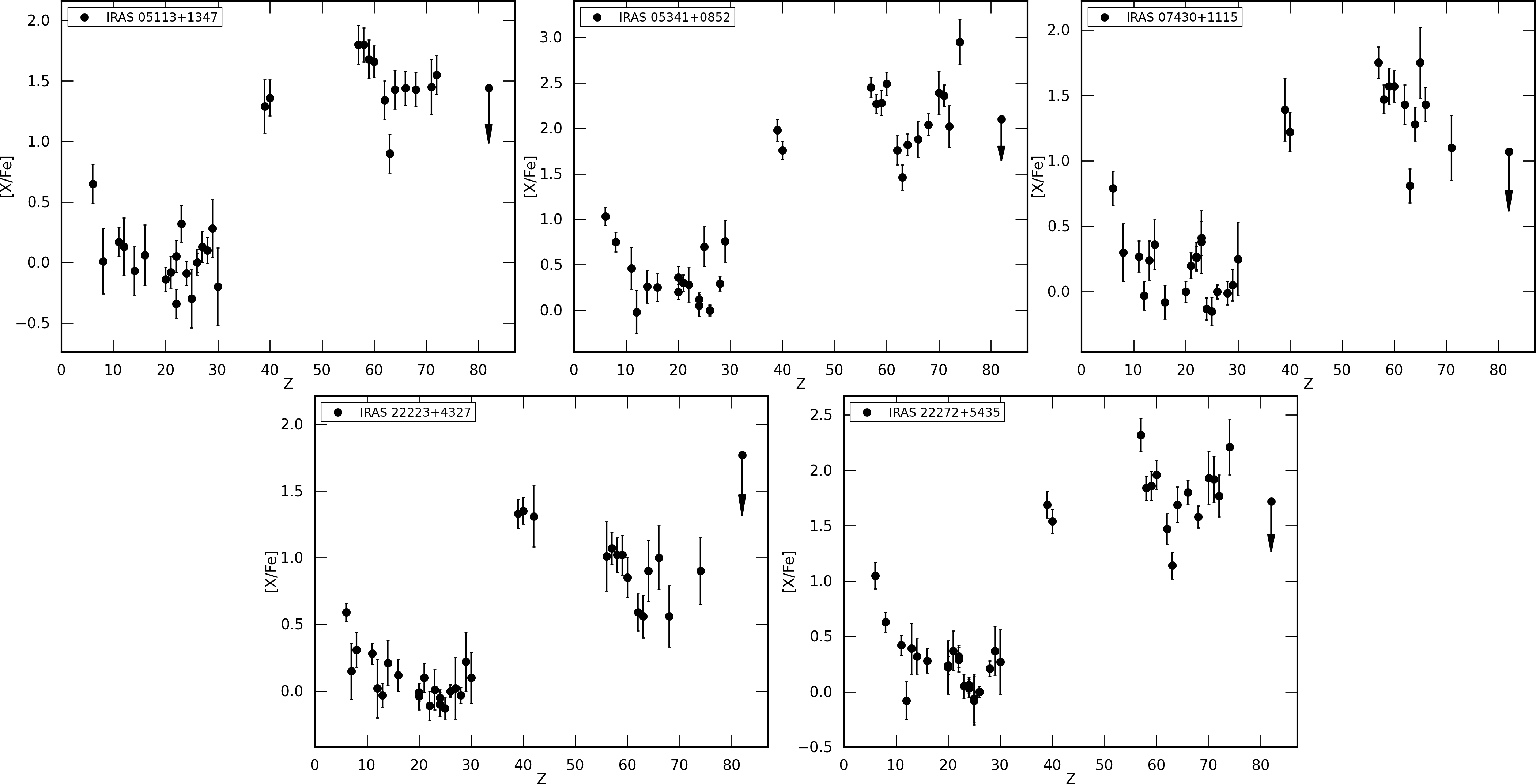}
\caption{Element-over-iron ratios for the programme stars with T$_{\textrm{eff}}$ < 7000 K. 
The down arrow at Z=82 displays the adopted Pb abundance upper limits.}\label{fig:xfe_cold}
\end{figure*}

\begin{figure*}[t!]
\centering
\includegraphics[width=18.5cm]{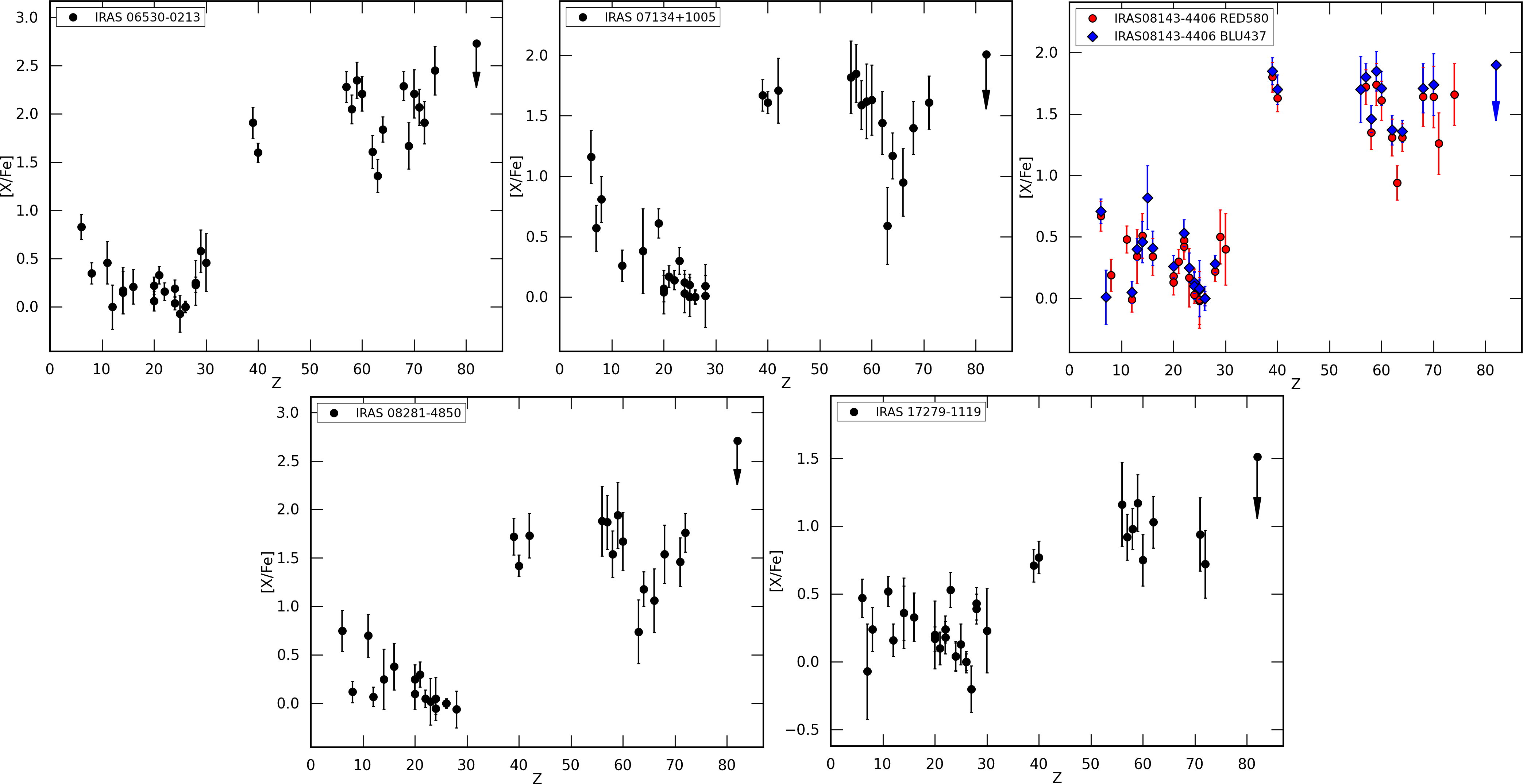}
\caption{Element-over-iron ratios for the programme stars with 7000 K $\leqslant$ T$_{\textrm{eff}}$ < 8000 K. 
The down arrow at Z=82 displays the adopted Pb abundance upper limits. For IRAS 08143-4406, the abundance 
results of both observational settings are shown.}\label{fig:xfe_middle}
\end{figure*}

\begin{figure*}[t!]
\centering
\includegraphics[width=12.5cm]{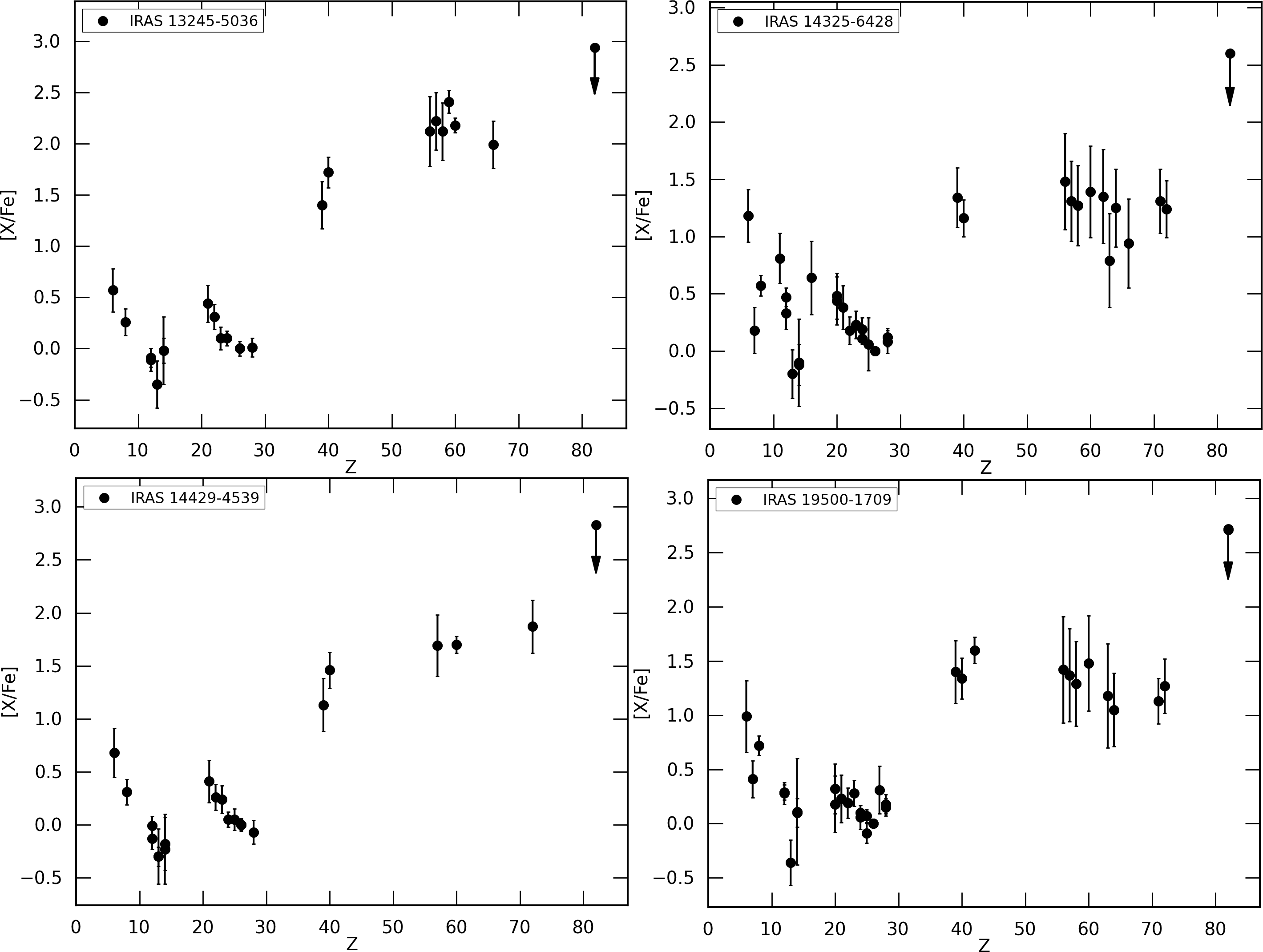}
\caption{Element-over-iron ratios for the programme stars with T$_{\textrm{eff}}$ $\geqslant$ 8000 K. 
The down arrow at Z=82 displays the adopted Pb abundance upper limits.}\label{fig:xfe_hot}
\end{figure*}

The abundance results of the different programme stars are listed in
Tables \ref{table:abun_cold} to \ref{table:abun_hot}; each table 
contains stars of similar effective temperatures. 
The corresponding [X/Fe] results are plotted in Figs. \ref{fig:xfe_cold}, \ref{fig:xfe_middle},
and \ref{fig:xfe_hot}. An overview of the lines used for these analyses can be found in 15 
catalogues available at CDS\footnote{A link to the
catalogues}.  The abundance results for Pb are discussed in
Sect. \ref{subsect:pb_abun}.

The total errors $\sigma_{\textrm{tot}}$ in Tables
\ref{table:abun_cold} to \ref{table:abun_hot} are calculated following
the methodology of \citet{deroo05a}.  The errors due to atmospheric
parameter uncertainties are calculated by determining the elemental
abundances for atmospheric models with an effective temperature
T$_{\textrm{eff}}$ $\pm$125 K, surface gravity $\log g$ $\pm$ 0.25 dex
and microturbulent velocity $\xi_t$ $\pm$ 0.25 km/s with respect to the
preferred value.  We include microturbulent velocity in
our error analysis despite the fact that for most ions, the 
associated error is only of the order of a few 0.01 dex. However, for
some ions for which only relatively strong spectral lines are
available like Y II, the error due to microturbulence can increase to
about 0.1 dex.

The total uncertainty $\sigma_{\textrm{tot}}$ on [X/Fe] is then the
quadratic sum of the error on the mean (the line-to-line scatter
$\sigma_{\textrm{l2l}}$ divided by the square root of the number of
lines used), atmospheric parameter
uncertainties ($\sigma_{\textrm{T}_{eff}}$, $\sigma_{\textrm{logg}}$,
$\sigma_{\xi_t}$), and the Fe abundance error ($\sigma_{\textrm{Fe}}$)
\begin{equation}
\sigma_{\textrm{tot}} = \sqrt{\left(\frac{\sigma_{\textrm{l2l}}}{\sqrt{N_{\textrm{ion}}}}\right)^2 + (\sigma_{\textrm{T}_{eff}})^2 + 
(\sigma_{\textrm{logg}})^2 + (\sigma_{\xi_t})^2  + \left(\frac{\sigma_{\textrm{Fe}}}{\sqrt{N_{\textrm{Fe}}}}\right)^2} .
\end{equation}

We assume an intrinsic uncertainty of 0.2 dex for all elemental
abundances, which are determined by only one line or via spectrum
synthesis. This error is a substitute of the line-to-line scatter
determined for elements with more useful lines.  The [X/Fe] abundances are calculated using
the Fe ion with an ionisation potential closest to the ionisation
potential of the studied ion. In other words, if the ionisation
potential of an ion is below the ionisation potential of Fe I, the Fe
I abundance is used for calculating [X/Fe]. If the ionisation
potential exceeds the ionisation potential of Fe I, the abundance of
Fe II is used instead.

In Tables \ref{table:abun_cold} to \ref{table:abun_hot}, 
all elemental abundances  were double-checked by comparing 
each spectral line with synthetic spectra with the
determined abundances. Examples of these comparisons are shown in
Figs. \ref{fig:iras22223_synth}, \ref{fig:iras06530_synth}, and
\ref{fig:iras13245_synth}. Fig. \ref{fig:iras22223_synth} shows a
spectrum synthesis check for the determined C abundance of IRAS
22223+4327. The red synthetic spectrum computed using the determined elemental abundances fits the
observed spectrum well, except for the strong spectral line at 6006
\AA{}. To investigate its identification, we artificially increased
the abundances of all elements with spectral lines in this spectral region, 
but none could produce the observed absorption. This shows that the linelists 
are not complete for our strongly s-process enriched stars, the spectra of
which are swamped with transitions of s-process elements. 
Also, for other stars and different spectral regions, we find
spectral lines not included in the linelists.

Fig. \ref{fig:iras06530_synth} shows a spectrum synthesis check of the
determined Nd abundance of IRAS 06530-0213 for the Nd II lines at 6031.270
and 6034.228 \AA{}, confirming the high Nd abundance of IRAS
06530-0213. The spectrum synthesis of the high excitation O I
triplet at 6156 \AA{} of IRAS 13245-5036 is shown in
Fig. \ref{fig:iras13245_synth} and shows a good agreement between the
observations and determined O abundances.

We briefly discuss the derived abundances followed by the comparison between our determined 
results and those of previous studies. In Sect. \ref{subsect:pb_abun}, we specifically discuss  the 
derived Pb abundances.

\subsection{CNO elements}

For all programme stars, we can determine the carbon and oxygen
abundances. The C/O ratios range from about 1.0 up to 2.5 
and are listed in the second column of Table \ref{table:ratios}. The
errors on the C/O ratios in Table \ref{table:ratios} include
line-to-line scatter and atmospheric model uncertainties. They 
can become significant because of the strong temperature sensitivity of
the available C and O lines within the temperature range of the sample
stars. The two new post-AGB stars, IRAS 13245-5036 and IRAS 14429-4539,
display C/O ratios of about 1.1 and 1.3, respectively. This
also confirms the observation that the 21 $\mu$m feature is only
detected around carbon-rich post-AGB stars.

For six of the stars, we could determine nitrogen abundances using
non-LTE sensitive spectral lines between 7000 and 9000 \AA{}. We
applied non-LTE corrections to the LTE N abundances following the
temperature-dependent corrections of Fig. 4 in \citet{lyubimkov11},
which shows a larger correction for higher temperatures. We adopted
the following corrections: for IRAS 22223+4327, we use a correction of
-0.3 dex; for IRAS 08143-4406, we apply a correction of -0.4 dex; for
IRAS 07143+1005 and IRAS 17279-1119, we apply corrections of -0.5 dex;
and for IRAS 14325-6428 and IRAS 19500-1709, we use corrections of
-0.65 dex and -0.7 dex, respectively. These corrections are already
included in the listed results of Tables \ref{table:abun_middle1},
\ref{table:abun_middle2}, and \ref{table:abun_hot}. Given the
uncertainties on the non-LTE correction, we assume a somewhat ad-hoc,
line-to-line scatter of 0.3 dex for all N abundance determinations.

\subsection{$S$-process elements}\label{subsect:s-proc}

The two new post-AGB stars, IRAS 13245-5036 and IRAS 14429-4539, show
strong s-process enrichments with a stronger overabundance of the hs
elements with respect to the ls elements. Both stars are warm and, hence,  intrinsically display a less rich spectrum. We could only determine a limited set
of \textit{s}-process elements for IRAS 14429-4539. The \textit{s}-process enrichments confirm the evolved nature of
both stars. This corroborates the finding that the stars displaying the
21 $\mu$m feature are all s-process rich and carbon-rich post-AGB stars, which are the
likely descendants of AGB carbon stars.

We confirm that all other programme stars show \textit{s}-process enhancements for both the ls
and hs elements with significant differences between the objects.
IRAS 17279-1119 displays a mild s-process enrichment and most of the
\textit{s}-process elements have an [X/Fe] < 1.0 dex, while the two
strongest enriched objects, IRAS 05341+0852 and IRAS 06530-0213, have
high \textit{s}-process [X/Fe] abundances reaching up to 2.5 dex.   The abundances derived from the two observational 
settings are similar for 
IRAS\,08143-4406 (upper right panel of Fig. \ref{fig:xfe_middle}).

Apart from the different levels of enrichment, we also find different
abundance distributions.  For IRAS 08143-4406, IRAS 14325-6428, IRAS
19500-1709, and IRAS 22223+4327, the enrichment of the ls
elements Y and Zr is higher or similar to that of the hs
elements, while for the other stars the situation is reversed. 
We go into more detail about the hs/ls behaviour in Sect. 
\ref{sect:discussion}.

For those objects with Mo determination, we find that Mo has similar
enhancements as Y and Zr. The elements with Z > 62 are also enriched
and their abundance distributions strongly vary from star to star 
to a level comparable to or less than that of the hs elements. 

\subsection{Comparison with previous studies}\label{subsect:prev_stud}

As online material in Appendix B, we provide  the comparison
between our abundance results and those of previous abundance studies
(see second to last column of Table \ref{table:obs}). For IRAS 13245-5036
and IRAS 14429-4539, there are no available data 
in literature.  We remark that previous studies used different
linelists and solar abundances, which can explain small differences
between our determinations and previous determinations. Especially, changes in oscillator
strength log $gf$ can significantly impact the individual spectral line 
abundances. 

\section{Pb abundance determination}\label{subsect:pb_abun}


For each star, we use an artificially high abundance to theoretically predict which 
optical Pb line (Pb I or Pb II) is the strongest for a specific set of atmospheric 
parameters. We then select the Pb line that gives the largest predicted EW. For 
some stars, both Pb lines result in a similar theoretical EW, so we can use both lines to
constrain the Pb abundance. 

Depending on the effective temperature of the sample stars, the Pb abundances are 
probed either using the strongest Pb I line at 4057.807 \AA{} or Pb II line at 
5608.853 \AA{}. As the Pb I line lies in a low S/N region with many 
spectral blends and the Pb II line is generally very small so 
that it can be confused with noise in the spectrum, none of the sample stars 
display a clear Pb line feature from which an accurate Pb abundance can be derived.
Therefore, we can only determine upper limits on the Pb abundance.

\begin{figure}[t!]
\resizebox{\hsize}{!}{\includegraphics{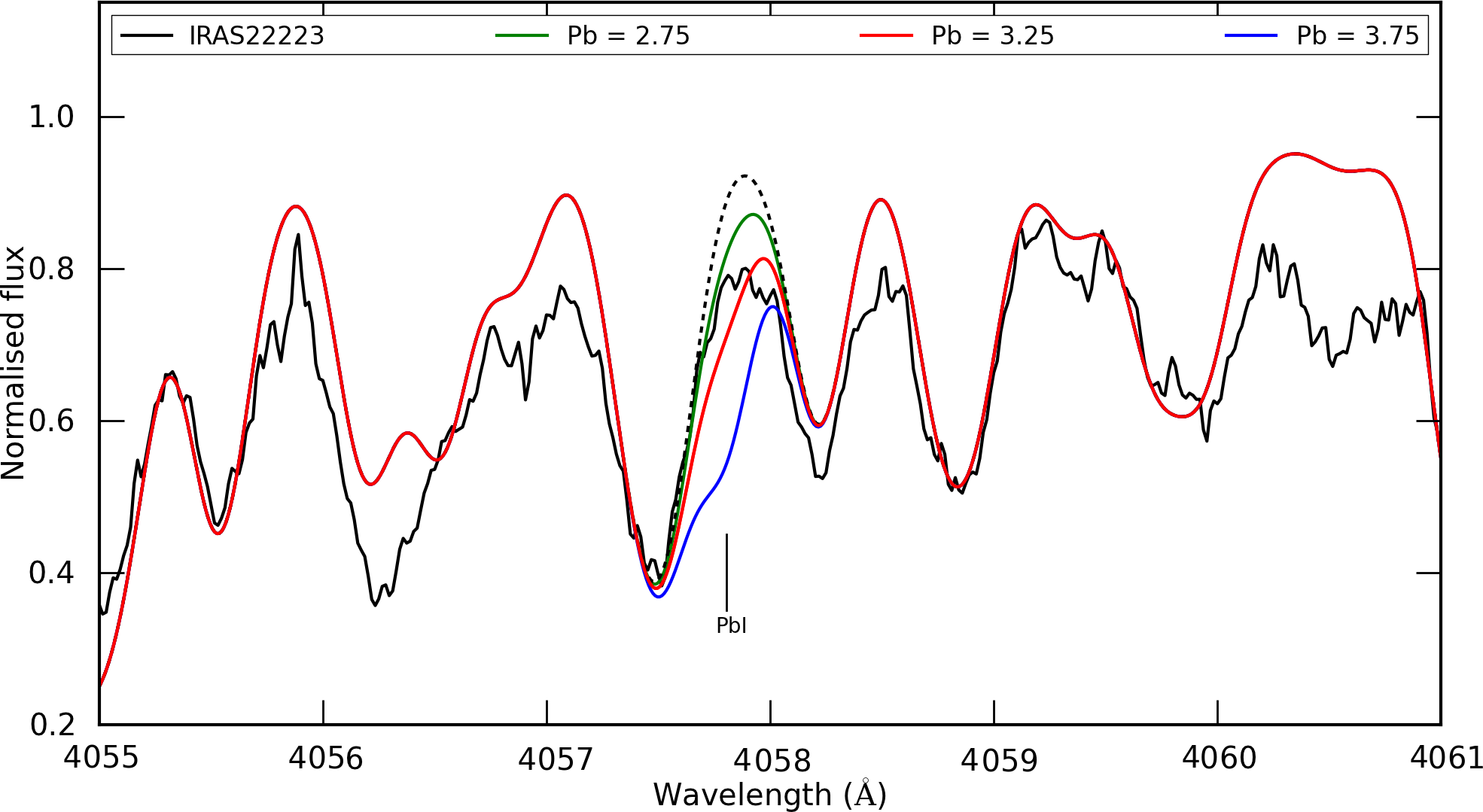}}
\caption{Spectrum synthesis of the Pb I line at 4057.807 \AA{} for IRAS 22223+4327. The black spectrum is the observed HERMES spectrum,
the coloured spectra represent synthetic spectra with different Pb abundances. The red line represents the adopted Pb abundance upper limit; 
the green and blue lines represent the adopted abundance upper limit -0.5 dex and +0.5 dex, respectively.
The dashed black line shows the synthetic spectrum if
no Pb is present. For more information, see text.}\label{fig:iras22223_Pb}
\end{figure}

\begin{figure}[t!]
\resizebox{\hsize}{!}{\includegraphics{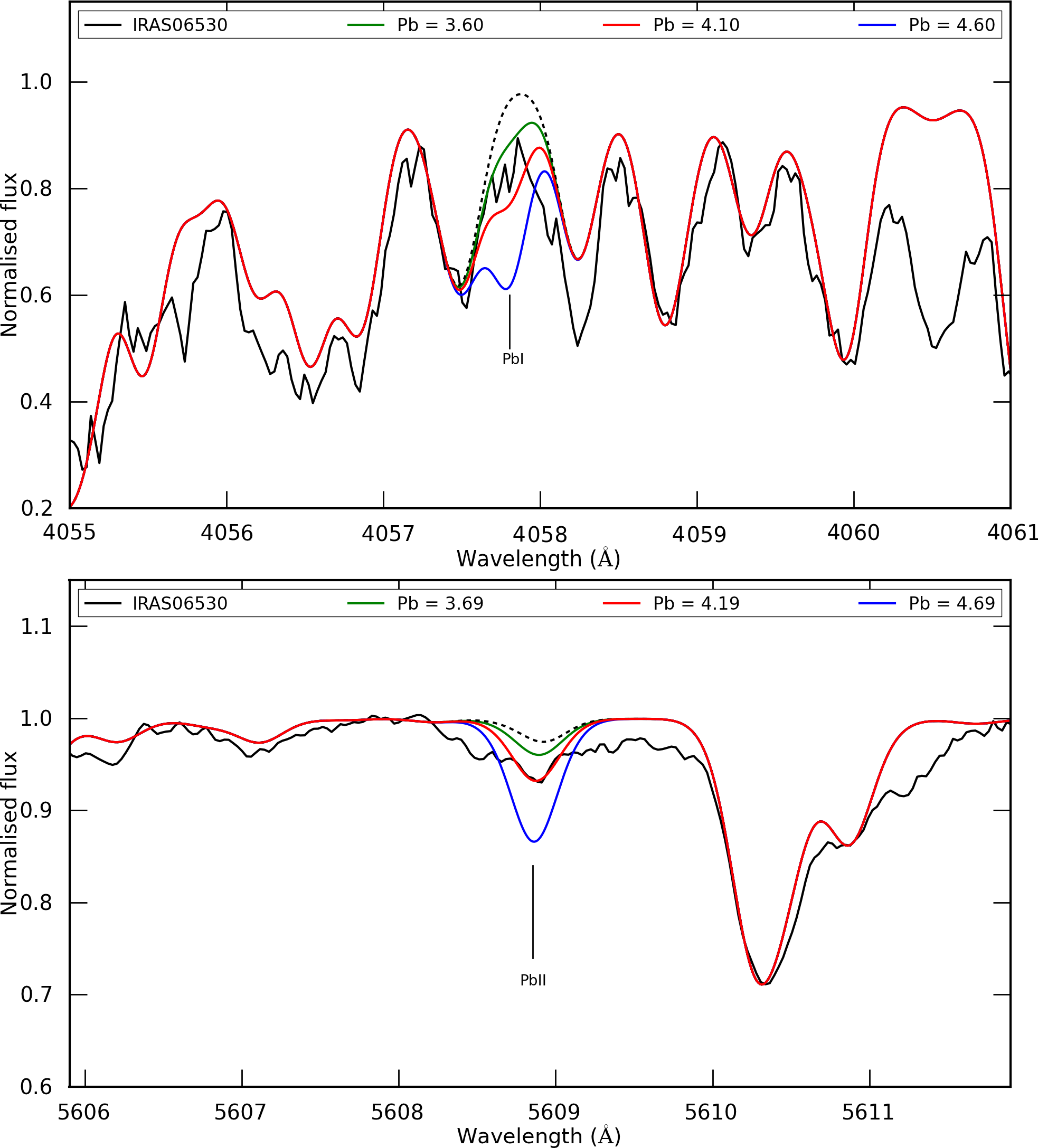}}
\caption{Spectrum synthesis of the Pb I line at 4057.807 \AA{} (upper panel) and of the Pb II line at 5608.853 \AA{} (lower panel) for 
the UVES spectra of IRAS 06530-0213. 
Lines and symbols are similar to Fig. \ref{fig:iras22223_Pb}.}\label{fig:iras06530_Pb}
\end{figure}

\begin{figure}[t!]
\resizebox{\hsize}{!}{\includegraphics{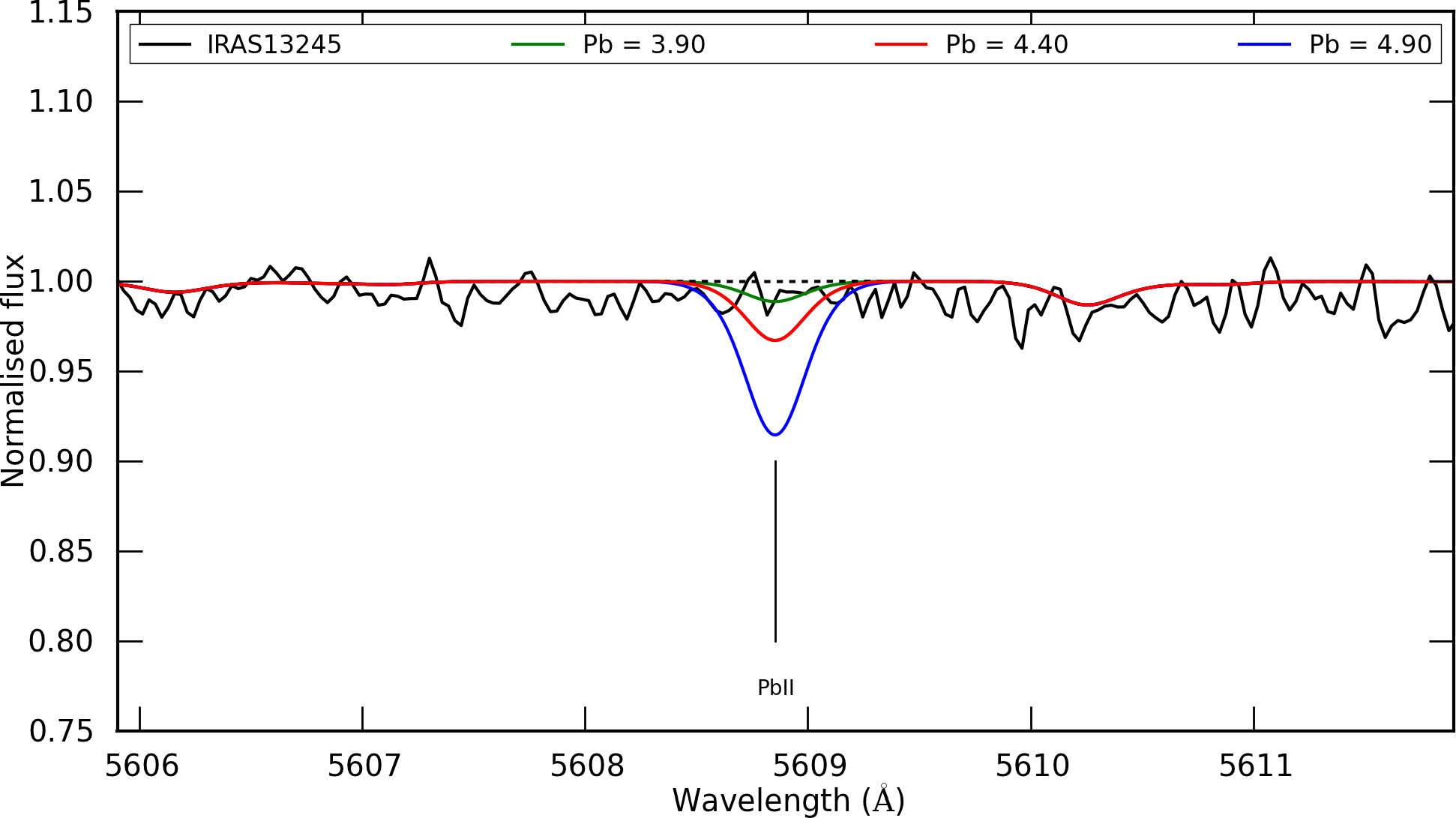}}
\caption{Spectrum synthesis of the Pb II line at 5608.853 \AA{} for the UVES spectrum of IRAS 13245-5036. 
Lines and symbols are similar to Figs. \ref{fig:iras22223_Pb} and \ref{fig:iras06530_Pb}.}\label{fig:iras13245_Pb}
\end{figure}

We generate synthetic spectra for the spectral region of the Pb
line, including all the derived abundances; these are listed in Tables 
\ref{table:abun_cold} to \ref{table:abun_hot}. 
We use the synthetic spectral lines to
estimate the local position of the continuum at the Pb line. 
The lowest Pb abundance for which the synthetic spectrum fully 
includes the observed spectral blend at the Pb line is adopted 
as upper limit of the Pb abundance.
Examples of our method are illustrated in
Figs. \ref{fig:iras22223_Pb}, \ref{fig:iras06530_Pb}, and
\ref{fig:iras13245_Pb} for three different sample stars.

Fig. \ref{fig:iras22223_Pb} shows the Pb abundance 
determination of IRAS 22223+4327 using the Pb I line at 4057.807
\AA{}. For stars with lower temperatures and/or high metallicities,
this spectral region contains many spectral blends. The S/N in this blue
region is generally only between 20 and 30. We use the identified spectral
blends at 4057.5 \AA{} and 4058.9 \AA{} to estimate the position of
the continuum and then use different synthetic Pb abundances to
determine the upper limit.  For each figure, the red synthetic spectra
represent our adopted abundance upper limit and the dashed black
lines show synthetic spectra without any Pb contribution included.  

In case of IRAS 22223+4327, there is no clear contribution from Pb. 
The blue synthetic spectrum
in Fig. \ref{fig:iras22223_Pb} provides a full observed feature
around 4057.807 \AA,{} but generates a large contribution to the blend with the
line at 4057.5 \AA\ that is not observed in the spectrum. Therefore,
we adopt an upper limit of the Pb abundance for IRAS 22223+4327, which only creates
a contribution to the blend that is still compliant with the
observed spectrum. Fig. \ref{fig:iras22223_Pb} also shows that there are
a number of spectral lines in this region that are not included in the
linelists. The line features at 4056.2 \AA{} and 4060.5
\AA{} are examples. Moreover, for stars with T$_{\textrm{eff}}$ < 7500 K and [Fe/H]
> 0.5 dex, we cannot reproduce the blended line at 4058.2 \AA{}
at the red side of the Pb I line, which would otherwise be a major
factor for constraining the local continuum position.

The Pb abundance determinations of IRAS 06530-0213 are
presented in Fig. \ref{fig:iras06530_Pb}. The upper panel shows the
determination via the Pb I line at 4057.807 \AA{}; the lower panel
shows the same for the Pb II line at 5608.853 \AA{}.  Also, for this
star, we do not detect a clear  Pb I line  and we use
the spectral blends at 4057.5 and 4058.9 \AA{} to estimate the
position of the continuum. We adopt a Pb abundance that
generates a synthetic Pb line, which fully incorporates the observed
line feature at 4057.807 \AA{}. For the Pb II line at 5608.853 \AA{},
we use the continuum between 5606.3 and 5608 \AA{} in combination with
the blend at 5610.3 \AA{} to estimate the position of the continuum at
5608.853 \AA{}. The Pb II line lies in a spectral region where the
continuum is affected by a blend of small unidentified spectral lines, ranging from
5608 up to 5610 \AA{}. The observed spectrum lies below the
continuum of the synthetic spectra. To make sure this feature is real,
we checked the individual observed spectra of IRAS 06530-0213 in Table
\ref{table:obs} and found the feature to be present in both
observations. We also compared our new spectra with the old observations from
\citet{reyniers04} and  this broad feature is present there as well. We
therefore expect this depression to be real. This is likely a set of
unresolved small lines. For the Pb II line, we adopt
a Pb abundance that fits the observed spectrum. However,
comparison with the spectra of \citet{reyniers04} suggest that the
small line at 5608.853 \AA{} in the broad feature between 5608 and
5610 \AA{} is not significant as it is not seen in the older spectra. As a  result of this possibility, in
combination with the lower upper limit found for the Pb I
line, we adopt the limit for the Pb I line as the upper limit of the
Pb abundance for IRAS\,06530-0213.

Fig. \ref{fig:iras13245_Pb} shows the Pb abundance
determination for IRAS 13245-5036 with the Pb II line at 5608.853
\AA{}.  We do not detect any Pb II line feature for this star, and
therefore we adopt an upper limit of the Pb abundance, which creates a
synthetic Pb II line that lies slightly below the observed spectrum. The Pb I
line at 4057.807 \AA{} would need a significantly higher Pb abundance
than the adopted upper limit based on the Pb II
line. Therefore, we only consider the upper limit based on the Pb II line.

The determinations of the upper limits on the Pb abundances of all other programme
stars are presented and discussed in Appendix A, which is provided as
online material for this paper. We do not detect strong Pb lines for
any of the 14 \textit{s}-process
enriched post-AGB stars. For the remaining part of the paper, we therefore refer to the Pb upper limit abundances 
as Pb$_{\textrm{up}}$. In Table \ref{table:pb}, we list the individual Pb upper limit abundance results for 
the Pb I and Pb II line of the programme stars. The element-over-iron ratios corresponding 
to the adopted Pb abundance are shown in Figs. \ref{fig:xfe_cold}, \ref{fig:xfe_middle}, and 
\ref{fig:xfe_hot}, and indicated with down arrows. The last six rows display the results 
of \citet{desmedt14,desmedt15}
for six carbon and \textit{s}-process enhanced post-AGB stars in the Magellanic Clouds and are 
used in the analysis of the neutron irradiation in Sect. \ref{sect:discussion}.


\begin{table*}[t!]
\begin{center}
\caption{\label{table:pb} Pb abundance upper limits Pb$_{\textrm{up}}$ of the programme stars, together with the Pb indices, [Pb$_{\textrm{up}}$/hs] and 
[Pb$_{\textrm{up}}$/ls], and the effective temperature. 
We adopt the solar Pb abundance of log $\epsilon$(Pb) = 1.75 dex from \citet{asplund09}. The abundances of Pb I and Pb II are determined via the 4057.807 \AA{} 
and 5608.853 \AA{} spectral lines, respectively. The last six rows indicate the Pb$_{\textrm{up}}$ results of six post-AGB stars in the Magellanic Clouds from 
\citet{desmedt14,desmedt15}.}
\begin{threeparttable}
\begin{tabular}{ l | c | cc | cc | c | c | c  } \hline\hline
Object                 &  [Fe/H]  &   log $\epsilon$(PbI$_{\textrm{up}}$)    &        [PbI$_{\textrm{up}}$/Fe]        &    log $\epsilon$(PbII$_{\textrm{up}}$)  
                       &        [PbII$_{\textrm{up}}$/Fe]      &  [Pb$_{\textrm{up}}$/hs]  &  [Pb$_{\textrm{up}}$/ls]  & T$_{\textrm{eff}}$ (K) \\   
\hline
IRAS 05113+1347        &  -0.49   &          < 2.65          &        <  1.44         &                          &                      
                       &   < -0.21   &  < 0.11    & 5500 \\
IRAS 05341+0852        &  -0.54   &          < 3.15          &        <  2.10         &                          &                      
                       &   < -0.14   &  < 0.23    & 6750 \\
IRAS 06530-0213        &  -0.32   &          < 4.10          &        <  2.73         &         < 4.19           &        < 2.76        
                       &  <  0.69   &   < 0.98    & 7375 \\
IRAS 07134+1005        &  -0.91   &          < 2.79          &        <  2.01         &                          &                      
                       &  < 0.38   &   < 0.37    & 7250 \\
IRAS 07430+1115        &  -0.31   &          < 2.50          &        <  1.07         &                          &                      
                       &  < -0.48   &  < -0.23    & 6000 \\
IRAS 08143-4406$^{a}$  &  -0.43   &          < 3.20          &        <  1.90         &                          &                      
                       &  < 0.32   &   < 0.13    & 7000 \\
IRAS 08281-4850        &  -0.26   &          < 4.20          &        <  2.74         &         < 4.20           &        < 2.71        
                       &  < 1.13   &   < 1.14    & 7875 \\
IRAS 13245-5036        &  -0.26   &                          &                        &         < 4.40           &        < 2.94        
                       &  < 0.91   &   < 1.38    & 9500 \\
IRAS 14325-6428        &  -0.56   &                          &                        &         < 3.79           &        < 2.60        
                       &   < 1.27   &   < 1.35    & 8000 \\
IRAS 14429-4539        &  -0.64   &                          &                        &         < 4.40           &        < 2.83        
                       &   < 1.36   &   < 1.54    & 9375 \\
IRAS 17279-1119        &  -0.64   &          < 2.75          &        <  1.51         &                          &                      
                       &   < 0.59   &   < 0.77    & 7250 \\
IRAS 19500-1709        &  -0.59   &          < 3.89          &        <  2.72         &                          &                      
                       &   < 1.38   &   < 1.35    & 8000 \\
IRAS 22223+4327        &  -0.30   &          < 3.25          &        <  1.82         &                          &                      
                       &   < 0.94   &   < 0.48    & 6500 \\
IRAS 22272+5435        &  -0.77   &          < 2.70          &        <  1.72         &                          &                      
                       &   < -0.18   &  < 0.11    & 5750 \\
\hline
J004441.04-732136.4    &  -1.34   &          < 3.00          &        <  2.58         &                          &                      
                       &   < 0.00   &  < 0.52    & 6250 \\
J050632.10-714229.8    &  -1.22   &          < 2.05          &        <  1.52         &                          &                      
                       &   < 0.45   &  < 0.10    & 6750 \\
J052043.86-692341.0    &  -1.15   &          < 2.00          &        <  1.40         &                          &                      
                       &   < -0.45   & < -0.27    & 5750 \\
J053250.69-713925.8    &  -1.22   &          < 2.23          &        <  1.70         &                          &                      
                       &   < -0.24   & < 0.19    & 5500 \\
J051213.81-693537.1    &  -0.56   &          < 3.30          &        <  2.11         &                          &                      
                       &   < 0.37   &  < 0.78    & 5875 \\
J051848.86-700246.9    &  -1.03   &          < 2.62          &        <  1.93         &                          &                      
                       &   < -0.19   &  < 0.47    & 6000 \\
\hline
\end{tabular}
    \begin{tablenotes}
      \small
      \item $^{a}$ Determined via the Blue437 spectra of IRAS 08143-4406.   
    \end{tablenotes}
\end{threeparttable}
\end{center}
\end{table*}

\section{Discussion}\label{sect:discussion}

The lack of accurate distance determinations for the 14 programme stars 
hinders accurate luminosity determinations, which prevents us from estimating the 
core masses and from constraining the initial masses. Therefore, the comparison of our abundance results 
with tailored, state-of-the-art theoretical AGB evolution and
nucleosynthetic models is difficult. 
Given the subsolar metallities we expect that the majority of the
14 sample stars will have low initial masses (M
$\leqslant$ 2.0 M$_{\odot}$). Future distance determinations will be 
essential to confirm this assumption.   

\subsection{Abundance results versus general predictions}

\begin{table*}[t!]
\caption{\label{table:ratios} Overview of the C/O ratio, the metallicity, and s-process indices for our sample of stars.}
\begin{center}
\begin{threeparttable}
\begin{tabular}{lcccccc} \hline\hline
Object                 &        C/O        &        [Fe/H]       &     [ls/Fe]     &      [hs/Fe]     &     [s/Fe]     &     [hs/ls]     \\   
\hline
IRAS 05113+1347        &  2.42 $\pm$ 0.40  &  -0.49  $\pm$ 0.15  & 1.33 $\pm$ 0.13 & 1.65 $\pm$ 0.07 & 1.54 $\pm$ 0.07 & 0.32 $\pm$ 0.15 \\
IRAS 05341+0852        &  1.06 $\pm$ 0.30  &  -0.54  $\pm$ 0.11  & 1.87 $\pm$ 0.08 & 2.24 $\pm$ 0.06 & 2.12 $\pm$ 0.05 & 0.37 $\pm$ 0.10 \\
IRAS 06530-0213        &  1.66 $\pm$ 0.39  &  -0.32  $\pm$ 0.11  & 1.75 $\pm$ 0.09 & 2.04 $\pm$ 0.08 & 1.94 $\pm$ 0.06 & 0.29 $\pm$ 0.13 \\
IRAS 07134+1005        &  1.24 $\pm$ 0.29  &  -0.91  $\pm$ 0.20  & 1.64 $\pm$ 0.13 & 1.63 $\pm$ 0.20 & 1.63 $\pm$ 0.14 & -0.01 $\pm$ 0.24  \\
IRAS 07430+1115        &  1.71 $\pm$ 0.30  &  -0.31  $\pm$ 0.15  & 1.30 $\pm$ 0.14 & 1.55 $\pm$ 0.06 & 1.47 $\pm$ 0.06 & 0.25 $\pm$ 0.15 \\
IRAS 08143-4406$^{a}$  &  1.66 $\pm$ 0.39  &  -0.43  $\pm$ 0.11  & 1.77 $\pm$ 0.08 & 1.58 $\pm$ 0.06 & 1.65 $\pm$ 0.05 & -0.19 $\pm$ 0.11 \\
IRAS 08281-4850        &  2.34 $\pm$ 0.42  &  -0.26  $\pm$ 0.11  & 1.57 $\pm$ 0.11 & 1.58 $\pm$ 0.12 & 1.58 $\pm$ 0.09 & 0.01 $\pm$ 0.17 \\
IRAS 13245-5036        &  1.11 $\pm$ 0.30  &  -0.30  $\pm$ 0.10  & 1.56 $\pm$ 0.14 & 2.03 $\pm$ 0.11 & 1.88 $\pm$ 0.09 & 0.47 $\pm$ 0.18 \\
IRAS 14325-6428        &  2.27 $\pm$ 0.40  &  -0.56  $\pm$ 0.10  & 1.25 $\pm$ 0.15 & 1.33 $\pm$ 0.19 & 1.30 $\pm$ 0.14 & 0.08 $\pm$ 0.24 \\
IRAS 14429-4539        &  1.29 $\pm$ 0.26  &  -0.18  $\pm$ 0.11  & 1.29 $\pm$ 0.15 & 1.47 $\pm$ 0.10 & 1.41 $\pm$ 0.08 & 0.18 $\pm$ 0.08 \\
IRAS 17279-1119        &  0.94 $\pm$ 0.22  &  -0.64  $\pm$ 0.12  & 0.74 $\pm$ 0.08 & 0.92 $\pm$ 0.09 & 0.86 $\pm$ 0.07 & 0.18 $\pm$ 0.12 \\
IRAS 19500-1709        &  1.02 $\pm$ 0.17  &  -0.59  $\pm$ 0.10  & 1.37 $\pm$ 0.29 & 1.34 $\pm$ 0.30 & 1.35 $\pm$ 0.21 & -0.03 $\pm$ 0.41 \\
IRAS 22223+4327        &  1.04 $\pm$ 0.22  &  -0.30  $\pm$ 0.11  & 1.34 $\pm$ 0.07 & 0.88 $\pm$ 0.07 & 1.03 $\pm$ 0.05 & -0.46 $\pm$ 0.10 \\
IRAS 22272+5435        &  1.46 $\pm$ 0.26  &  -0.77  $\pm$ 0.12  & 1.61 $\pm$ 0.08 & 1.90 $\pm$ 0.07 & 1.80 $\pm$ 0.05 & 0.28 $\pm$ 0.11 \\
\hline
\end{tabular}
    \begin{tablenotes}
      \small
      \item $^{a}$ C/O ratio is calculated from the results of the Red580 observations of IRAS 08143-4406; all other results are calculated from the 
                   results of the Blue437 observations.   
    \end{tablenotes}
\end{threeparttable}
\end{center}
\end{table*}

\begin{figure}[t!]
\resizebox{\hsize}{!}{\includegraphics{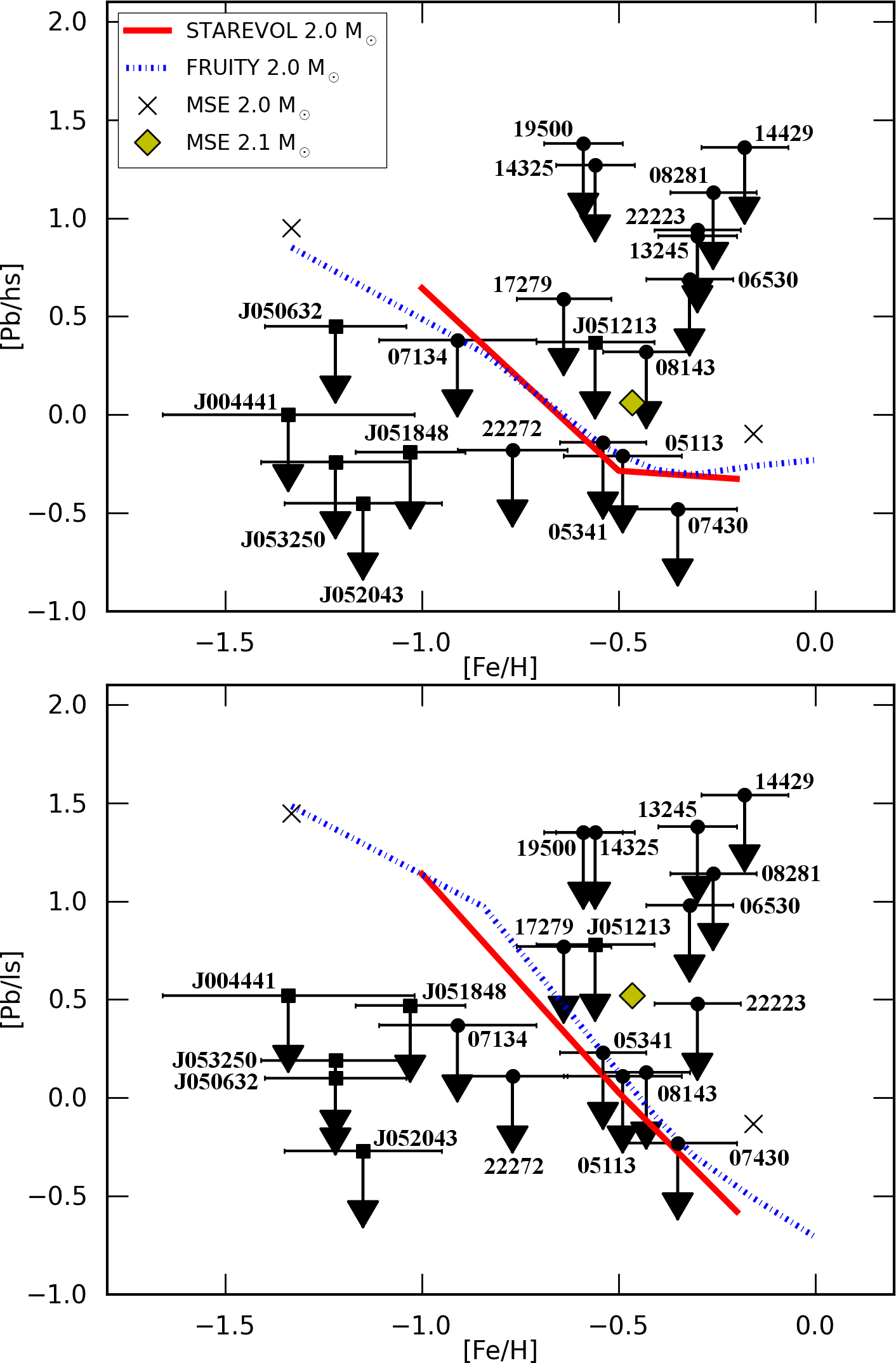}}
\caption{The observed [Pb$_{\textrm{up}}$/hs] (upper panel) and the [Pb$_{\textrm{up}}$/ls] (lower panel) versus [Fe/H] results of the 
Galactic sample stars of this study (black circles) and the Magellanic Cloud stars in \citet{desmedt14,desmedt15} (black squares). 
The observed abundance upper limits are plotted together with the [Pb/hs] and [Pb/ls] predictions 
of the 2.0 M$_{\odot}$ STAREVOL models (red full line), the 2.0 M$_{\odot}$ and 2.1 M$_{\odot}$
Mount Stromlo models (MSE; black crosses and yellow diamond, respectively), and the 2.0 M$_{\odot}$ FRUITY models (blue dot-dashed line). The black horizontal 
lines represent the [Fe/H] uncertainty of the displayed stars. 
The position of each star from this study is indicated with the first numbers of its IRAS name; the first part of the 2MASS name is used for Magellanic Cloud stars from \citet{desmedt14} and 
\citet{desmedt15}.}\label{fig:Pb_index}
\end{figure}

\begin{figure}[t!]
\resizebox{\hsize}{!}{\includegraphics{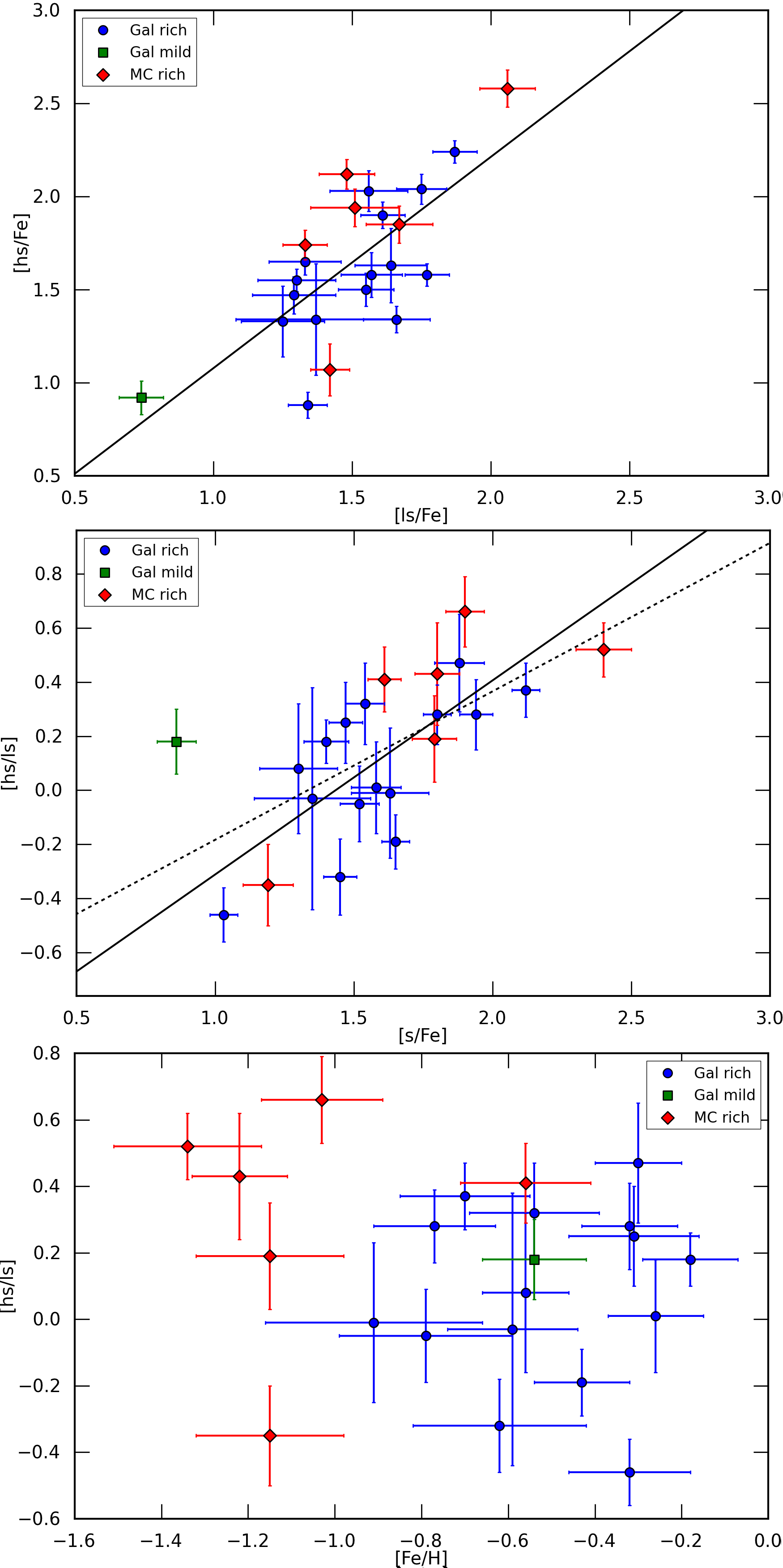}}
\caption{Upper panel: correlation between the \textit{s}-process indices [hs/Fe] and [ls/Fe] for our sample stars and Magellanic Cloud
stars from \citet{desmedt14,desmedt15}. Also included are the results of two Galactic \textit{s}-process enriched stars from 
\citet{reyniers04}: IRAS 04296+3429 and IRAS 23304+6147. Galactic objects are represented by blue circles (= strongly-enriched) and 
a green square (= mildly-enriched), 
Magellanic Cloud objects are represented by red diamonds. 
Middle panel: similar to upper panel but for the correlation between the total enrichment 
in \textit{s}-process elements [s/Fe] and the [hs/ls] index. The dashed line shows the least-squares fit to all results, the full line shows the fit 
for all stars except for IRAS 17279-1119 (green square).
Lower panel: similar to upper panel but for the correlation between the metallicity [Fe/H] and the [hs/ls] index.
For more information, see text.}\label{fig:index}
\end{figure}

In this section, we compare 
our derived Pb$_{\textrm{up}}$ results with AGB model predictions
obtained for a star of 2 M$_{\odot}$. The \textit{s}-process in present AGB models is 
almost insensitive to the initial stellar mass for these low-mass AGB stars.

Different observational indices are commonly used for representing \textit{s}-process
distributions and \textit{s}-process overabundances, namely [ls/Fe], [hs/Fe],
[s/Fe], and [hs/ls].  The elements that 
are included for the calculation of these indices vary between authors. Here, we calculate
the indices with the same elements as in \citet[][and references therein]{desmedt15}. For the ls index, we take the 
mean of the relative abundances of Y and Zr. For hs, we compute the mean of the relative 
abundances of La, Ce, Nd, and Sm. The [hs/Fe] and [ls/Fe] of the programme stars 
are listed in Table \ref{table:ratios}. The index [s/Fe] is the mean of the relative abundances of the
elements for the ls and hs indices and the index [hs/ls] is calculated according to 
[hs/Fe]-[ls/Fe]. The  index [hs/ls] is an indicator of the
neutron irradiation when the \textit{s}-process takes place in radiative shells 
\citep{gallino98,busso01}. Larger neutron irradiation results in the creation of 
heavier \textit{s}-process elements, hence, [hs/ls] > 0.
Indices [Pb/hs] ([Pb/Fe] - [hs/Fe]) and [Pb/ls] ([Pb/Fe] - 
[ls/Fe]) then represent the overabundance of Pb with respect to the other \textit{s}-process 
overabundances. The third and second last column of Table \ref{table:pb} show the 
[Pb$_{\textrm{up}}$/hs] and [Pb$_{\textrm{up}}$/ls]. 

We were unable to determine the abundances of Ce for IRAS 14429-4539
and Sm in four hotter stars. To estimate the Ce and Sm abundances for
these objects, we scale the abundances of Ce and Sm  to the
abundances of La and Nd, which are the elements with an atomic mass closest to Ce and
Sm. To scale the abundances, we make use of the AGB nucleosynthesis models from
the online database FRUITY\footnote{http://fruity.oa-teramo.inaf.it/}
(Franec Repository of Upgraded Isotopic Tables and
Yields; \citet{cristallo11}), which covers a wide metallicity range,
and we use the models with metallicities closest to the metallicity of
the star. We use an initial mass of 1.5 M$_\odot$ for the
FRUITY models. 

The [s/Fe] results in Table \ref{table:ratios} indeed confirm the
strong \textit{s}-process enhancements of the sample stars with [s/Fe]
$\geqslant$ 1.0 except for IRAS 17279-1119 with [s/Fe] = 0.86 dex, which
is mildly \textit{s}-process enriched (0.0 < [s/Fe] < 1.0 dex).  IRAS
22223+4327 and IRAS 08143-4406 have [hs/ls] < 0. 
IRAS 07134+1005, IRAS 08281-4850, IRAS 14325-6428, and IRAS 19500-1709 
have [hs/ls] $\approx$ 0, while the 
remaining eight stars, including the mildly enriched IRAS 17279-1119 
have [hs/ls] > 0.

The dominant neutron source in low-mass AGB stars is expected to be the
$^{13}$C($\alpha$,n)$^{16}$O reaction during the radiative interpulse phase,
leading to [hs/ls] > 0 at subsolar metallicities $-1.2$ dex $\la$ [Fe/H] $\la -0.1$ dex 
\citep{goriely00}. In intermediate-mass AGB stars (M > $\sim$ 4 M$_{\odot}$), 
the $^{22}$Ne($\alpha$,n)$^{25}$Mg reaction 
within the convective thermal pulse may dominate the neutron production if the 
temperature at the base of the pulse reaches some 3.5 $\times$ $10^8$ K. In this 
case, the production of light \textit{s}-process elements is favoured because of 
the convective nature of the region where the neutron irradiation takes place, 
and the resulting [hs/ls] index tends to be negative \citep[e.g.][]{goriely00,goriely04,karakas12}.

In Fig. 14, we compare the adopted Pb$_{\textrm{up}}$ results with the [Pb/hs] and [Pb/ls] 
predictions of AGB evolution and nucleosynthesis models, as in \citet{desmedt15}. The determined abundances are 
compared to the 2.0 M$_{\odot}$ model predictions of the STAREVOL code \citep[][and references therein]{goriely04,siess07},
the 2.0 M$_{\odot}$ and 2.1 M$_{\odot}$ Mount-Stromlo Evolutionary (MSE) predictions 
\citep[][and references therein]{fishlock14,karakas10a}, and the 2.0 M$_{\odot}$ model predictions of 
FRUITY \citep{cristallo11}. The MSE, STAREVOL, and FRUITY predictions are rather insensitive to the initial stellar 
mass and show the global expected behaviour of [Pb/hs] and [Pb/ls] as a function of [Fe/H].
For the clarity of Fig. \ref{fig:Pb_index}, we have omitted the 1.5 M$_{\odot}$ FRUITY model 
data since the differences between the 1.5 M$_{\odot}$ and 2.0 M$_{\odot}$ models are small.

For higher metallicity stars with [Fe/H] > -0.7 dex, the model
predictions are well within the observed Pb$_{\textrm{up}}$ results,
while for stars with [Fe/H] < -0.7 dex the discrepancy between
observations and predictions become larger as the metallicity
decreases. This trend is also confirmed by stars with well-constrained
Pb$_{\textrm{up}}$ results, such as IRAS 05113+1347, IRAS 05341+0852,
IRAS 06530-0213, and IRAS 22223+4327, covering an [Fe/H] range from -0.8
dex to -0.3 dex. For IRAS 07430+1115, the [Pb$_{\textrm{up}}$/hs] result
falls within the predicted limits. 

In addition, it is not surprising to see that the Pb$_{\textrm{up}}$ results 
of the hottest stars in the sample lie high above the model predictions in Fig. \ref{fig:Pb_index}.
As expected from Fig. \ref{fig:teff_Pb}, the very high Pb$_{\textrm{up}}$ results for these 
stars are more related to temperature than to visible Pb line features. Therefore, we consider these 
Pb$_{\textrm{up}}$ as poorly constrained. We conclude that stars with T$_{\textrm{eff}}$ > 
7500 K without significantly high Pb abundances are not useful for comparison with
theoretical Pb abundance predictions. However, these hot stars remain useful for
C/O ratio and other \textit{s}-element comparisons.

\subsection{Correlations}\label{subsect:corr}

Previous studies \citep[e.g.][]{vanwinckel03} have shown that there is no correlation between
metallicity and neutron irradiation. According to the models, the [hs/ls] index
is expected to increase with decreasing metallicity up to [Fe/H] $\sim$ -0.4 dex 
after which it strongly decreases towards higher metallicities 
\citep[e.g.][]{goriely00}. With the addition of our 
new results, we confirm that this is not observed: in the lower panel of 
Fig. \ref{fig:index}, we illustrate the absence of this expected 
correlation between metallicity and [hs/ls].  This implies that
the total neutron irradiation is not (only) dependent on metallicity
in our metallicity range.

In the upper panel of Fig. \ref{fig:index}, we  show the correlation 
between [hs/Fe] and [ls/Fe] for our sample stars, the Magellanic Cloud 
post-AGB stars in \citet{vanaarle13,desmedt14,desmedt15}, and for two additional 
Galactic \textit{s}-process enriched post-AGB stars from \citet{reyniers04}.
It is clear that both \textit{s}-process indices are correlated and a high [ls/Fe] generally 
also implies a strong enhancement in [hs/Fe]. 
The spread is, however, large and this indicates that the different objects 
were subject to a variable neutron irradiation, irrespective of the dilution of the atmosphere.

In previous studies, \citet{reyniers04}, \citet{vanaarle13}, and
in \citet{desmedt15} identified a strong correlation between [s/Fe] 
and the [hs/ls] index in Galactic and Magellanic
Cloud objects. Strongly enriched objects, for which there has
been a low dilution of the enriched material with the envelope
material, were also subject to strong neutron irradiation and, hence, showed a
high [hs/ls].  Here, we add/update the results of our 14
programme stars to the previously used sample.
The resulting [s/Fe] and [hs/ls] correlation is shown in the middle panel of
Fig. \ref{fig:index} where two correlations are shown.  The dashed
line represents the correlation between the [s/Fe] vs [hs/ls] results
for all stars in the plot; the full line represents the correlation
without the inclusion of IRAS 17279-1119.  We choose these two
correlations as IRAS 17279-1119 has [s/Fe] < 1 dex whereas the other
stars have [s/Fe] $\geqslant$ 1 dex.  The correlation of the full
and dashed lines in the middle panel of Fig. \ref{fig:index} are 0.75 and
0.65, respectively. The second result shows that the inclusion of the mildly enhanced 
IRAS 17279-1119 moderately lowers the correlation for all objects. 
This suggests a different \textit{s}-process history for IRAS 17279-1119 with
respect to the other programme stars (see Sect. \ref{sect:iras17279}).

\section{IRAS 17279-1119}\label{sect:iras17279}

\begin{figure}[t!]
\resizebox{\hsize}{!}{\includegraphics{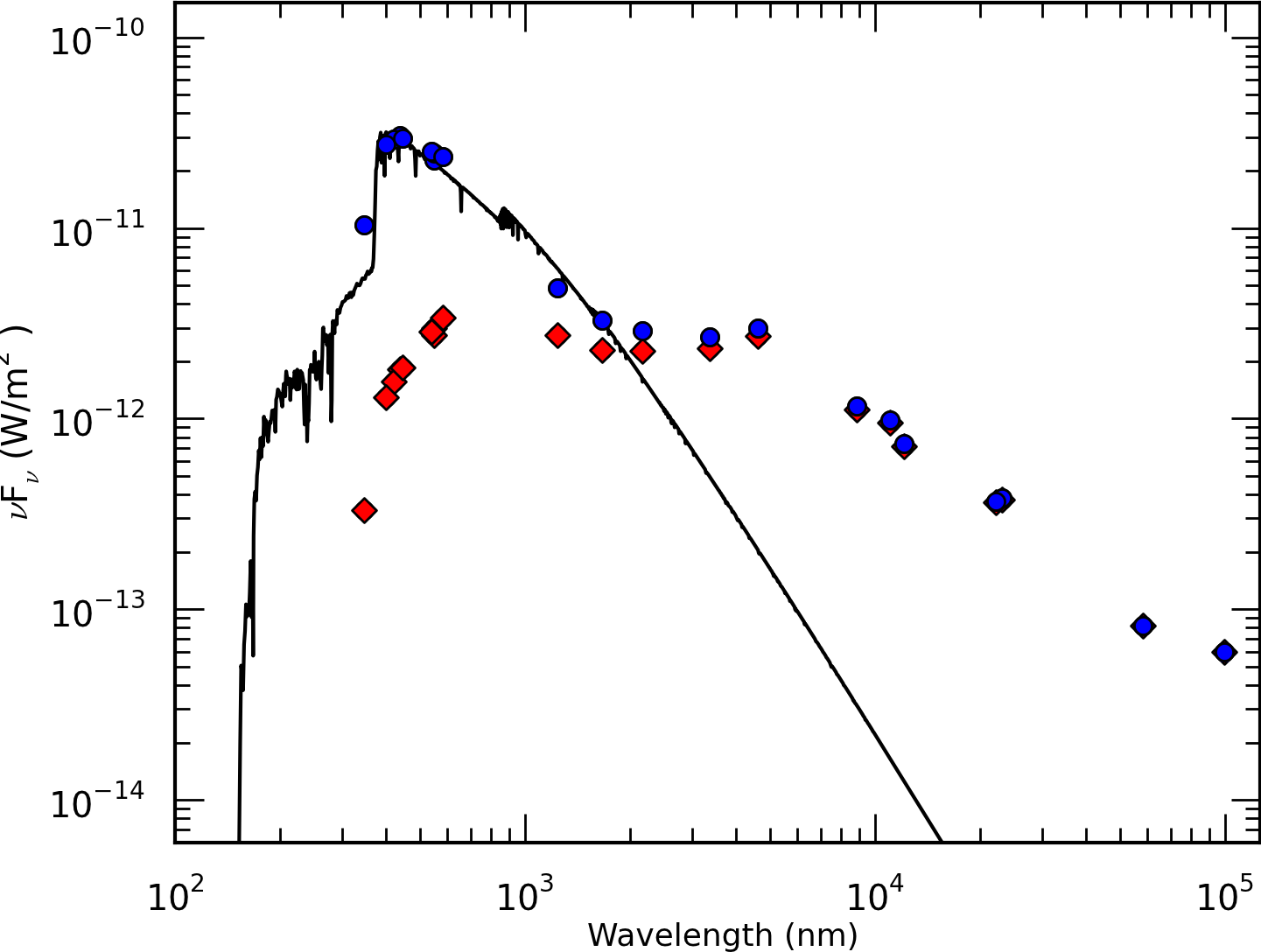}}
\caption{Spectral energy distribution of IRAS 17279-1119. Red dots represent the red, original photometry, blue dots
represent the dereddened photometry. The black line is the scaled Kurucz atmosphere model.}\label{fig:sed_hd158616}
\end{figure}

\begin{figure}[t!]
\resizebox{\hsize}{!}{\includegraphics{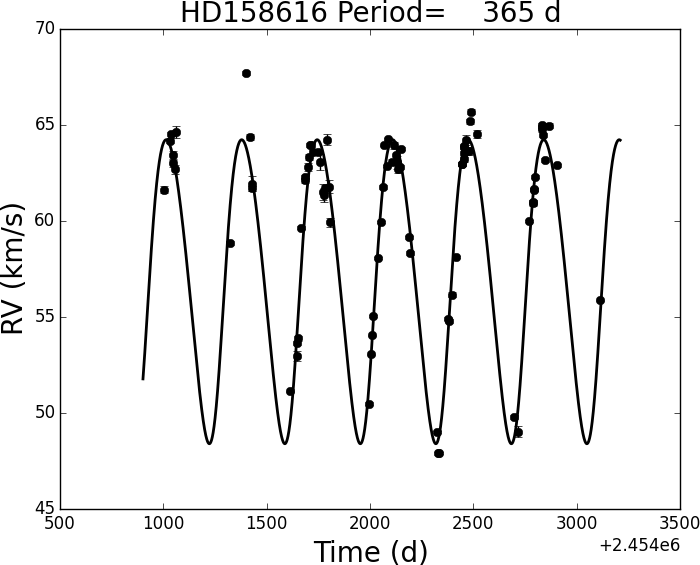}}
\caption{Radial velocity orbit of IRAS 17279-1119. Black dots represent the observed radial velocities;
the black curve is the sine-fit to the orbit. For more information, see text.}\label{fig:orbit_hd158616}
\end{figure}

IRAS 17279-1119 has only a mild s-process enrichment (see lower right panel of Fig. \ref{fig:xfe_middle}) and is 
an outlier in the middle panel of Fig. \ref{fig:index}, so in this section we focus specifically on this object.

Fig. \ref{fig:sed_hd158616} shows the spectral energy
distribution (SED) of IRAS 17279-1119.  The photometric data for
constructing the SED is retrieved from the following catalogues:
the General Catalogue of Photometric Data \citep[GCPD;][]{GCPD},
the All-sky compiled catalogue of 2.5 million stars
\citep{kharchenko01}, the 2MASS All-Sky Catalog of Point Sources
\citep{cutri03}, the WISE All-Sky Data catalogue \citep{cutri12},
the AKARI/IRC mid-infrared all-sky survey \citep{ishihara10}, and
the IRAS catalogue of Point Sources \citep{IRAS}.

Fig. \ref{fig:sed_hd158616} shows both the original raw photometric points (red
diamonds) as well as the dereddened data (blue dots). We consider two possible reddening sources for Galactic
objects. The first
source is reddening by interstellar dust in the Galaxy towards
IRAS 17279-1119. The second source is reddening by the
circumstellar dust of IRAS 17279-1119 itself. 
We use the Galactic extinction curves of \citet{cardelli89} to
determine the reddening for both sources. The total reddening is determined by
applying a $\chi^2$ minimisation on the fit between the
dereddened broadband fluxes and our preferred Kurucz model
atmosphere, which we used for the abundance study of IRAS
17279-1119. The error on the total reddening E(B-V) is determined
by a Monte Carlo simulation of 100 arrays with a normal
distribution of the original flux. We find a total reddening of
E(B-V) = 0.76 $\pm$ 0.02 for IRAS 17279-1119.

The SED shown in Fig.~\ref{fig:sed_hd158616} indicates an excess starting 
at 2 micron indicating the thermal emission of dust near the dust-sublimation
temperature and therefore close to the central star.
It is now well established that this traces the presence of a
stable compact circumbinary disc \citep[e.g.][]{deruyter06,deroo07,gielen11,hillen13,hillen14,bujarrabal15,gezer15}. 
The specific characteristics of the SED are closely related to the binary nature of the central
star \citep{vanwinckel09}.

We obtained radial velocity monitoring data with the HERMES spectrograph
\citep{raskin11} mounted on the 1.2 meter Mercator telescope in the
framework of our large programme on evolved binaries \citep{vanwinckel10,
gorlova13}. In total we obtained 85 radial velocity measurements
starting on 19/06/2009 covering a total time span of 2211d.
We also used the ASAS \citep{pojmanski04} photometry to study the photometric
behaviour. The photometry consists of 788 data points covering a
time span of 3156d. Many of the variables near the Population II Cepheid
instability strip have complex light curves in which several modes are
excited with periods close to each other so that beating effects are
easily detectable \citep{kiss06}. IRAS 17279-1119 is no exception and
periods of 90, 87, and 83 days are found in the photometry using a S/N criterion of 4 in
the Fourier transform. Unfortunately, our photometric series do not overlap in time 
with the radial velocity series.

\begin{table}
\caption{Orbital elements of IRAS 17279-1119.}
\centering
\label{table:orbital}     
\begin{tabular}{lcc}    
\hline\hline            
                                   &                 & $\sigma$  \\ \hline
Period (d)                         & 365.0           & 1.0       \\ 
e                                  & 0.0 (forced)    &           \\ 
K (km s$^{-1}$)                    & 7.92            & 0.44      \\
$\gamma$ (km s$^{-1}$)             & 56.8            & 0.5       \\
a\,$\sin{\rm i}$ (A.U.)            & 0.27            & 0.02      \\
f(M) (M$_{\odot}$)                 & 0.026           & 0.004     \\ 
RMS residuals  ( km\,s$^{-1}$)     & 1.8             &           \\ \hline\hline
\end{tabular}    
\end{table}

We detect orbital motion (see Fig~\ref{fig:orbit_hd158616}) with an
orbital period of exactly one year.
The orbital solution is clear with a fractional variance reduction of 91\% in the fit.
In Table~\ref{table:orbital} we list the orbital data. We cover in total
close to six cycles, but as the period is close to one year, the phase coverage
is not ideal. Nonetheless, the orbital detection is clear.

With an 'a $\sin i$' of only 0.27 AU, it is clear that the current
orbit is too small to accommodate a TP-AGB star whose radius is
expected to be 1-2 AU. The object must have experienced strong binary
interaction when on the AGB. The mass function gives a constraint on
the mass of the companion provided the mass of the primary and the
inclination are known. As the primary is a post-AGB star, it is fair
to assume the mass is a typically white dwarf mass of 0.6 M$_{\odot}$.
Assuming a most probable inclination of 60 degrees, the mass of the
companion is only 0.3 M$_{\odot}$. With an inclination of 30 degrees,
this becomes 0.6 M$_{\odot}$.  As there are no
symbiotic activities, we suggest that the companion is non-evolved.

To our knowledge, IRAS 17279-1119 is the first \textit{s}-process rich
post-AGB star discovered to be in a spectroscopic binary. If, during
the binary interaction phase, there was mass transfer of enriched
material to the secondary, the companion accreted \textit{s}-process
rich gas. This object seems therefore a precursor of extrinsically
\textit{s}-process enriched stars. These are binaries in which the observed
\textit{s}-process rich component was polluted by material expelled
from their companion stars when they passed the AGB evolutionary phase and
which are now  white dwarfs \citep[see][]{jorissen98}. IRAS 17279-1119
may very well be the first detected precursor of Ba dwarfs, given
the low mass of the companion.

\section{Conclusions}\label{sect:conclusion}

In this paper, we presented a homogeneous abundance study of a sample
of 14 Galactic \textit{s}-process enriched post-AGB stars with the aim
of determining Pb abundances. Our results confirm that two newly
identified 21 $\mu$m sources, IRAS 13245-5036 and IRAS 14429-4539, are
indeed carbon and \textit{s}-process enriched and we confirm the post-AGB
nature of both objects. Furthermore, we find a strong chemical
diversity for the sample stars, ranging from only a mild \textit{s}-process
enrichment up to strong \textit{s}-process enrichment.

None of the sample objects display clear distinctive Pb
lines in their spectra and therefore we can only determine 
Pb abundance upper limits (Pb$_{\textrm{up}}$). These Pb
abundance upper limits are determined using synthetic
spectra. A comparison with theoretical spectra reveals that some of the
cooler stars in the sample display visible contributions of Pb in
the unresolved blends. Because of the low S/N, it is impossible to
determine the contribution and hence abundance accurately. 

For the hotter stars with T$_{\textrm{eff}}$ > 7500 K,
there are no visible Pb line contributions in the spectra. 
Therefore, these Pb$_{\textrm{up}}$ results are poorly
constrained, which makes hot post-AGB stars without drastically high Pb
abundances not useful for the Pb comparison with theoretical AGB
nucleosynthesis predictions.

A comparison of observed Galactic Pb$_{\textrm{up}}$ abundances with 
theoretical \textit{s}-process calculations of $\sim$ 2.0 M$_{\odot}$ 
stellar models 
confirm the results of Magellanic Cloud stars from \citet{desmedt14}
and \citet{desmedt15}. For metallicities [Fe/H] > -0.7 dex, the predicted 
Pb abundances are well within the observed Pb$_{\textrm{up}}$ results, while 
there exists an increasing discrepancy towards lower metallicities. Additional 
research is necessary to explain this trend.

We confirm that there is no correlation
between neutron irradiation and metallicity in the metallicity range
probed here.
These results show that a wide spread in neutron irradiation is needed at all metallicities to explain these results. 
In comparative studies of extrinsically \textit{s}-process enhanced objects, similar conclusions were obtained, even at lower metallicity \citep[e.g.][]{bisterzo12}.

We confirm the correlation between [hs/ls] and [s/Fe]
indicating the connection between strong neutron irradiation and overall photospheric enrichments.  The mildly \textit{s}-process enriched
IRAS 17279-1119 does not follow this correlation because the orbital
elements suggest that an evolutionary cut-off from the AGB was forced
prematurely because of binary interaction: the current orbit is too small
to accommodate an AGB star with a typical luminosity. In addition, 
IRAS 17279-1119 is to our knowledge the first confirmed \textit{s}-process 
rich post-AGB star in a binary system.

In summary, we find an increasing discrepancy between observed and
predicted Pb abundances towards lower metallicities. All objects with
a metallicity [Fe/H] < $-$0.7 dex show Pb upper limits that are lower
than the predicted Pb abundances. Contrary to the Magellanic Cloud
samples, the distances to the Galactic sample stars are poorly known
and hence it is difficult to constrain their current core masses and their initial masses.  Distance estimates are hence necessary to
connect the determined abundance results with tailored theoretical predictions
in detail.  The GAIA distance determinations will be crucial in this
respect.

\begin{acknowledgements}
The authors thank S. Cristallo for the public availability of the FRUITY models.
KDS, HVW, RM, and DK acknlowledge support of the KU Leuven contract GOA/13/012.
DK acknowledges support of the FWO grant G.OB86.13. 
SG and LS are FNRS research associates. AIK was supported through an
Australian Research Council Future Fellowship (FT110100475).
Based on observations obtained with the HERMES spectrograph, which is supported by the Research Foundation - Flanders (FWO), Belgium, the Research Council of KU Leuven, Belgium, the Fonds National de la Recherche Scientifique (F.R.S.-FNRS), Belgium, the Royal Observatory of Belgium, the Observatoire de Gen\`eve, Switzerland and the Th\"ringer Landessternwarte Tautenburg, Germany.

\end{acknowledgements}

\bibliographystyle{aa}

\Online

\begin{appendix}

\section{Pb abundances of individual stars}
In this appendix, we give an overview of the Pb abundance upper limit determinations for the different 
programme stars together with a brief discussion of the Pb$_{\textrm{up}}$ results.

\subsection{IRAS 05113+1347}
\begin{figure}
\resizebox{\hsize}{!}{\includegraphics{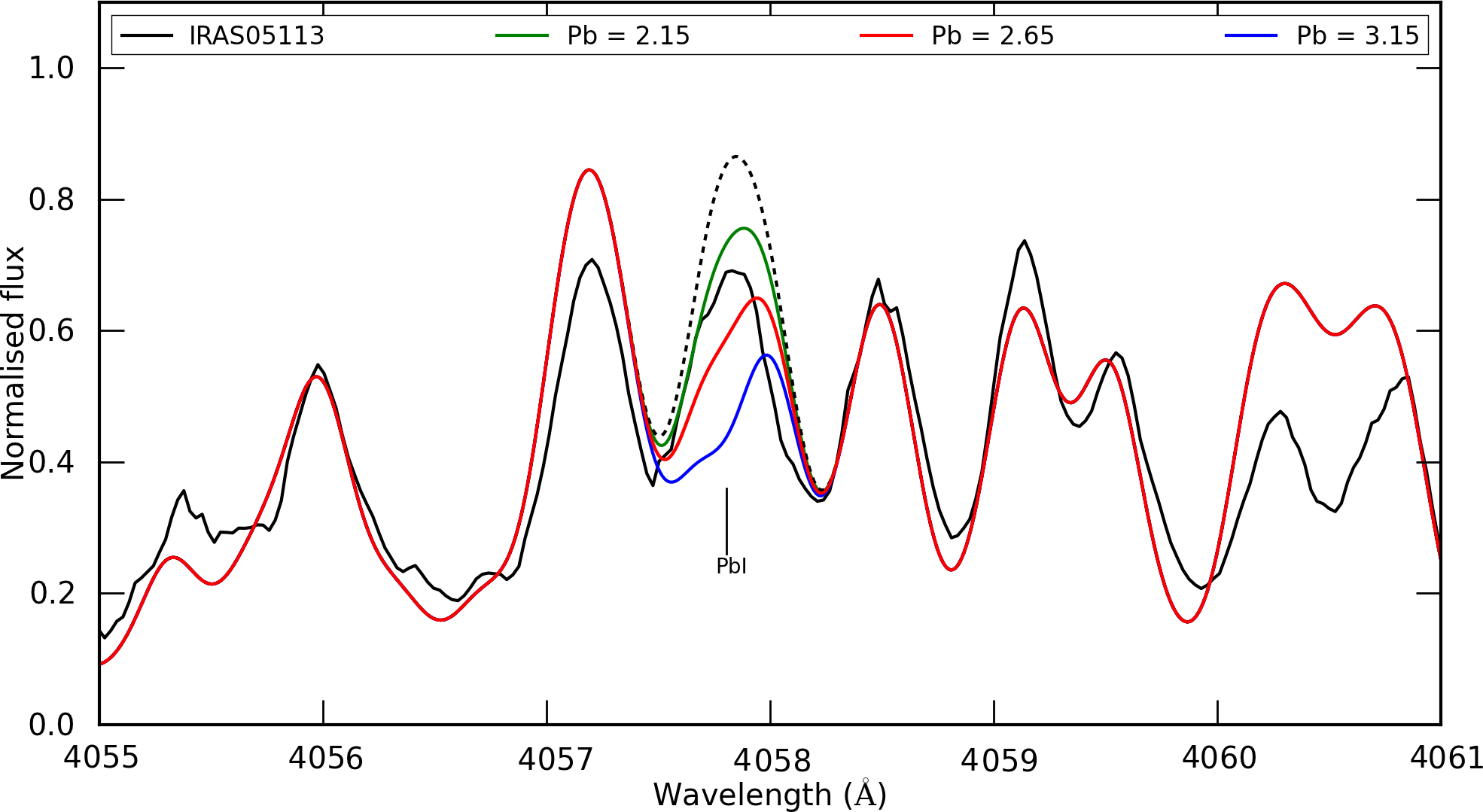}}
\caption{Spectrum synthesis of the Pb I line at 4057.807 \AA{} for IRAS 05113+1347. The black 
spectrum is the observed UVES spectrum;
the coloured spectra represent synthetic spectra with different Pb abundances. The red line 
represents the adopted Pb abundance upper limit; 
the green and blue lines represent the adopted abundance upper limit -0.5 dex and +0.5 dex, respectively.
The dashed black line shows the synthetic spectrum if
no Pb is present.}\label{fig:iras05113_Pb}
\end{figure}

The Pb abundance upper limit determination of IRAS 05113+1347 is presented in Fig. \ref{fig:iras05113_Pb}. 
The determination of the continuum is difficult as the region is dominated by large spectral blends. 
We use the spectral lines at 4057.5 and 4058.2 \AA{} to optimally estimate the position of the continuum.
Thereafter, we adopted a Pb abundance upper limit, which fully incorporates the line feature 
at 4057.807 \AA{}.

The adopted Pb$_{\textrm{up}}$ of IRAS 05113+1347 results in a
[Pb$_{\textrm{up}}$/Fe] which is about 0.2 dex lower than the 
value for [hs/Fe] (left upper of panel Fig.  \ref{fig:xfe_cold})
and similar to the element-over-iron ratios of the elements beyond 
the Ba-peak.

\subsection{IRAS 05341+0852}
\begin{figure}[t!]
\resizebox{\hsize}{!}{\includegraphics{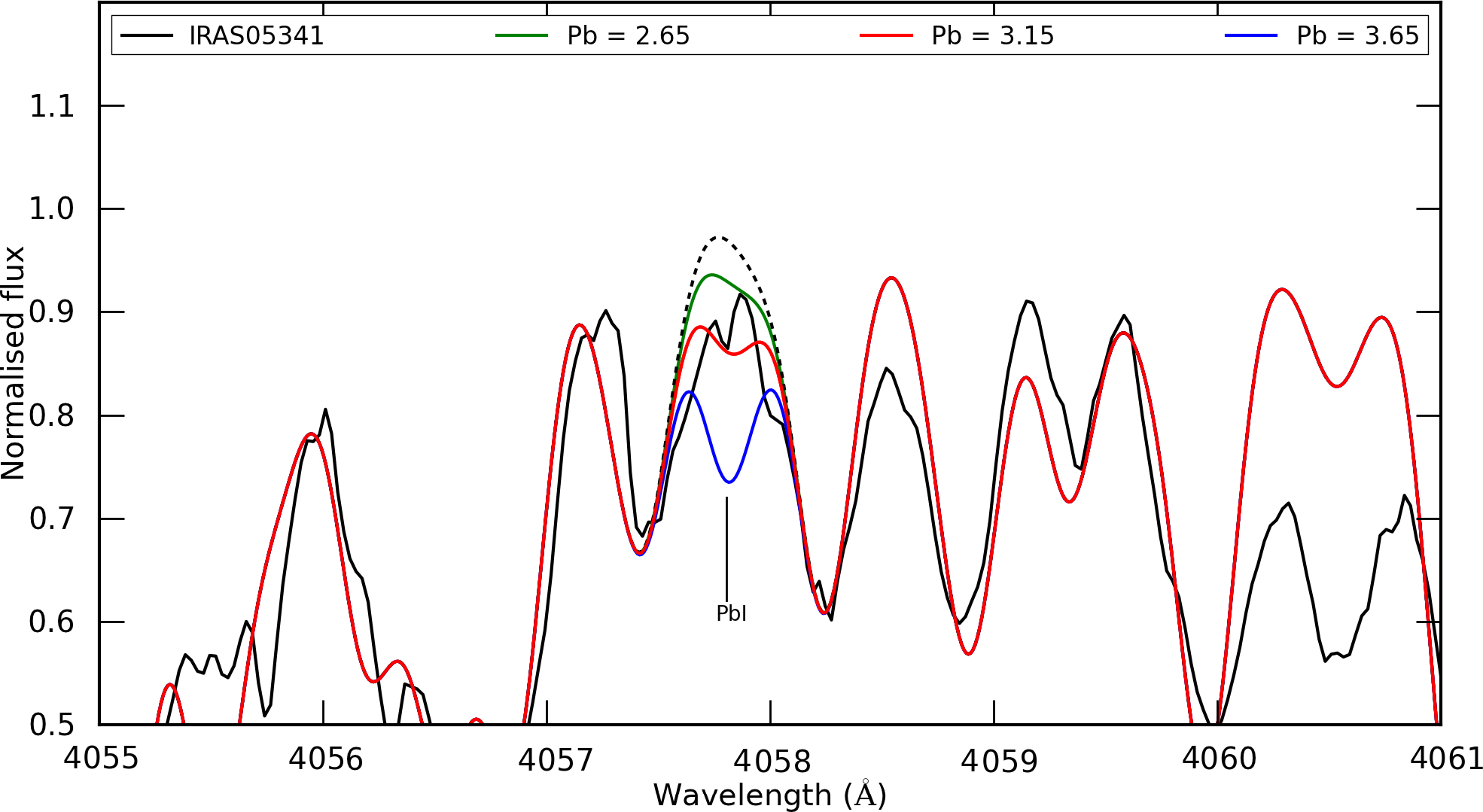}}
\caption{Spectrum synthesis of the Pb I line at 4057.807 \AA{} for 
the UVES spectrum of IRAS 05341+0852. Lines and symbols are similar to Fig. \ref{fig:iras05113_Pb}.
}\label{fig:iras05341_Pb}
\end{figure}

For the strongly \textit{s}-process enriched IRAS 05341+0852, we estimate the position of the continuum 
by finding a compromise between 
the spectral blends at 4057.5 and 4058.2 \AA{} in Fig. \ref{fig:iras05341_Pb}. We adopt a Pb 
abundance upper limit, which 
fully includes the small visible line feature.

Although it is one of the most \textit{s}-process enriched Galactic post-AGB
stars studied to date, IRAS 05341+0852 displays a
[Pb$_{\textrm{up}}$/Fe] that is about 0.15 dex lower than [hs/Fe]
(middle upper panel of Fig. \ref{fig:xfe_cold}). Like in IRAS
05113+1347, the [Pb$_{\textrm{up}}$/Fe] seems similar to the
element-over-iron ratios of the elements beyond the Ba-peak.

\subsection{IRAS 06530-0213}
The Pb abundance determination of IRAS 06530-0213 is described in
detail in Sect. \ref{subsect:pb_abun}. The spectral synthesis of the Pb regions are depicted in Fig.~\ref{fig:iras06530_Pb}. We use the Pb$_{\textrm{up}}$ result of the Pb I
line for this strongly \textit{s}-process enriched post-AGB star.  
The resulting [Pb$_{\textrm{up}}$/Fe] is about 0.7 dex
higher than [hs/Fe] (left upper panel of Fig. \ref{fig:xfe_middle})
and the element-over-iron ratios of the elements beyond the Ba-peak.

\subsection{IRAS 07143+1005}
\begin{figure}[t!]
\resizebox{\hsize}{!}{\includegraphics{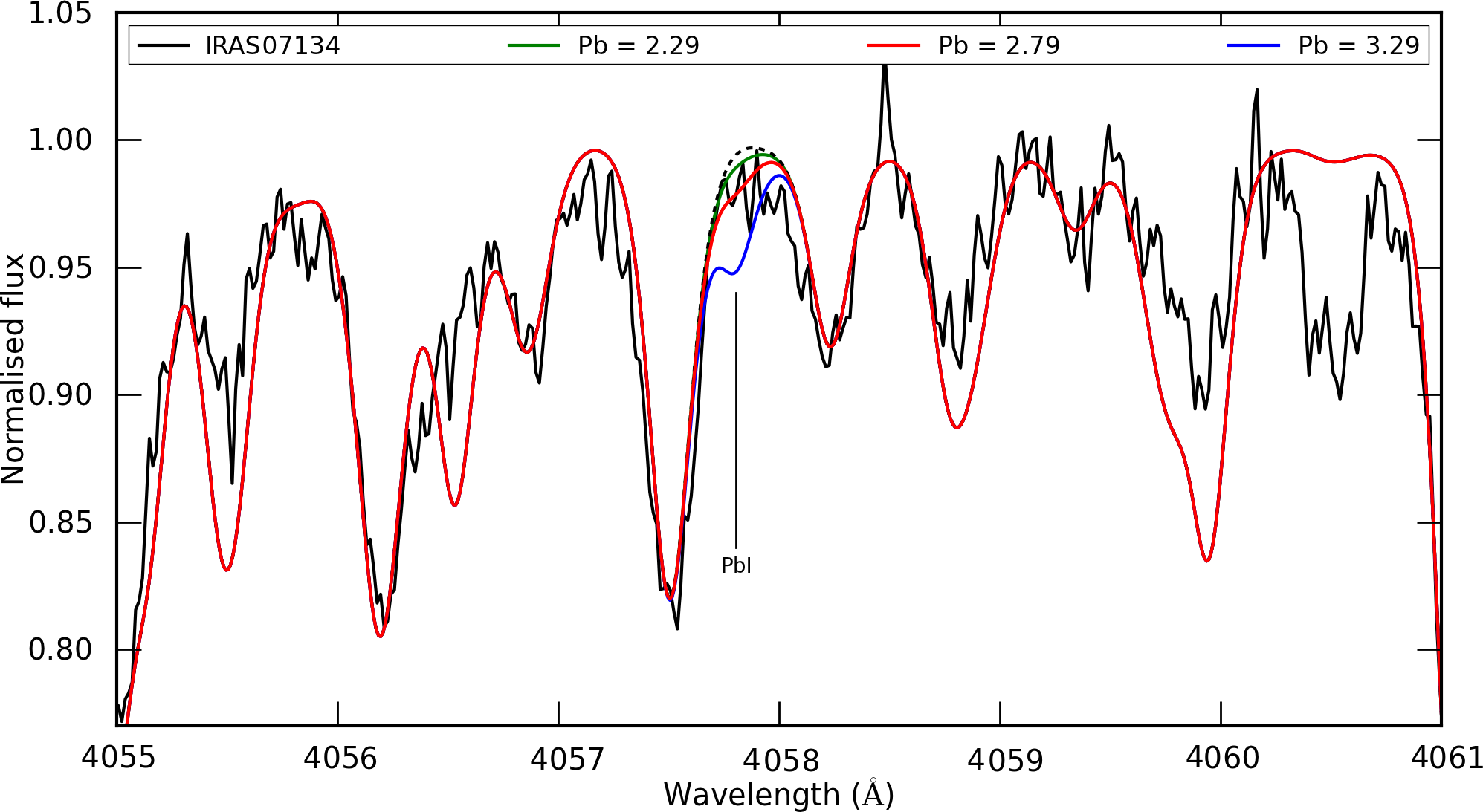}}
\caption{Spectrum synthesis of the Pb I line at 4057.807 \AA{} for 
the HERMES spectrum of IRAS 07143+1005. Lines and symbols are similar to Fig. \ref{fig:iras05113_Pb}.
}\label{fig:iras07134_Pb}
\end{figure}

We estimate the position of the continuum of IRAS 07143+1005 using the spectral blends at 4056.2, 
4056.9 and 4057.5 \AA{} at the blue side of the Pb I line, 
and the blends at 4058.2 and 4059.3 \AA{} at the red side of the Pb I line in Fig. \ref{fig:iras07134_Pb}. 
The spectral region of the Pb I line is dominated by noise and we cannot detect any visible 
contribution from the Pb line. 
We therefore adopt a Pb abundance upper limit, which includes the noise at 4057.807 \AA{}.

The adopted [Pb$_{\textrm{up}}$/Fe] is about 0.4
dex higher than [hs/Fe] and [ls/Fe], which are both similar in
enrichment (middle upper panel of Fig. \ref{fig:xfe_middle}).

\subsection{IRAS 07430+1115}
\begin{figure}[t!]
\resizebox{\hsize}{!}{\includegraphics{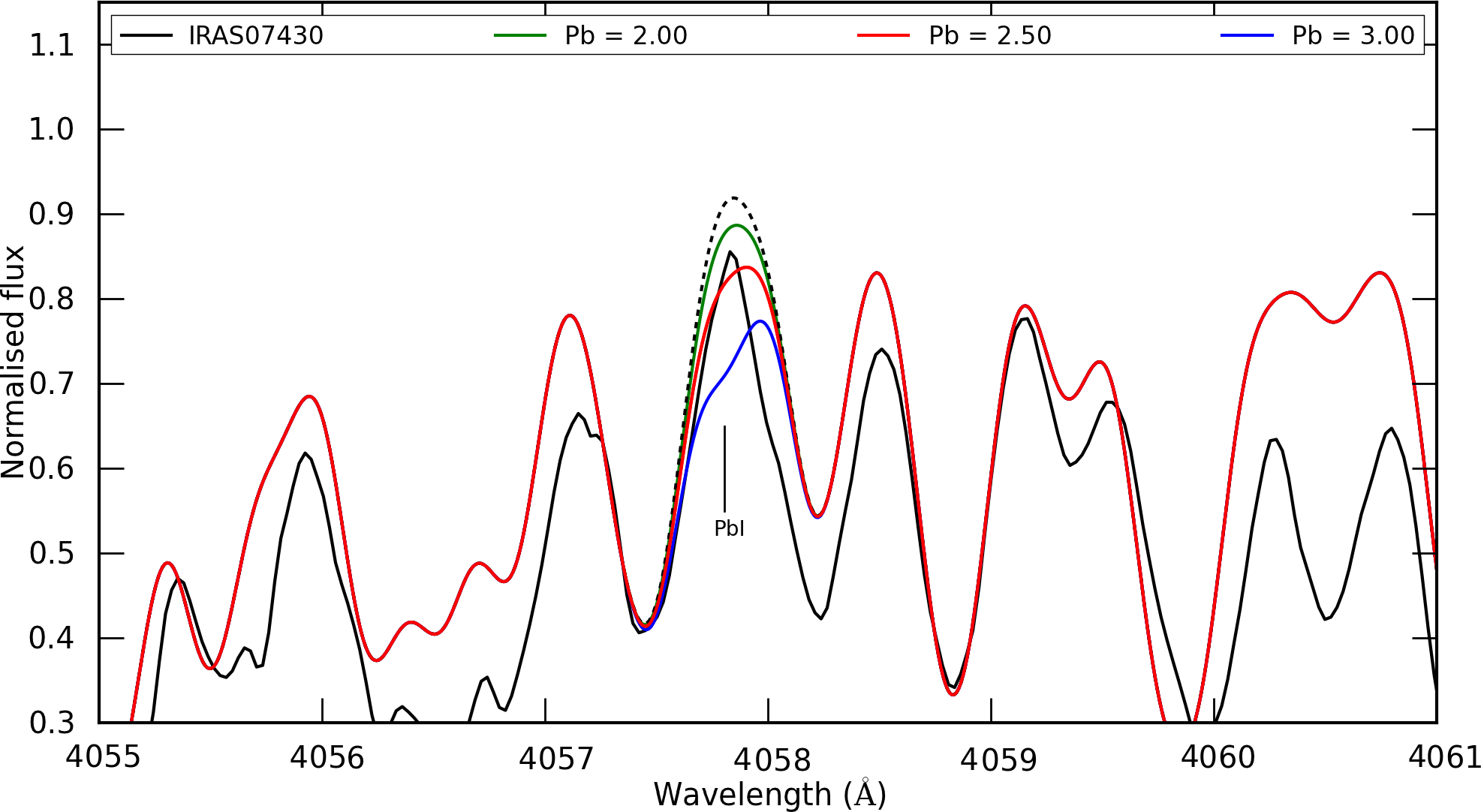}}
\caption{Spectrum synthesis of the Pb I line at 4057.807 \AA{} for 
the UVES spectrum of IRAS 07430+1115. Lines and symbols are similar to Fig. \ref{fig:iras05113_Pb}.
}\label{fig:iras07430_Pb}
\end{figure}

In Fig. \ref{fig:iras07430_Pb}, we use the spectral blends at 4057.5 and 4058.8 \AA{} to estimate the 
position of the continuum. Although not clearly visible, 
the synthetic spectrum without the Pb I line suggests a possible small contribution of Pb at 4057.807 \AA{}. 
Therefore, we adopt a Pb abundance upper limit that incorporates the peak of the Pb I line feature.

The adopted Pb$_{\textrm{up}}$ of IRAS 07430+1115 gives a
[Pb$_{\textrm{up}}$/Fe], which is significantly lower than [hs/Fe] with
a difference of about 0.5 dex (right upper panel of
Fig. \ref{fig:xfe_cold}).

\subsection{IRAS 08143-4406}
\begin{figure}[t!]
\resizebox{\hsize}{!}{\includegraphics{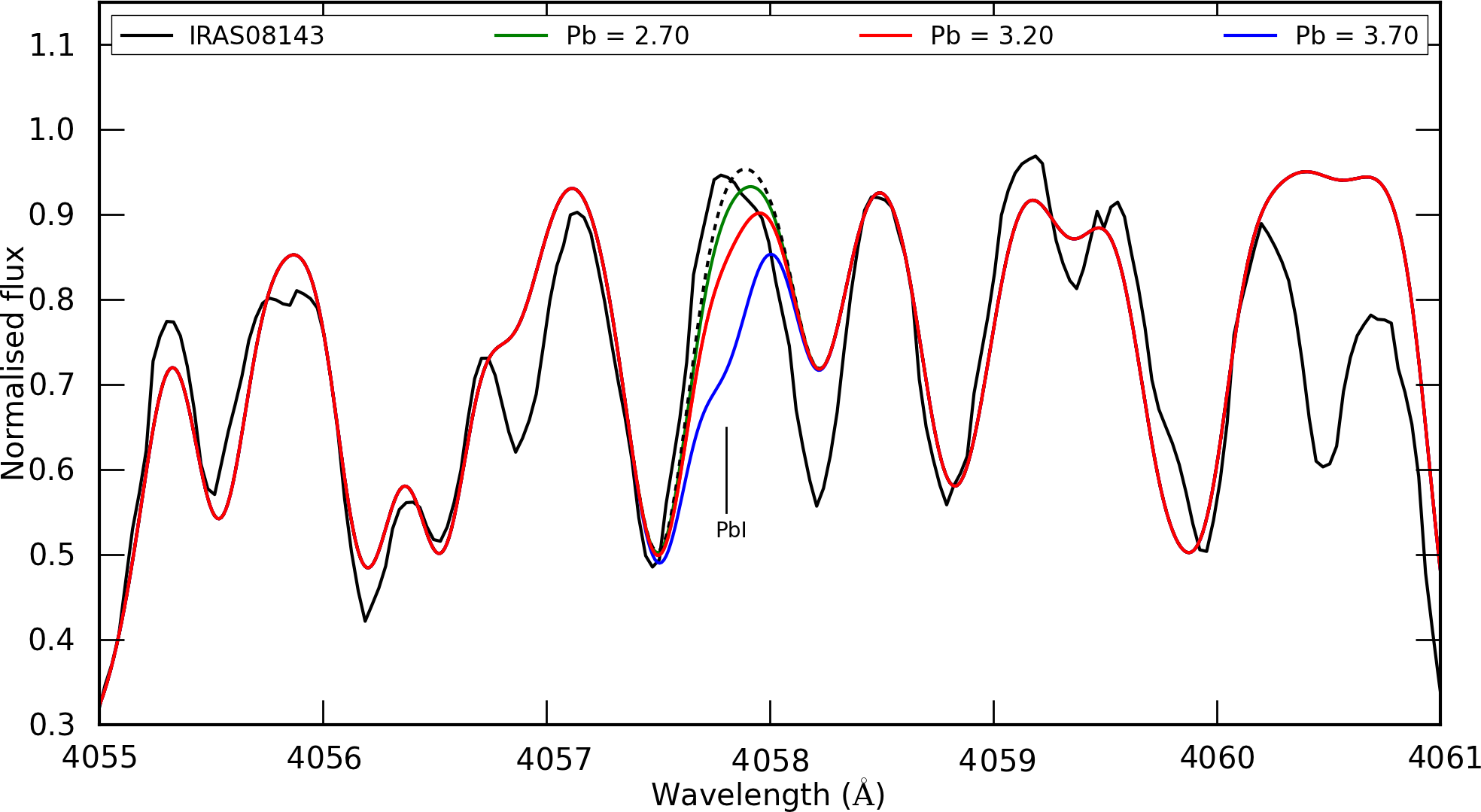}}
\caption{Spectrum synthesis of the Pb I line at 4057.807 \AA{} for 
the BLUE437 UVES spectrum of IRAS 08143-4406. Lines and symbols are similar to Fig. \ref{fig:iras05113_Pb}.
}\label{fig:iras08143_Pb}
\end{figure}

The position of the continuum for IRAS 08143-4406 is estimated using the spectral blends at 4057.5 and 
4058.8 \AA{} in Fig. \ref{fig:iras08143_Pb}. 
The spectrum of IRAS 08143-4406 does not show visible contribution of Pb at 4057.807 \AA,{} and we 
see that with the estimated continuum position, 
the spectral feature at the blue side of the the Pb I line is not even contained within the 
boundaries of the synthetic spectrum without Pb I line, which could  possibly be attributed to noise. 
As a result of these uncertainties, we adopt a Pb abundance upper limit
that fully incorporates the spectral line  in the spectral region of the Pb I line. 

We find that [Pb$_{\textrm{up}}$/Fe] is
about 0.1 higher than [ls/Fe] and 0.3 dex higher than [hs/Fe],
respectively (right upper panel of Fig. \ref{fig:xfe_middle}).

\subsection{IRAS 08281-4850}
\begin{figure}[t!]
\resizebox{\hsize}{!}{\includegraphics{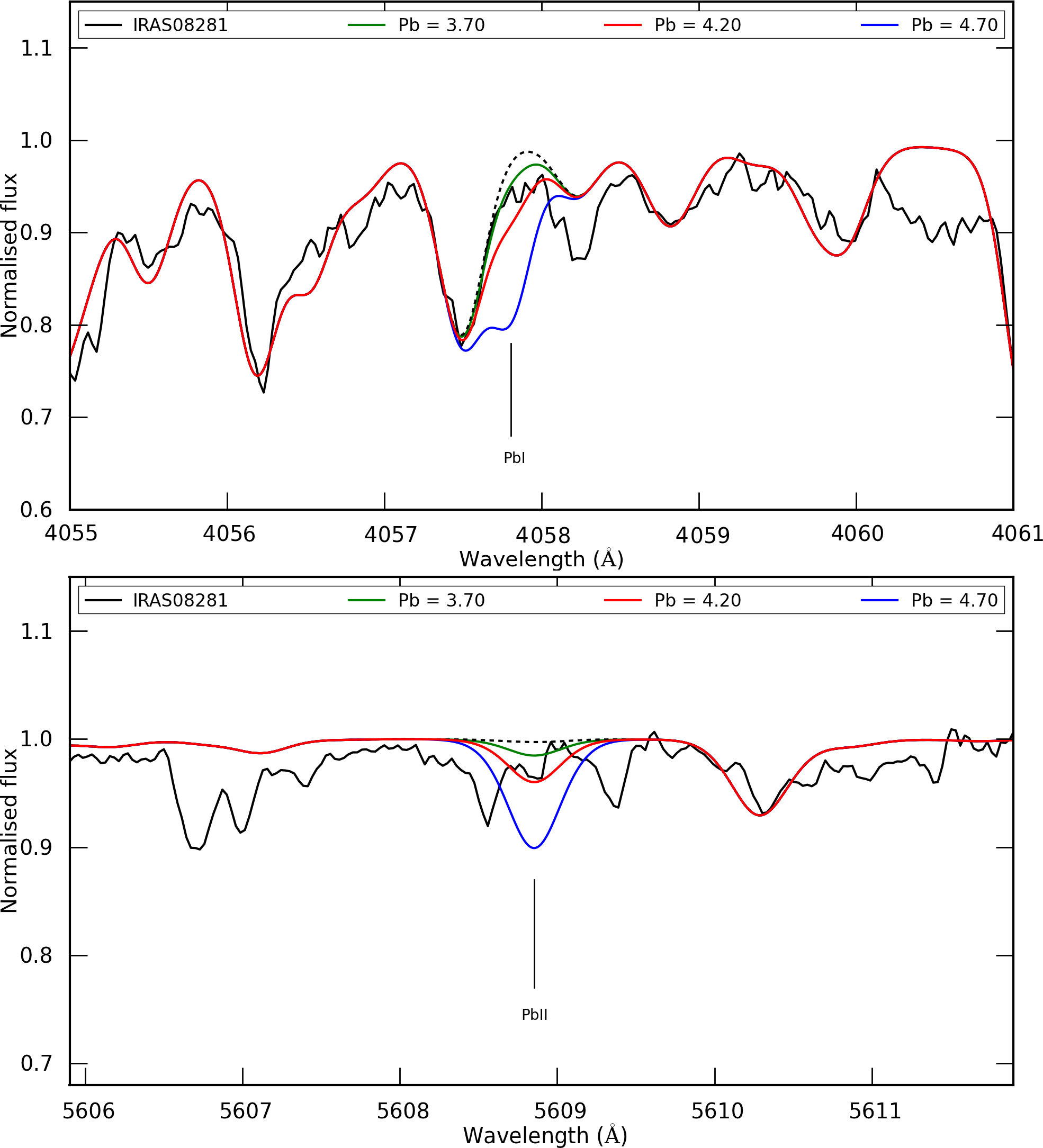}}
\caption{Spectrum synthesis of the Pb I line at 4057.807 \AA{} (upper panel) and the Pb II line at 
5608.853 \AA{} (lower panel) for
the UVES spectra of IRAS 08281-4850. Lines and symbols are similar to Fig. \ref{fig:iras05113_Pb}.
}\label{fig:iras08281_Pb}
\end{figure}

The spectra of IRAS08281-4850 reveal possible Pb line features in both panels of Fig. \ref{fig:iras08281_Pb}, 
but we assume these features are more likely due to noise than a real abundance of Pb.
Nevertheless, we adopt a similar Pb abundance upper limit for both Pb lines that fully incorporates the 
visible spectral features. For the Pb I line, we use the spectral blends at 4056.2, 4057.5, and 
4058.8 \AA{} to estimate 
the position of the continuum. For the Pb II line, we see unidentified spectral lines at 5606.7 
and 5607 \AA{} and also 
at both sides of the Pb II line. These spectral lines are not included in the linelists and we 
can only use the continuum in 
the region 5606 - 5606.5 \AA{} and 5611.5 - 5612 \AA,{} in combination with the spectral line at 
5610.2 \AA,{} to estimate 
the continuum position.

We find a similar Pb$_{\textrm{up}}$ for both the
Pb I and Pb II lines. This results in a [Pb$_{\textrm{up}}$/Fe] that
is about 1.2 dex higher than the [hs/Fe] and [ls/Fe] (left lower panel of Fig. \ref{fig:xfe_middle}).

\subsection{IRAS 13245-5036}
The Pb abundance determination of IRAS 13245-5036 is described in detail in Sect. \ref{subsect:pb_abun}.  The spectral synthesis of the Pb II region is depicted in Fig.~\ref{fig:iras13245_Pb}.
For this hottest star in the sample, we only use
the Pb II line to determine Pb$_{\textrm{up}}$. The adopted
[Pb$_{\textrm{up}}$/Fe] is about 0.9 dex higher than [hs/Fe] (left
upper panel of Fig. \ref{fig:xfe_hot}).

\subsection{IRAS 14325-6428}
\begin{figure}[t!]
\resizebox{\hsize}{!}{\includegraphics{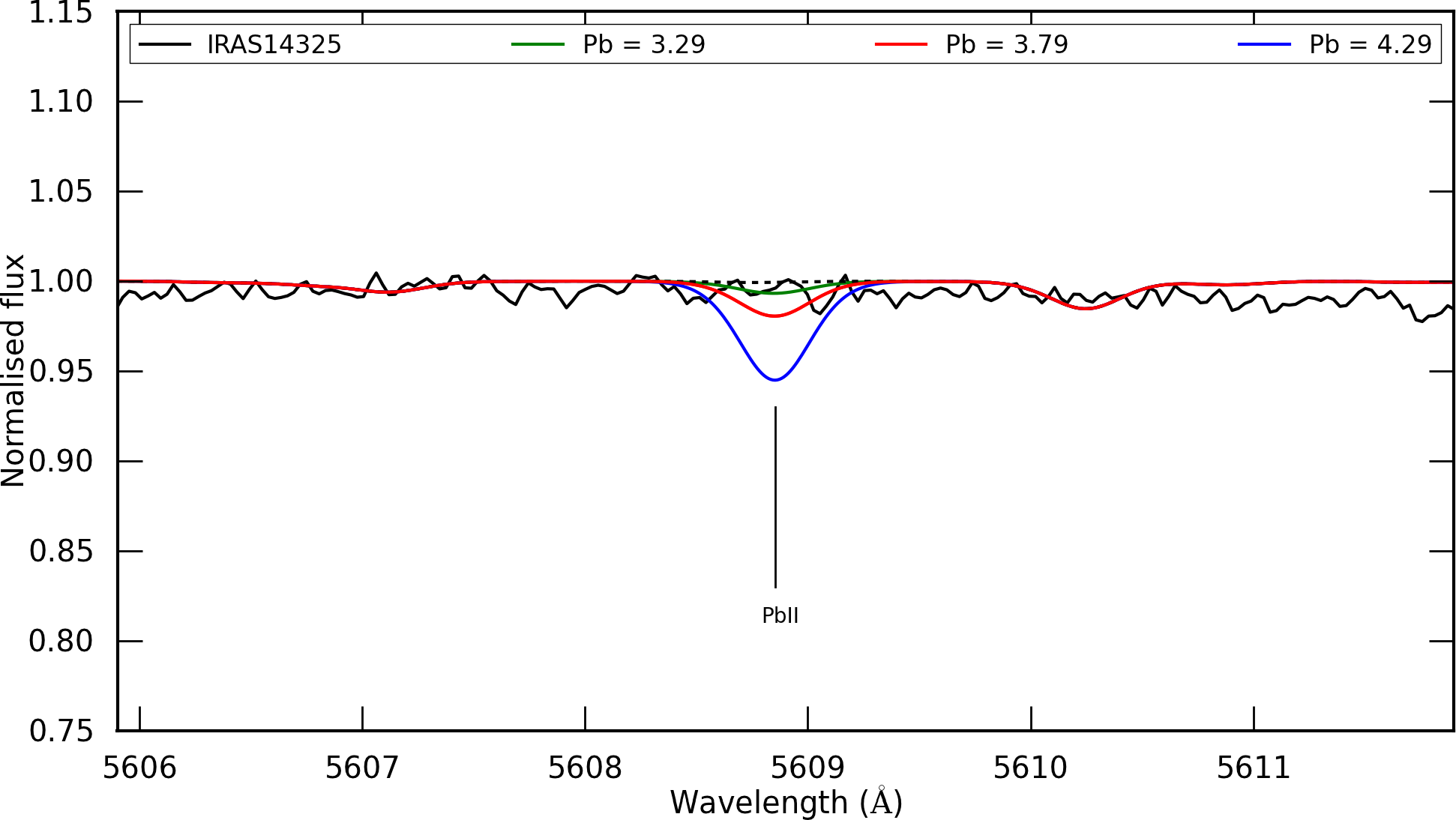}}
\caption{Spectrum synthesis of the Pb II line at 5608.853 \AA{} for
the UVES spectrum of IRAS 14325-6428. Lines and symbols are similar to Fig. \ref{fig:iras05113_Pb}.
}\label{fig:iras14325_Pb}
\end{figure}

The Pb II line spectral region of IRAS 14325-6428 is dominated by noise and does not show a visible 
Pb line contribution in Fig. \ref{fig:iras14325_Pb}.
We estimate the position of the continuum by fitting to the average continuum in the shown spectral 
region. Around 
5610.2 \AA{}, a small line appears  in the synthetic spectra that is not present in the 
observed spectra, so we do not use this
line for the continuum estimate. We then adopt a Pb abundance upper limit, which incorporates the 
continuum at 5608.853 \AA{}.

The adopted [Pb$_{\textrm{up}}$/Fe] is only determined  via the 
Pb II line, and is about 1.3 dex higher than [hs/Fe] 
(right upper panel of Fig. \ref{fig:xfe_hot}).

\subsection{IRAS 14429-4539}
\begin{figure}[t!]
\resizebox{\hsize}{!}{\includegraphics{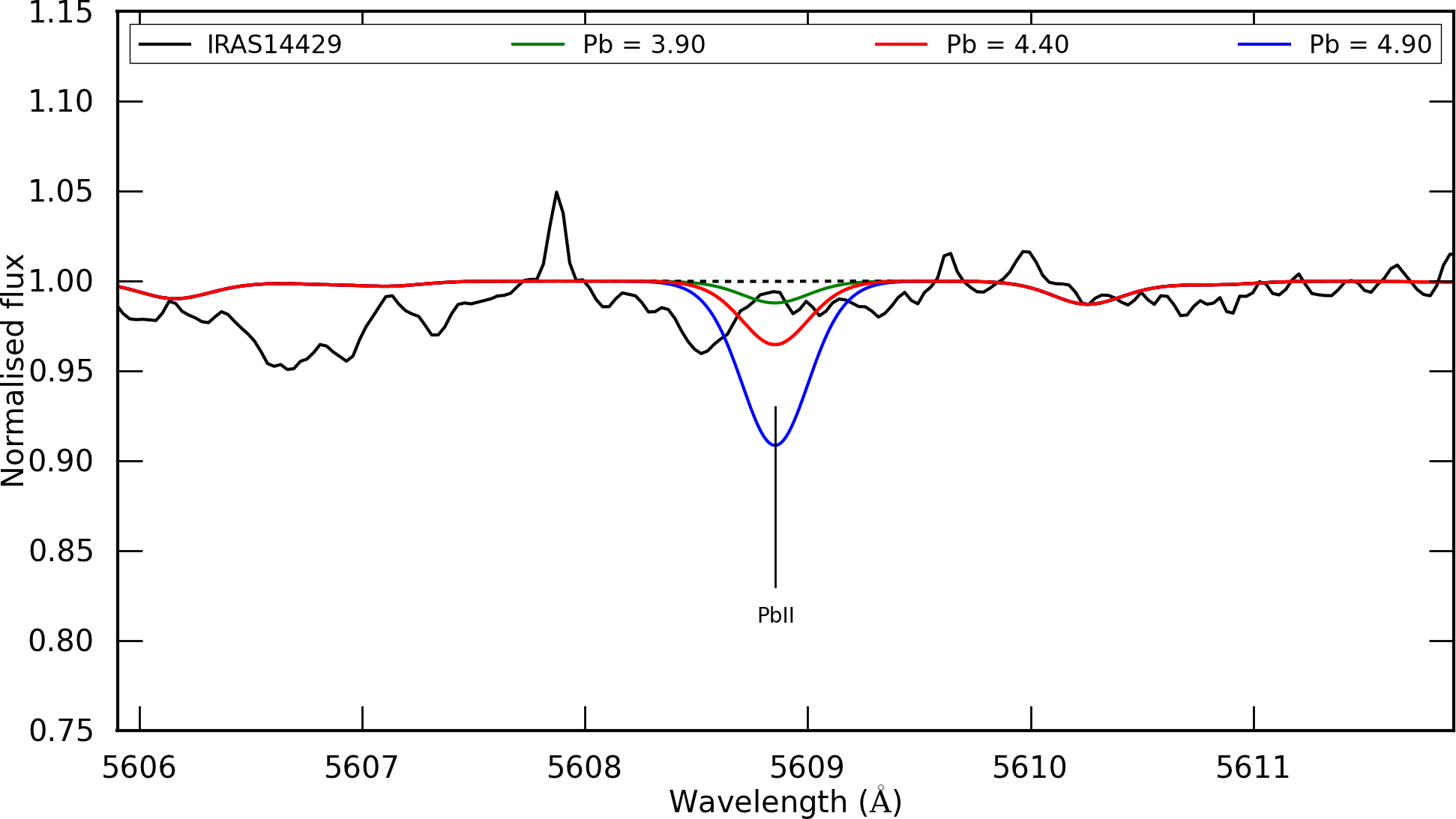}}
\caption{Spectrum synthesis of the Pb II line at 5608.853 \AA{} for
the UVES spectrum of IRAS 14429-4539. Lines and symbols are similar to Fig. \ref{fig:iras05113_Pb}.
}\label{fig:iras14429_Pb}
\end{figure}

Fig. \ref{fig:iras14429_Pb} shows that the spectral region of the Pb II at 5608.853 \AA{} of IRAS 
14429-4539 is totally dominated by noise 
and that there is no visible contribution of the Pb line. 
There is a visible line in the spectrum at 5606.6 \AA,{} but it is not clear whether this feature is 
real or due to noise. For this reason, it is not included 
in the linelists. We therefore estimate the position of the continuum by fitting the average continuum 
in the shown spectral region. We then adopt a  
Pb abundance upper limit, which incorporates the estimated continuum at 5608.853 \AA{}.

For the second hottest star in the sample, we 
only use the Pb II line for the Pb$_{\textrm{up}}$ determination. 
[Pb$_{\textrm{up}}$/Fe] exceeds [hs/Fe] by almost 
1.4 dex (left lower panel of Fig. \ref{fig:xfe_hot}).

\subsection{IRAS 17279-1119}
\begin{figure}[t!]
\resizebox{\hsize}{!}{\includegraphics{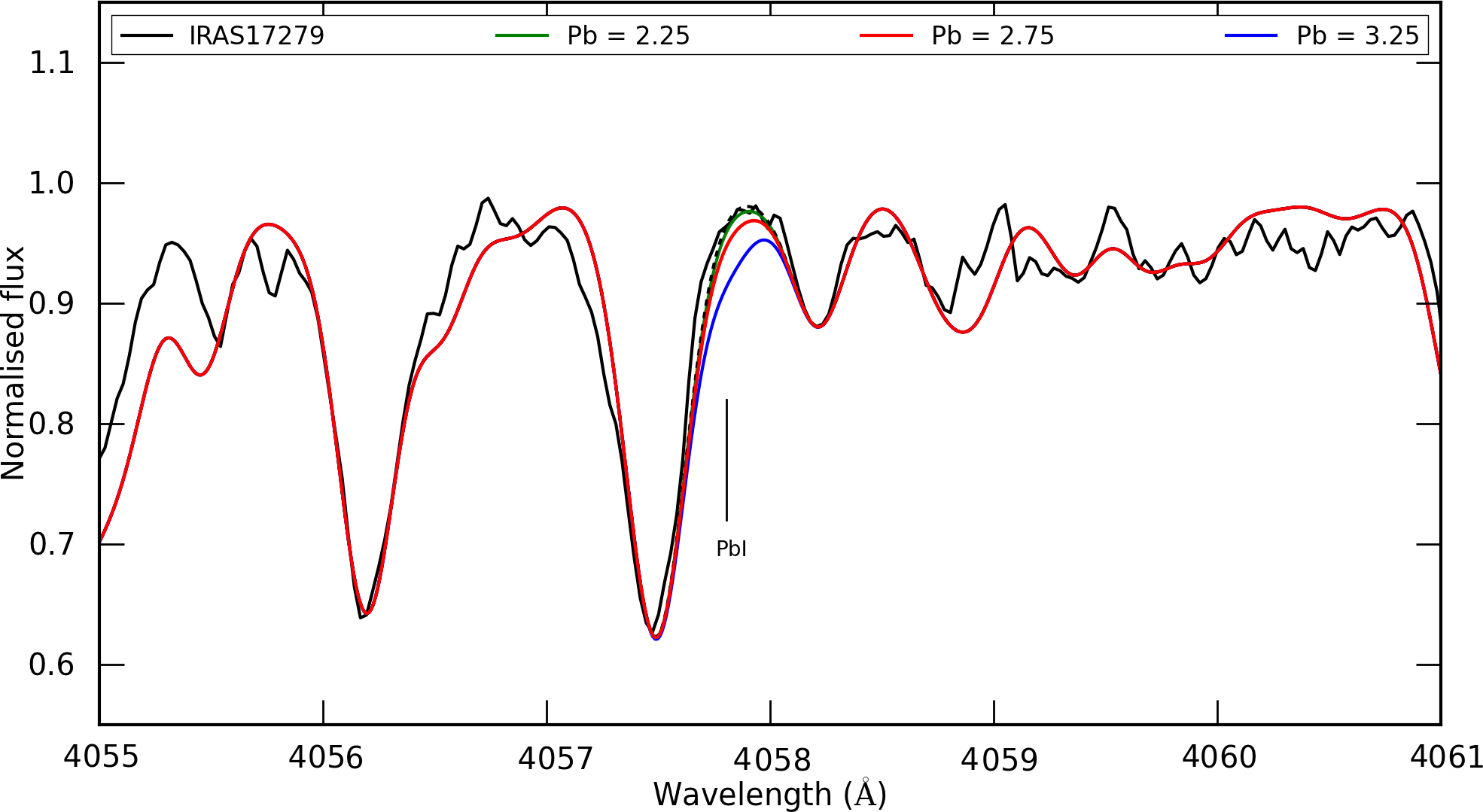}}
\caption{Spectrum synthesis of the Pb I line at 4057.807 \AA{} for 
the UVES spectrum of IRAS 17279-1119. Lines and symbols are similar to Fig. \ref{fig:iras05113_Pb}.
}\label{fig:iras17279_Pb}
\end{figure}

The mildly IRAS 17279-1119 does not show any visible Pb line features in Fig. \ref{fig:iras17279_Pb}. 
The local continuum is estimated using the 
spectral blends at 4056.2, 4057.5 and 4058.2 \AA{}. We adopt a Pb abundance upper limit, which fully 
incorporates the spectral region at 4057.807 \AA{}.

The [Pb$_{\textrm{up}}$/Fe] for the mildly \textit{s}-process enriched
post-AGB star IRAS 17279-1119 is 0.6 dex higher than [hs/Fe] (right
lower panel of Fig. \ref{fig:xfe_middle}).

\subsection{IRAS 19500-1709}
\begin{figure}[t!]
\resizebox{\hsize}{!}{\includegraphics{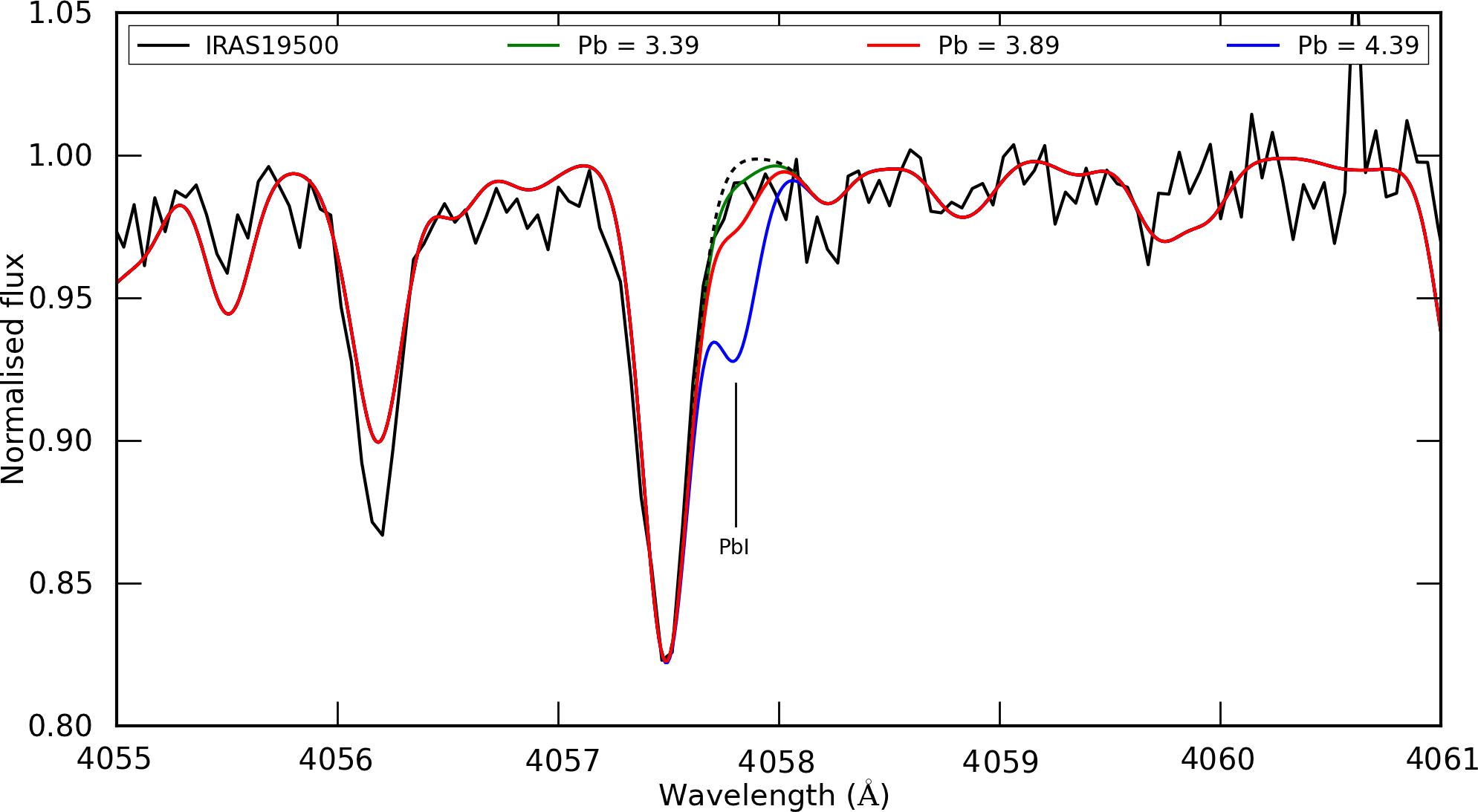}}
\caption{Spectrum synthesis of the Pb I line at 4057.807 \AA{} for
the HERMES spectrum of IRAS 19500-1709. Lines and symbols are similar to Fig. \ref{fig:iras05113_Pb}.
}\label{fig:iras19500_Pb}
\end{figure}

For the atmospheric parameter set of IRAS 19500-1709 to appear, the Pb I line at 4057.807 \AA{} requires 
a lower Pb abundance than the Pb II line at 5608.853 \AA{}.
Therefore, we only use the Pb I line for our analysis. Fig. \ref{fig:iras19500_Pb} shows that the 
spectral region of the Pb I line has a poor S/N. We use the spectral
blend at 4057.5 \AA{} to estimate the position of the continuum and then adopt a Pb abundance upper 
limit, which fully incorporates the noise at 4057.807 \AA{}.

For IRAS 19500-1709, [hs/ls] $\approx$ 0  and the derived 
[Pb$_{\textrm{up}}$/Fe] is about 1.35 dex higher 
than [ls/Fe] and [hs/Fe] (right lower panel of Fig. \ref{fig:xfe_hot}).

\subsection{IRAS 22223+4327}
The Pb abundance determination of IRAS 22223+4327 is described in detail in Sect. \ref{subsect:pb_abun}. The spectral synthesis of the Pb I region is depicted in Fig.~\ref{fig:iras22223_Pb}.
For IRAS 22223+4327, the adopted Pb$_{\textrm{up}}$ results in a
[Pb$_{\textrm{up}}$/Fe], which lies about 1.0 dex above [hs/Fe] (left
lower panel of Fig. \ref{fig:xfe_cold}). For IRAS 22223+4327, the ls
elements have higher enrichments than the hs elements and our estimate 
for [Pb$_{\textrm{up}}$/Fe] lies about 0.5 dex above [ls/Fe].

\subsection{IRAS 22272+5435}
\begin{figure}[t!]
\resizebox{\hsize}{!}{\includegraphics{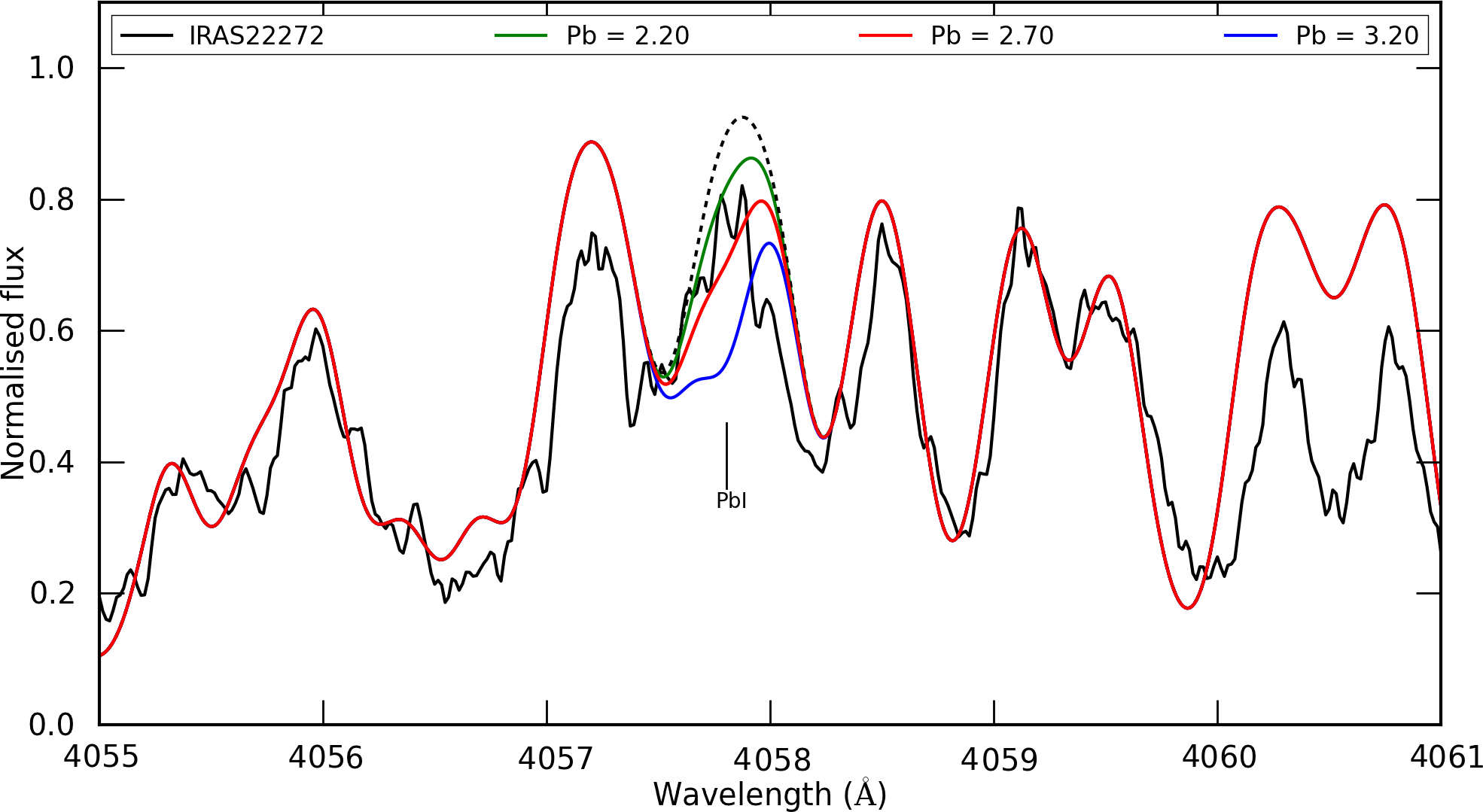}}
\caption{Spectrum synthesis of the Pb I line at 4057.807 \AA{} for 
the HERMES spectrum of IRAS 22272+5435. Lines and symbols are similar to Fig. \ref{fig:iras05113_Pb}.
}\label{fig:iras22272_Pb}
\end{figure}

The spectral region of the Pb I line for IRAS 22272+5435 is dominated by strong blends in combination 
with low S/N in Fig. \ref{fig:iras22272_Pb}. We use the 
spectral blends at 4057.5 and 4058.8 \AA{} to estimate the position of the continuum. We then adopt a 
Pb abundance upper limit that fully includes the visible 
line feature at 4057.807 \AA{}.

The adopted [Pb$_{\textrm{up}}$/Fe] of IRAS 22272+5435 lies between
the results of [hs/Fe] and [ls/Fe] (right lower panel of
Fig. \ref{fig:xfe_cold}).  The element-over-iron ratios of the
elements beyond the Ba-peak are also slightly higher than
[Pb$_{\textrm{up}}$/Fe].

\section{Comparison to previous studies}

In this appendix, we present the comparisons of our derived abundances
with those of previous studies. In this paper we rely on high-quality data and perform  strictly 
homogeneous abundance analyses based on the same methods and atomic data. The comparison with the literature is overall satisfactory
but some noticeable differences occur, especially when the original data were of inferior quality to our UVES and Hermes data.

\begin{figure}[t!]
\resizebox{\hsize}{!}{\includegraphics{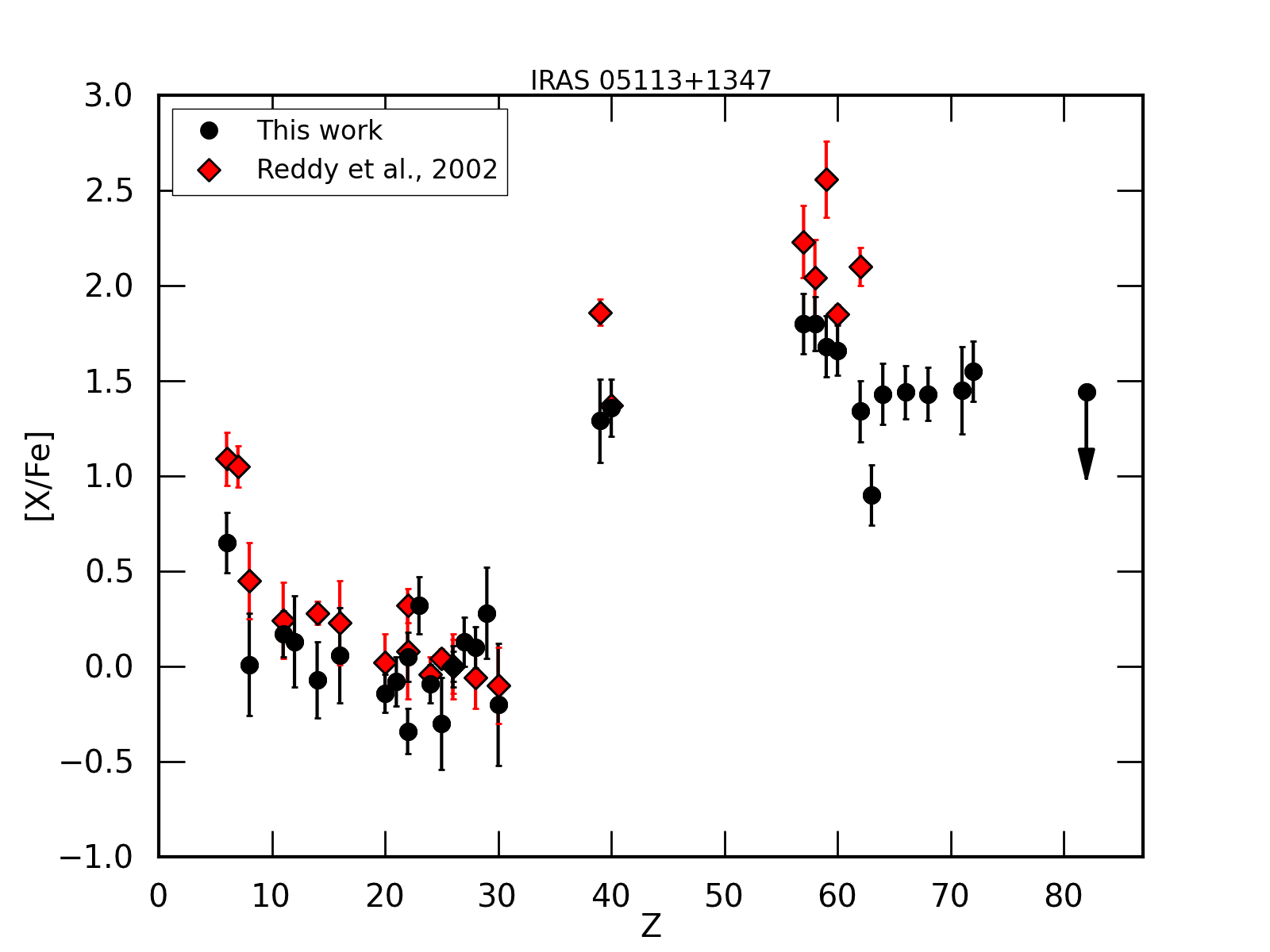}}
\caption{Comparison between our derived abundances and the results of \citet{reddy02} for IRAS 05113+1347.
}\label{fig:iras05113_comp}
\end{figure}

\begin{figure}[t!]
\resizebox{\hsize}{!}{\includegraphics{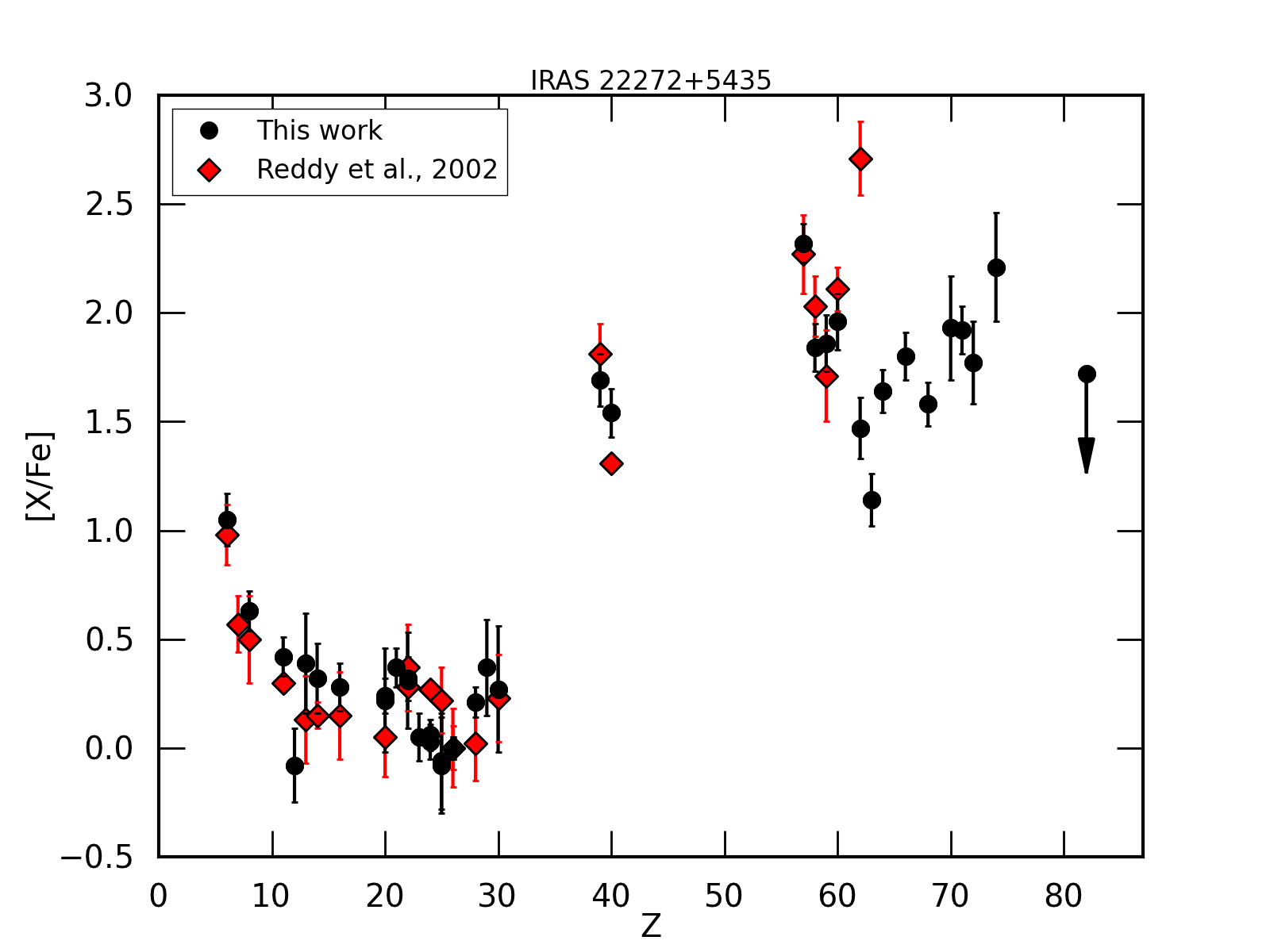}}
\caption{Comparison between our derived abundances and the results of \citet{reddy02} for IRAS 22272+5435.
}\label{fig:iras22272_comp}
\end{figure}

Objects IRAS 05113+1347 and IRAS 22272+5435 have been previously
studied by \citet{reddy02}; the comparisons between the abundance
results are presented in Figs. \ref{fig:iras05113_comp} and
\ref{fig:iras22272_comp}.  We find similar sets of atmospheric
parameters although our derived metallicity for IRAS 05113+1347 is
approximately 0.2 dex higher. We find a higher C/O ratio for IRAS
05113+1347 but lower abundances for Y and the heavy
s-elements. In particular,  Pr and Sm differ strongly, by almost 0.8
dex.  For Sm, the same occurs with IRAS 22272+5435, for which
\citet{reddy02} find an abundance that is about 1.25 dex higher than
our Sm abundance. For all other elements, the abundance differences
are significantly lower. Given the consistent distribution of the abundances presented here, and the superior S/N of our UVES spectra, we evaluate our new abundances to be more reliable.

\begin{figure}[t!]
\resizebox{\hsize}{!}{\includegraphics{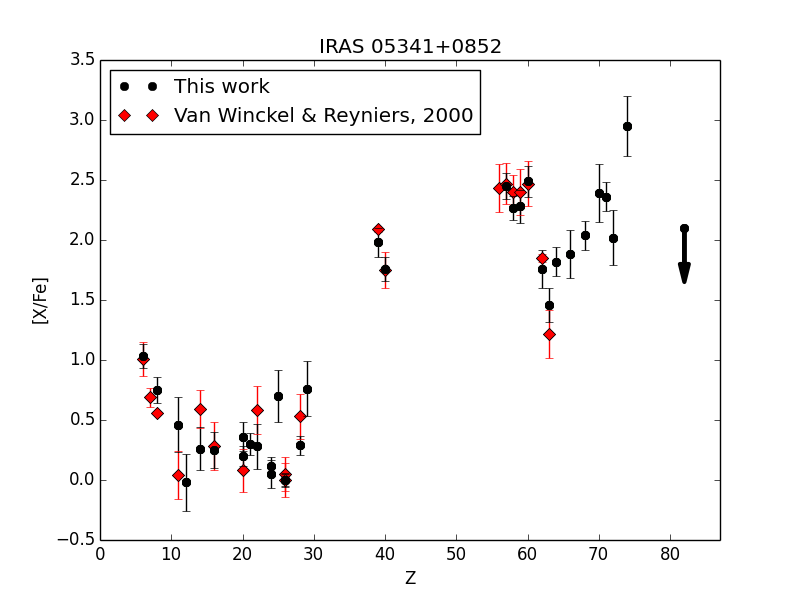}}
\caption{Comparison between our derived abundances and the results of \citet{vanwinckel00} for IRAS 05341+0852.
}\label{fig:iras05341_comp}
\end{figure}

\begin{figure}[t!]
\resizebox{\hsize}{!}{\includegraphics{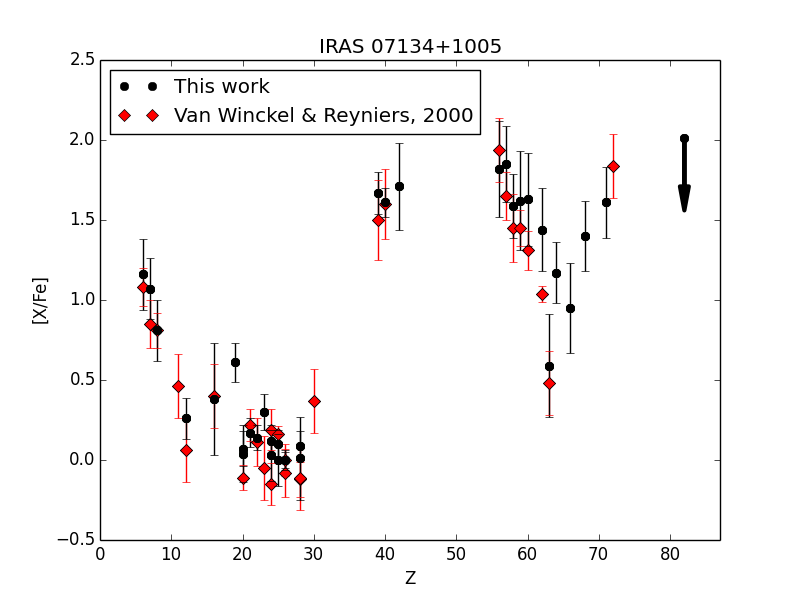}}
\caption{Comparison between our derived abundances and the results of \citet{vanwinckel00} for IRAS 07134+1005.
}\label{fig:iras07134_comp}
\end{figure}

\begin{figure}[t!]
\resizebox{\hsize}{!}{\includegraphics{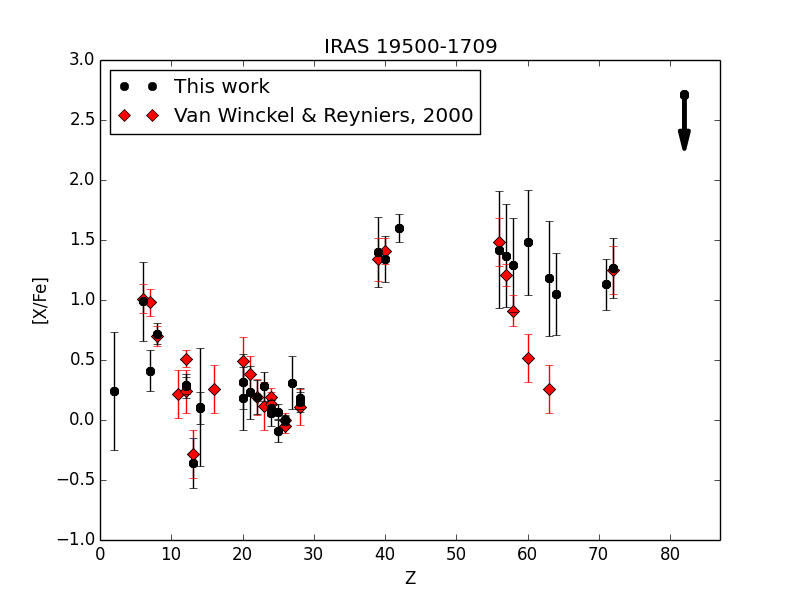}}
\caption{Comparison between our derived abundances and the results of \citet{vanwinckel00} for IRAS 19500-1709.
}\label{fig:iras19500_comp}
\end{figure}

For IRAS 05341+0852, IRAS 07134+1005, IRAS 19500-1709, and IRAS
22223+4327, we find similar atmospheric parameters as
\citet{vanwinckel00}. The strongest difference is the derived
temperature of IRAS 05341+0852, which is 250 K hotter in our study. 
We find similar results for IRAS 05341+0852
(Fig. \ref{fig:iras05341_comp}) and IRAS 07134+1005
(Fig. \ref{fig:iras07134_comp}), and only  our Sm result differs by about
0.4 dex from that of \citet{vanwinckel00}. For IRAS 19500-1709
(Fig. \ref{fig:iras19500_comp}), all results are similar except for Nd
and Eu, for which we find abundances that are a remarkable 0.9 dex higher. 
The large differences in those two elements are attributed to the use of a
different set of spectral lines to determine the
abundances. \citet{vanwinckel00}  used only one very small line in the blue ($\lambda$ 3863.408)
with a measured EW $\approx$ 3 m\AA{} to probe the Nd abundance and also used a line of  3 m\AA{}
for the Eu abundance.  The analyses of Nd  presented here is based on five lines 
with EWs ranging from about 5 up to 24 m\AA{}. We devised a much more systematic approach to our 
earlier work and find  the low Nd abundance of our earlier work, \citet{vanwinckel00}, to be suspicious
given the very small and maybe spurious line. Also, in our new abundance of Eu presented here, the value  
is only based on one line; this is different than our earlier work with an EW $\approx$ 16 m\AA{}.  

\begin{figure}[t!]
\resizebox{\hsize}{!}{\includegraphics{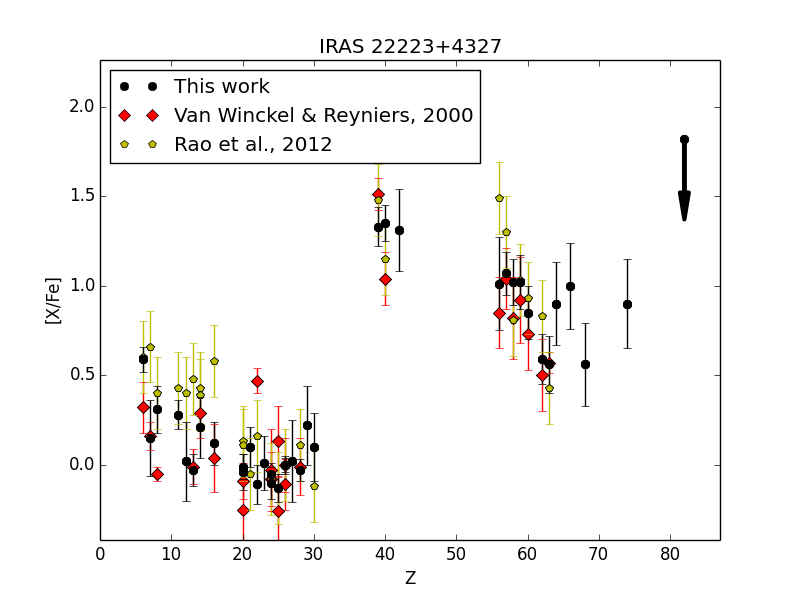}}
\caption{Comparison between our derived abundances and the results of \citet{vanwinckel00} and \citet{rao11} for IRAS 22223+4327.
We adopt a standard deviation of 0.2 dex for all results of \citet{rao11}.
}\label{fig:iras22223_comp}
\end{figure}

Our abundance
results of IRAS 22223+4327 (Fig. \ref{fig:iras22223_comp}) are similar
to the determined abundances from \citet{vanwinckel00} and
\citet{rao11}. All three independent studies confirm the overabundance
of ls elements with respect to the hs elements of IRAS 22223+4327.

\begin{figure}[t!]
\resizebox{\hsize}{!}{\includegraphics{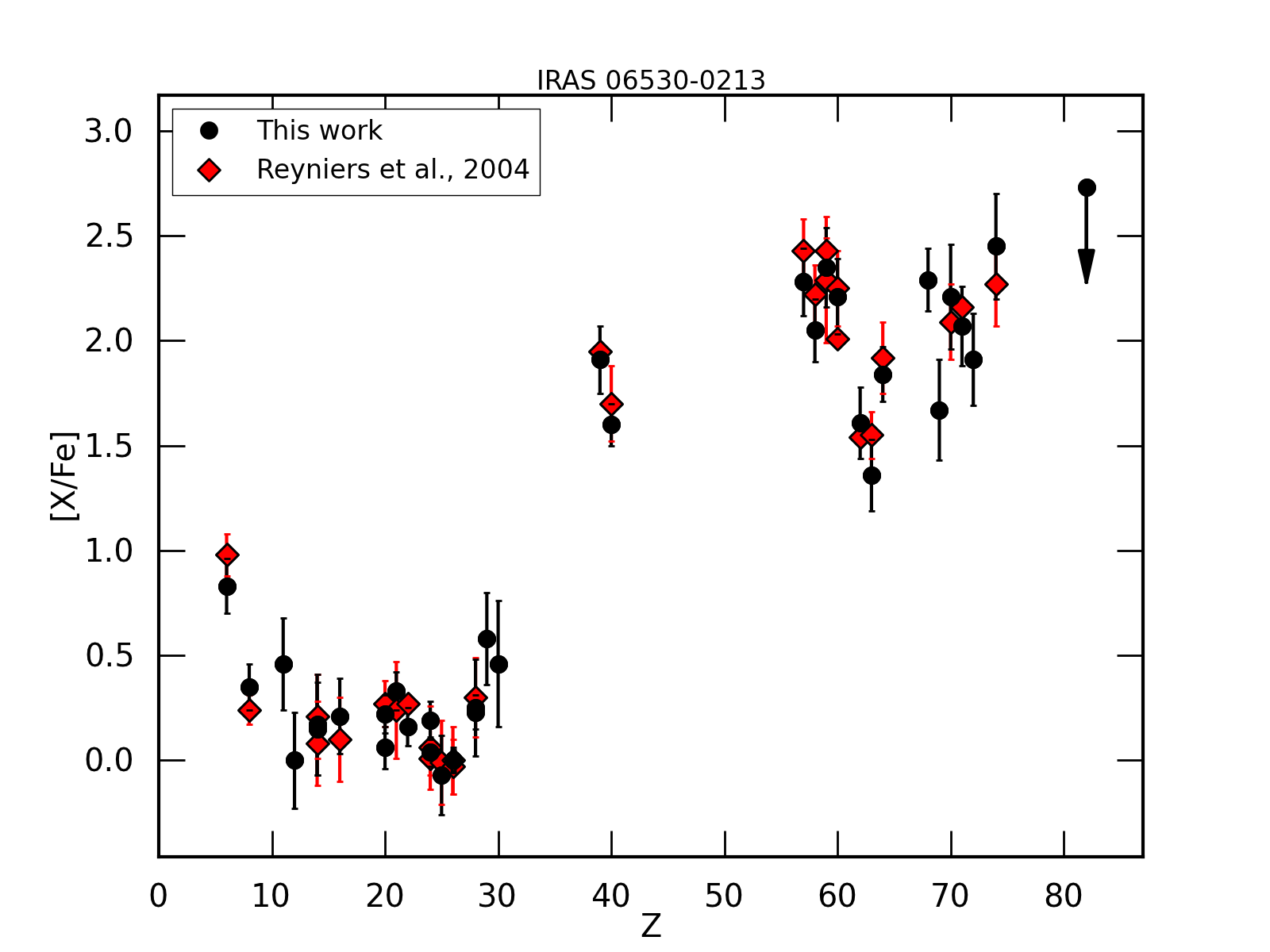}}
\caption{Comparison between our derived abundances and the results of \citet{reyniers04} for IRAS 06530-0213.
}\label{fig:iras06530_comp}
\end{figure}

\begin{figure}[t!]
\resizebox{\hsize}{!}{\includegraphics{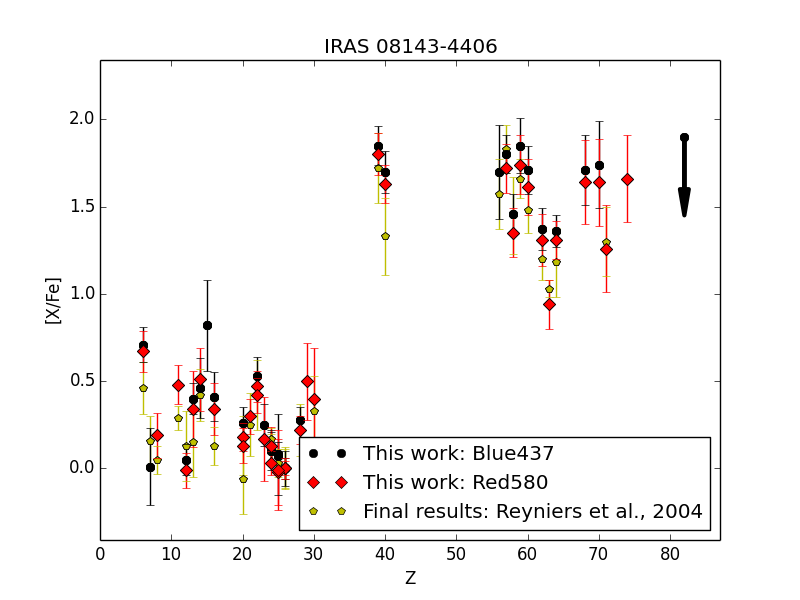}}
\caption{Comparison between our derived abundances for both observational settings of IRAS08143-4406 and the final results of \citet{reyniers04}.
}\label{fig:iras08143_comp}
\end{figure}

Our atmospheric parameter results for IRAS 06530-0213 and IRAS
08143-4406 (Figs. \ref{fig:iras06530_comp} and
\ref{fig:iras08143_comp}) confirm the derived atmospheric parameter
sets of \citet{reyniers04}. Since only the combined [X/Fe] results 
for the two observational settings for IRAS 08143-4406 are provided in 
\citet{reyniers04}, we compare our individual results for the two settings
with these combined results. For both stars, our results are similar. The
strongest difference is our higher Zr abundance of about +0.3 dex
with respect to \citet{reyniers04}.

\begin{figure}[t!]
\resizebox{\hsize}{!}{\includegraphics{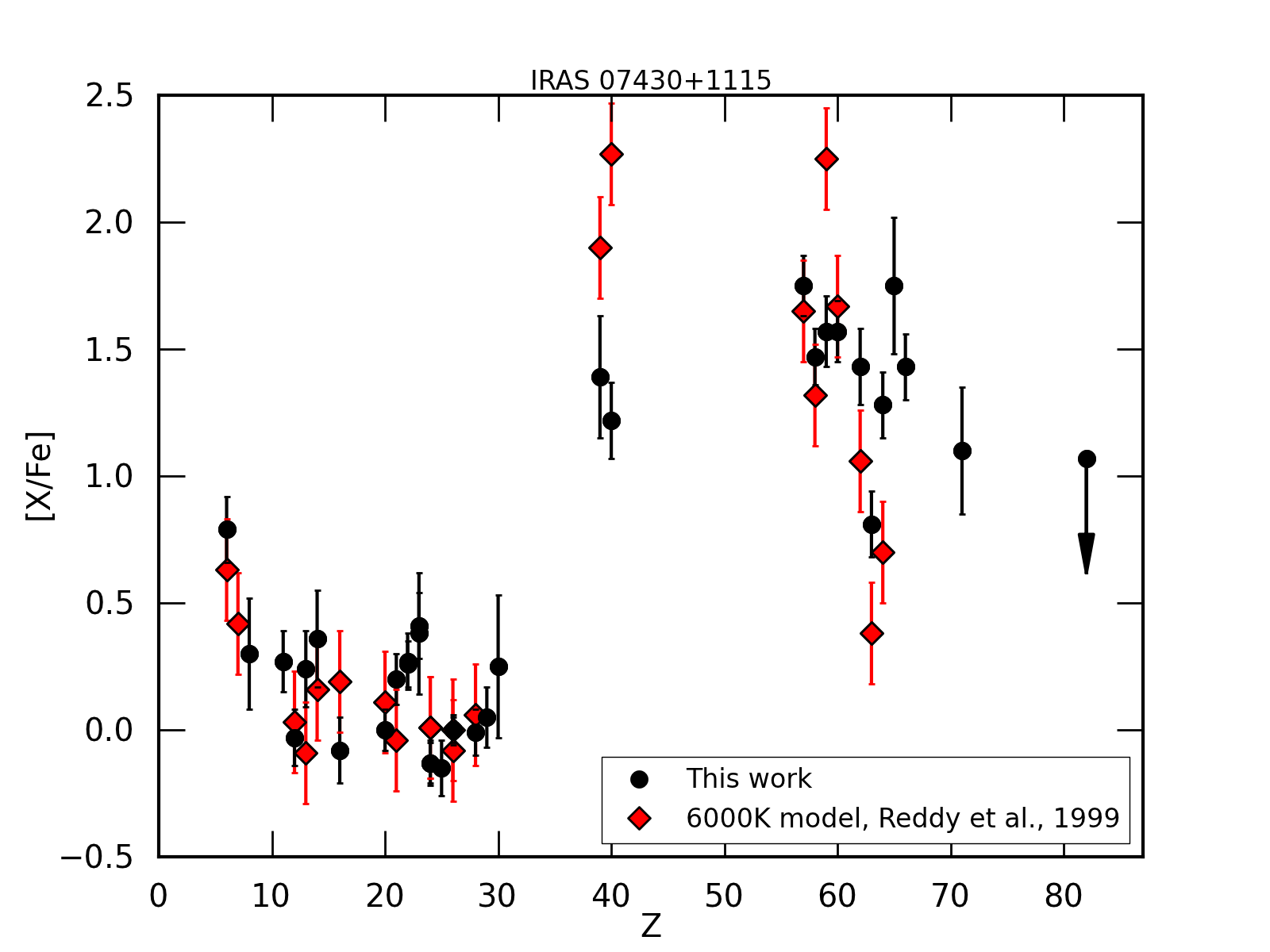}}
\caption{Comparison between our derived abundances and the results of \citet{reddy99} for IRAS 07430+1115.
We adopt a standard deviation of 0.2 dex for all results of \citet{reddy99}.
}\label{fig:iras07430_comp}
\end{figure}

For IRAS 07430+1115 (Fig. \ref{fig:iras07430_comp}), we find large
abundance differences with respect to \citet{reddy99} for almost all
\textit{s}-process elements, even though we used almost exactly the same
atmospheric parameter set. \citet{reddy99} find Y and Zr abundances
that are  $\pm$ 0.5 and $\pm$ 1.0 dex higher than our
results, respectively. \citet{reddy99} find lower abundances of about
0.3 dex for Sm and Eu . Except for one Eu line, we have used a different set of spectral 
lines than \citet{reddy99} for the analysis of Y, Zr, Sm, and Eu because 
many of these lines are part of a spectral blend in our UVES spectra. 
If these blends are included in the EW calculation, they result in EWs 
that are much higher than the actual EW, which may explain the difference between our abundance results 
and those of \citet{reddy99}. For the only common Eu, we find 
a difference of -0.07 dex between our log gf value and those of 
\citet{reddy99}, which may partly explain the abundance difference.

\begin{figure}[t!]
\resizebox{\hsize}{!}{\includegraphics{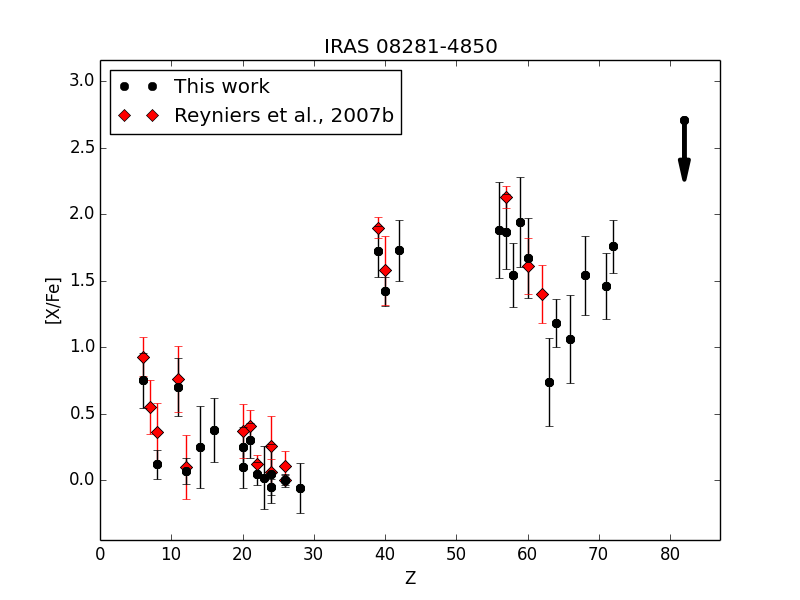}}
\caption{Comparison between our derived abundances for both observations of IRAS 08281-4850 and the final results of \citet{reyniers07c}.
}\label{fig:iras08281_comp}
\end{figure}

\begin{figure}[t!]
\resizebox{\hsize}{!}{\includegraphics{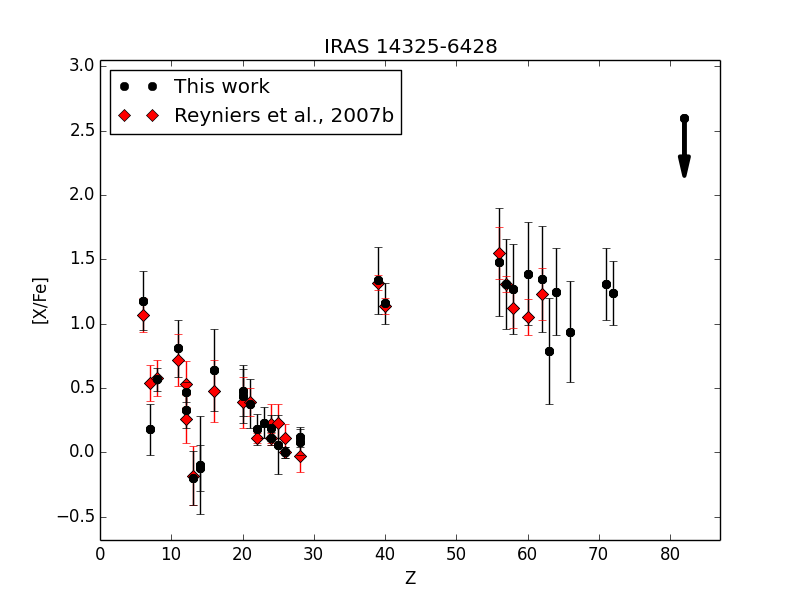}}
\caption{Comparison between our derived abundances for both observations of IRAS 14325-6428 and the final results of \citet{reyniers07c}.
}\label{fig:iras14325_comp}
\end{figure}

The atmospheric parameter results of \citet{reyniers07c} for IRAS
08281-4850 and IRAS 14325-6428 are similar to our results, although
they find a microturbulent velocity, which is 4 km/s higher than our
preferred result for IRAS 14325-6428. We find similar results for IRAS
08281-4850 (Fig. \ref{fig:iras08281_comp}) and also for IRAS
14325-6428, (Fig. \ref{fig:iras14325_comp}) despite the large microturbulence
difference.

\begin{figure}[t!]
\resizebox{\hsize}{!}{\includegraphics{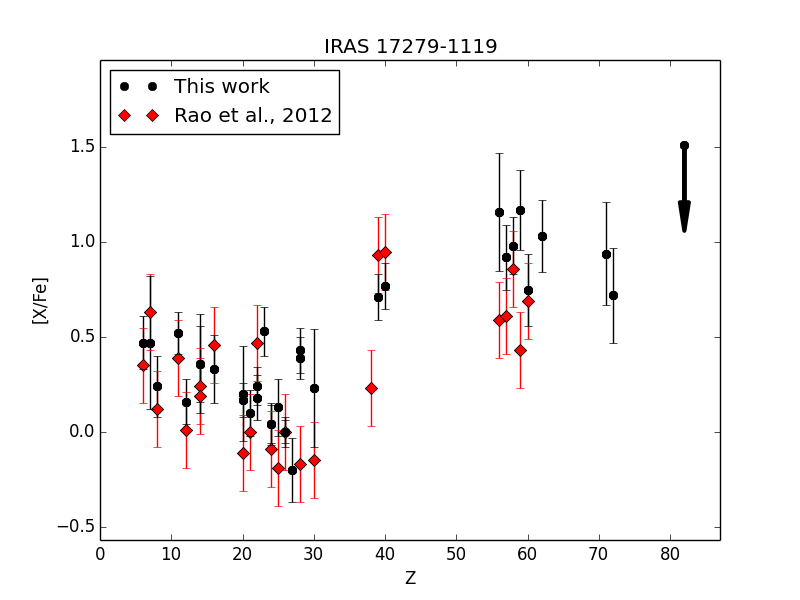}}
\caption{Comparison between our derived abundances and the results of \citet{rao11} for IRAS 17279-1119.
We adopt a standard deviation of 0.2 dex for all results of \citet{rao11}.
}\label{fig:iras17279_comp}
\end{figure}

As discussed above, our results for IRAS 22223+4327 correspond to those
of \citet{rao11}. In case of IRAS 17279-1119
(Fig. \ref{fig:iras17279_comp}), we find similar atmospheric
parameters as \citet{rao11} except for our $\log g,$ which differs by +1.0
dex from that of \citet{rao11}. We find a stronger enrichment of the hs
elements resulting in a mean [hs/Fe] of about 0.9 dex, while
\citet{rao11} find [hs/Fe] $\approx$ 0.6 dex. Nevertheless, we still
classify this object as mildly \textit{s}-process enriched ([s/Fe] < 1) based on
our abundance results.

\end{appendix}

\end{document}